\documentclass[journal]{IEEEtran}

\usepackage{color,graphicx,amsmath,amssymb,amsthm,epsfig,mathrsfs,cite,bm,graphics}

\usepackage{caption}\usepackage{algpseudocode}

\usepackage{algorithmicx}
\usepackage{algorithm} 
\usepackage[utf8]{inputenc}
\usepackage{amsmath}
\usepackage{amsfonts}
\usepackage{amssymb}
\usepackage{placeins}
\usepackage{graphicx,tikz}

\usetikzlibrary{patterns,arrows,decorations.pathreplacing}
\usepackage{graphicx,graphics,subcaption}

\usepackage{tikz}
\usetikzlibrary{shapes.geometric, arrows}
\usetikzlibrary{shapes.multipart}
\tikzstyle{startstop} = [rectangle, rounded corners, minimum width=3cm, minimum height=1cm,text centered, draw=black, fill=gray!30]
\tikzstyle{io} = [trapezium, trapezium left angle=70, trapezium right angle=110, minimum width=3cm, minimum height=1cm, text centered, draw=black, fill=blue!30]
\tikzstyle{connector} = [circle,draw=black,fill=gray!30]
\tikzstyle{process} = [rectangle, minimum width=0.5cm, minimum height=1cm, text centered, text width=6.5cm,  draw=black, fill=gray!30]
\tikzstyle{decision} = [diamond, minimum width=1cm, minimum height=1cm, text centered, text width=2.5cm,draw=black, fill=gray!30]

\tikzstyle{arrow} = [thick,->,>=stealth]

\usetikzlibrary{positioning,shadows,arrows}
\usetikzlibrary{calc,positioning,shapes,decorations.pathreplacing}

\usetikzlibrary{patterns}
\usetikzlibrary{shapes,arrows}

\usetikzlibrary{positioning}
\usetikzlibrary{shapes.geometric}
\usetikzlibrary{shapes.misc}
\usetikzlibrary{automata,positioning}

\newcommand{\revOneDel}[1]{}

\title{Spectrum Access In Cognitive Radio Using A Two Stage Reinforcement Learning Approach}

\author{ Vishnu Raj \hspace{16pt} Irene Dias \hspace{16pt} Thulasi Tholeti \hspace{16pt} Sheetal Kalyani\\
    \hspace{-0 cm}Department of Electrical Engineering \\ Indian Institute of Technology, Madras \\
    Chennai, India 600 036 \\
    \texttt{\{ee14d213,ee15m068,ee15d410,skalyani\}@ee.iitm.ac.in}
}

\begin{document}
	\maketitle
	\begin{abstract}
	    With the advent of the 5th generation of wireless standards and an increasing demand for higher throughput, methods to improve the spectral efficiency of wireless systems have become very important. In the context of cognitive radio, a substantial increase in throughput is possible if the secondary user can make smart decisions regarding which channel to sense and when or how often to sense. Here, we propose an algorithm to not only select a channel for data transmission but also to predict how long the channel will remain unoccupied so that the time spent on channel sensing can be minimized. Our algorithm learns in two stages - a reinforcement learning approach for channel selection and a Bayesian approach to determine the optimal duration for which sensing can be skipped. Comparisons with other learning methods are provided through extensive simulations. We show that the number of sensing is minimized with negligible increase in primary interference; this implies that lesser energy is spent by the secondary user in sensing and also higher throughput is achieved by saving on sensing.
    \end {abstract}
    
    \begin{IEEEkeywords} 
        Cognitive Radio, Multi-armed Bandit, Opportunistic Spectrum Sensing, Reinforcement Learning, Bayesian Learning
    \end{IEEEkeywords}
	
	\section{Introduction}      \label{sec:introduction}
	
	Cognitive radio(CR) was one of the key ideas introduced to overcome spectrum scarcity, back in 1999  \cite{Mitola1999}. It has drawn a lot of attention ever since its inception as it promises a huge increase in spectrum utilization which is crucial for supporting higher data rates as well as hosting more number of  users. Spectrum utilization gains more context now than ever with the advent of 5G wireless technology and Internet of Things. Maximizing spectrum utilization is imperative for satisfying the ambitious demands of 5G systems - immense capacity, high reliability and the perception of limitless connectivity. To achieve this, we need to leverage the information available in the network using algorithms that can learn from the available data. CR improves spectrum utilization by tapping into licensed spectrum when it is underutilized. Secondary users (SUs) attempt to transmit their data by sensing holes in the primary traffic. Each SU is required to sense the available channels, typically using an energy detector \cite{digham2007energy,atapattu2011energy,ghasemi2007optimization,liu2006sensing}, to ensure that it is not occupied before transmitting so as to minimize the interference caused to the primary user (PU). Therefore, one of the crucial aspects of a cognitive radio system is to attain maximum secondary user throughput whilst keeping the interference caused to the PU below a pre-specified threshold. Often, strict constraints are imposed on the maximum interference that the primary user can tolerate. 
	
	Initially, a SU needs to select a channel for transmission. Naively, the SU could pick a channel to sense at random and transmit data if it is found to be idle. On the other hand, multi-slot sensing can also be performed; if a channel sensed is busy, another channel is sensed until one is found to be idle. We make the observation that if a pattern could be associated with the primary traffic, the predictability of primary traffic is improved.
	
	Various methods based on learning, which keep track of the channel occupancy, have been proposed to ensure that the channel that is most likely to be idle is chosen. One of the methods of posing the optimal channel selection problem is as a multi-armed bandit(MAB) problem where the different channels to be picked act as arms of the bandit \cite{Jouini2009}. In \cite{jiang2009optimal}, the authors use a dynamic programming approach to choose the optimal channel sensing order when adaptive modulation is employed. Reinforcement Learning (RL) algorithms like Q Learning \cite{Li2009} and Artificial Neural Networks \cite{hou2016throughput} have also been shown to improve CR performance. Some of the other methods for spectrum sensing suggested in \cite{xing2013spectrum} include Multi-layer Perceptron based neural networks, Bayesian inference methods, Auto-Regressive model based approach, static neighbour graphs and moving average based methods. In \cite{Xu2013a}, a survey of various decision-theoretic approaches such as game theory, Markov Decision Processes, MABs, etc. for performing channel selection is given. From our survey of the papers that use learning in cognitive radio\cite{Anandkumar2011,Zhu2016,Gai2010}, we note that they typically address the effect of their algorithms on throughput and regret but do not focus on time spent on sensing. A quiet period is assumed in the beginning of every frame that is used for sensing the channel\cite{Liu2010}. A recurrent sensing period for every frame not only affects the throughput and the energy efficiency (since the SU expends considerable energy in sensing the channel) of the SU but also results in the under-utilization of the spectrum which is a major issue, with the ever-increasing demand for capacity.
	
	Judicious sensing is quite important for achieving maximum throughput as time utilized for channel sensing is time lost for data transmission. In other traditional cognitive radio papers like \cite{liang2008sensing}, the authors formulate an optimization problem to maximize throughput given a constraint on the probability of detection and prove that an optimal sensing time exists for a fixed frame size. On the other hand, \cite{pei2007sensing} considers a transmission scheme where the sensing duration is fixed, and an expression for the optimal frame size that maximizes normalizes throughput is derived; this determines how often sensing is carried out. While in \cite{liang2008sensing,pei2007sensing}, the time spent on sensing is taken into consideration, the expressions for optimal frame size/sensing time depend on the parameters of the primary traffic which might not be available to SU or might be changing with time. Further, the above mentioned works assume that the ON/OFF times are sampled from an exponential distribution in both their derivations and simulations. Work in \cite{Oksanen2015} does consider primary traffic with unknown parameters; however, this work again concentrates exclusively on channel selection and not on time spent on sensing.
	
	Summarizing, the works that adopt reinforcement learning techniques in cognitive radio to learn the primary traffic behaviour do not consider optimizing the sensing time; they assume a sensing period is available for every frame. The works that aim to optimize sensing time assume knowledge of the primary user statistics and mostly consider an Exponential PU traffic model.  We believe that to maximize the throughput of the SU, algorithms that learn which channel to sense should also learn how often to sense. Addressing both issues simultaneously will lead to better performance of the cognitive radio system and this is the key focus of this work.
	 
	It is obvious that if channel sensing can be skipped in a frame or more, the benefit in throughput and energy efficiency would be remarkable; but it might adversely affect the interference caused to the primary user. In that case, we raise the following question: how often should we sense? In our work, we introduce a novel method to skip sensing which is based on learning the behaviour of the primary traffic over time through a Bayesian approach. Our work proposes a novel algorithm to maximally exploit the spectrum by picking the channel to sense using a multi-armed bandit, and then sensing the selected channel at specific intervals using a Bayesian method to decide the interval duration between two sensing operations. We believe that introducing two layers of learning will ensure higher spectrum utilization than the existing methods. 
	When compared to existing RL methods, our approach significantly reduces the amount of sensing required and leads to increase in both throughput and energy efficiency. We provide extensive comparison results with other existing learning algorithms through simulations and show that the proposed approach is superior in terms of its performance. Further, for simulations, we have considered the popular DTMC model for discrete time primary user traffic and both Generalized Pareto Model and Exponential model for continuous traffic since in \cite{lopez2012spectrum} it has been shown that the generalized Pareto distribution captures the primary user traffic variations more accurately.
	
	This paper is organized as follows: Section \ref{sec:SystemModel} describes the system model that is considered for our experiments. Section \ref{sec:Background} gives a brief overview of  reinforcement learning algorithms. The proposed method for channel sensing is described in section \ref{sec:PropApproach}. The simulations performed and the results observed therein are discussed in section \ref{sec:SimRes}. Finally, section \ref{sec:Conc} presents the concluding remarks.

	\section{System Model} \label{sec:SystemModel}
	
	In our system model, we consider N channels that can be accessed by the secondary user. The secondary user wishes to transmit data packets so as to maximize its own throughput whilst causing the least interference to the licensed users. The channel is sensed to check whether it is idle or busy before the data is sent, unless we are positive that the channel is available for transmission. This depends on the channel model that is used. We assume that the sensing operation is repeated until we find a channel that is idle or we exhaust all available channels. The aim is to reduce the number of sensing operations that need to be performed so that a larger part of the frame is available for data transmission.

    To do so, we employ learning algorithms that predict the current state of the channel based on the previously observed states. The learning algorithm returns a ranked list of channels in the order of estimated probability of occupancy. The sensor then senses the channels in that order. Once a channel is found to be idle, the secondary user transmits its data. Upon transmission, we encounter one of the two scenarios: the transmission is unsuccessful either due to collision or error in the channel, or the packet is delivered successfully. If a packet drop/collision occurs, we achieve a zero throughput at the receiver end, else, high throughput is achieved. In the rest of the section, we focus on  describing the frame structure used by the secondary users and the primary user traffic model.

   The data is transmitted by the secondary user in frames of fixed size, $T$. Each frame that is transmitted has a sensing duration and a transmission duration. During the sensing duration the secondary user senses the channels either randomly or based on input from the learning algorithms. It is assumed that each channel sensing takes time $\tau$. Once a channel is sensed to be free, then data is transmitted for the rest of the frame. The sensing in each frame can be done in two ways: single slot sensing and multi-slot sensing as shown in Figures 1 and 2. In the first frame type, only one channel is sensed in each frame(for a duration $\tau$) and if this channel is found to be vacant, the user transmits for a duration of $T - \tau$; else, transmission does not happen for that frame. But this method denies the secondary user a chance to find another free channel that might have been vacant during that time. 
   
   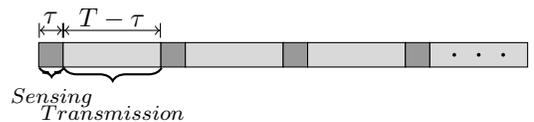
\begin{figure}[h]
       \centering
\ifCLASSOPTIONonecolumn
        \begin{tikzpicture}[scale=1.55]
\else
        \begin{tikzpicture}[scale=0.65]
\fi
            [
            node distance=0.5mm
            ]
            \draw [fill = gray!80!white](1.0,1.0) rectangle (1.5,1.5);
            \draw [fill = gray!30!white](1.5,1.0) rectangle (3.5,1.5);
            \draw [fill = gray!80!white](3.5,1.0) rectangle (4.0,1.5);
            \draw [fill = gray!30!white](4.0,1.0) rectangle (6.0,1.5);
            \draw [fill = gray!80!white](6.0,1.0) rectangle (6.5,1.5);
            \draw [fill = gray!30!white](6.5,1.0) rectangle (8.5,1.5);
            \draw [fill = gray!80!white](8.5,1.0) rectangle (9.0,1.5);
            \draw [fill = gray!30!white](9.0,1.0) rectangle (11.0,1.5);
            \node[fill,circle,inner sep = 0.5pt] at (9.5,1.25){};
            \node[fill,circle,inner sep = 0.5pt] at (10.0,1.25){};
            \node[fill,circle,inner sep = 0.5pt] at (10.5,1.25){};
            \draw [|<->|] (1.0,1.75) -- (1.5,1.75);
            \node at (1.25,2.00) {$\tau$};
            \draw [|<->|] (1.5,1.75) -- (3.5,1.75);
            \node at (2.5,2.00) {$T - \tau$};
            \draw [thick,decorate,decoration={brace,amplitude=3.5pt,mirror},xshift=0.4pt,yshift=-0.4pt](1,1.0) -- (1.5,1.0) node[black,midway,yshift=-0.4cm] {\footnotesize $Sensing$};
            \draw [thick,decorate,decoration={brace,amplitude=5pt,mirror},xshift=0.4pt,yshift=-0.4pt](1.5,1.0) -- (3.5,1.0) node[black,midway,yshift=-0.6cm] {\footnotesize $Transmission$};
            
        \end{tikzpicture} \\
       \caption{Single slot sensing frame}
       \label{fig:Frame1}
   \end{figure}
    
    \begin{figure}[h]
        \centering
        \captionsetup{justification=centering}
\ifCLASSOPTIONonecolumn
        \begin{tikzpicture}[scale=1.25]
\else
        \begin{tikzpicture}[scale=0.65]
\fi
            [
            node distance=0.5mm
            ]
            \draw [fill = gray!80!white](1.0,1.0) rectangle (1.4,1.5) node[pos=.5] {\tiny{$S_1$}};
            \draw [fill = gray!80!white](1.4,1.0) rectangle (1.8,1.5) node[pos=.5] {\tiny{$S_2$}};
            \draw [fill = gray!80!white](1.8,1.0) rectangle (2.6,1.5);
            \node[fill,circle,inner sep = 0.5pt] at (2.15,1.25){};
            \node[fill,circle,inner sep = 0.5pt] at (2.25,1.25){};
            \node[fill,circle,inner sep = 0.5pt] at (2.35,1.25){};
            \draw [fill = gray!80!white](2.6,1.0) rectangle (3.0,1.5) node[pos=.5] {\tiny{$S_{k_1}$}};
            \draw [fill = gray!30!white](3.0,1.0) rectangle (7.0,1.5);
            \draw [|<->|] (1.0,1.75) -- (3.0,1.75);
            \node at (2.0,2.00) {$k_1\tau$};
            \draw [|<->|] (3.0,1.75) -- (7.0,1.75);
            \node at (5.0,2.00) {$T - k_1\tau$};
            \draw [thick,decorate,decoration={brace,amplitude=5pt,mirror},xshift=0.4pt,yshift=-0.4pt](1,1.0) -- (3.0,1.0) node[black,midway,yshift=-0.6cm] {\footnotesize $Sensing$};
            \draw [thick,decorate,decoration={brace,amplitude=5pt,mirror},xshift=0.4pt,yshift=-0.4pt](3.0,1.0) -- (7.0,1.0) node[black,midway,yshift=-0.6cm] {\footnotesize $Transmission$};

            \draw [fill = gray!80!white](7.0,1.0) rectangle (7.4,1.5) node[pos=.5] {\tiny{$S_1$}};
            \draw [fill = gray!80!white](7.4,1.0) rectangle (7.8,1.5) node[pos=.5] {\tiny{$S_2$}};
            \draw [fill = gray!80!white](7.8,1.0) rectangle (8.6,1.5);
            \node[fill,circle,inner sep = 0.5pt] at (8.15,1.25){};
            \node[fill,circle,inner sep = 0.5pt] at (8.25,1.25){};
            \node[fill,circle,inner sep = 0.5pt] at (8.35,1.25){};
            \draw [fill = gray!80!white](8.6,1.0) rectangle (9.0,1.5) node[pos=.5] {\tiny{$S_{k_2}$}};
            \draw [fill = gray!30!white](9.0,1.0) rectangle (13.0,1.5);
            \draw [|<->|] (7.0,1.75) -- (9.0,1.75);
            \node at (8.0,2.00) {$k_2\tau$};
            \draw [|<->|] (9.0,1.75) -- (13.0,1.75);
            \node at (11.0,2.00) {$T - k_2\tau$};
            
        \end{tikzpicture} \\
        \caption{Multi slot sensing frame}
        \label{fig:Frame2}
    \end{figure}
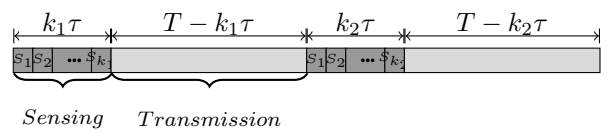
    In the multi-slot sensing frames the secondary user keeps on sensing the channels till it finds a vacant channel to transmit. Assume that for the $i_{th}$ frame, $k_i$ channels are sensed. Therefore, the total sensing time is $k_i \tau$. Data is transmitted for $T- k_i \tau$. In case sensing is entirely skipped for a frame $i$, $k_i = 0$; data is transmitted for the entire frame duration $T$. Multi-slot sensing allows the secondary user to transmit its data more often than the single-slot scheme as the SU can sense different channels until an idle channel is found.
    
    We now move on to another important component of the system model: the structure of the primary traffic. The primary user traffic on $N$ channels is modeled to be independent. We take two approaches to modeling the primary traffic: the discrete-time model and the continuous-time model. This is done to show that the proposed algorithm performs well in a wide range of primary traffic scenarios.

    In the Discrete-Time Markov Chain(DTMC) model, the time index set is discrete. \cite{zhao2007decentralized} model their primary traffic as a Markov chain with $2^N$ states and show that we can consider a simplified 2-state (idle and busy) for a given channel that evolves independently. The behaviour of the channel can be expressed by a transition probability matrix as given below.
    \begin{align}
        P = 
        \begin{bmatrix}
            p_{00} & p_{01} \\
            p_{10} & p_{11}
        \end{bmatrix}
    \end{align}
    where $p_{ij}$ represents the probability that the system transitions from state $s_i$ to $s_j$. The discrete formulation of the primary traffic is also adopted in \cite{Filippi2011,G2013,Oksanen2015}.
    
    Duty Cycle($\Psi$)  is defined as the probability that the channel is busy and can be written as
    $P(S=s_0) = 1 - \Psi$ and $P(S=s_1) = \Psi$. The DTMC model can be used to reproduce any arbitrary
    DC, $\Psi$, by selecting the transition probabilities as $p_{01} = p_{11} = \Psi$ and $p_{00} =
    p_{10} = 1 - \Psi$, which yields,
    \begin{align}
        P =
            \begin{bmatrix}
                1 - \Psi & \Psi \\
                1 - \Psi & \Psi
            \end{bmatrix}
    \end{align}
    For channels with varying load, the $P$ matrix will also vary with time and the duty cycle will
    become a time varying quantity, $\Psi(t)$. The channel load under this model is considered to be a random variable, which can be characterised by its PDF. The empirical PDFs of $\Psi$ in real systems can be accurately fitted with the Beta distribution(Eqn.\ref{eqn:beta_pdf}) or the Kumaraswamy distribution(Eqn.\ref{eqn:k_pdf}) as suggested by \cite{lopez2012spectrum}. The Beta distribution is given by
    \begin{align}    \label{eqn:beta_pdf}
        f_{X}^{B}(x;\alpha,\beta) &= \frac{1}{B(\alpha,\beta)} x^{\alpha-1} (1-x)^{\beta-1}, \qquad x \in (0,1) 
    \end{align}
    where $\alpha > 0$, $\beta > 0$ and $B(\alpha,\beta)$ is the Beta function.
        
    The Kumaraswamy distribution is given by
    \begin{align}    \label{eqn:k_pdf}
        f_{X}^{K}(x;a,b) &= a b x^{a-1} (1-x^a)^{b-1}, \qquad x \in (0,1), a,b>0
    \end{align}

    The work in \cite{stabellini2010quantifying} which studies modelling the spectrum opportunities suggests that Generalized Pareto Distribution(GPD) is a good approximation to the observed measurements.  In the continuous time model, GPD is a heavy-tailed distribution that allows us to model a wider range of traffic when compared with an exponential distribution. Hence, for the continuous-time model, we assume that the on and off times are distributed according to GPD. The probability density function of a GPD is given by 
    \begin{equation}
        f(x |k,\sigma,\theta) = \dfrac{1}{\sigma} \left( 1+ k \dfrac{x - \theta}{\sigma} \right) ^{-1 - \frac{1}{k}} \hspace{0.5cm} \theta <x ; k > 0
    \end{equation}
    where $k$, $\sigma$ and $\theta$ are the shape, scale and location parameters respectively. Varying traffic load can be generated by varying the parameters of GPD. For example, fast-varying traffic can be modelled by increasing the scale parameter whereas percentage of occupancy can be modelled by varying the location parameter of the ON and OFF times.

    \begin{figure}
        \includegraphics[width=1.\linewidth]{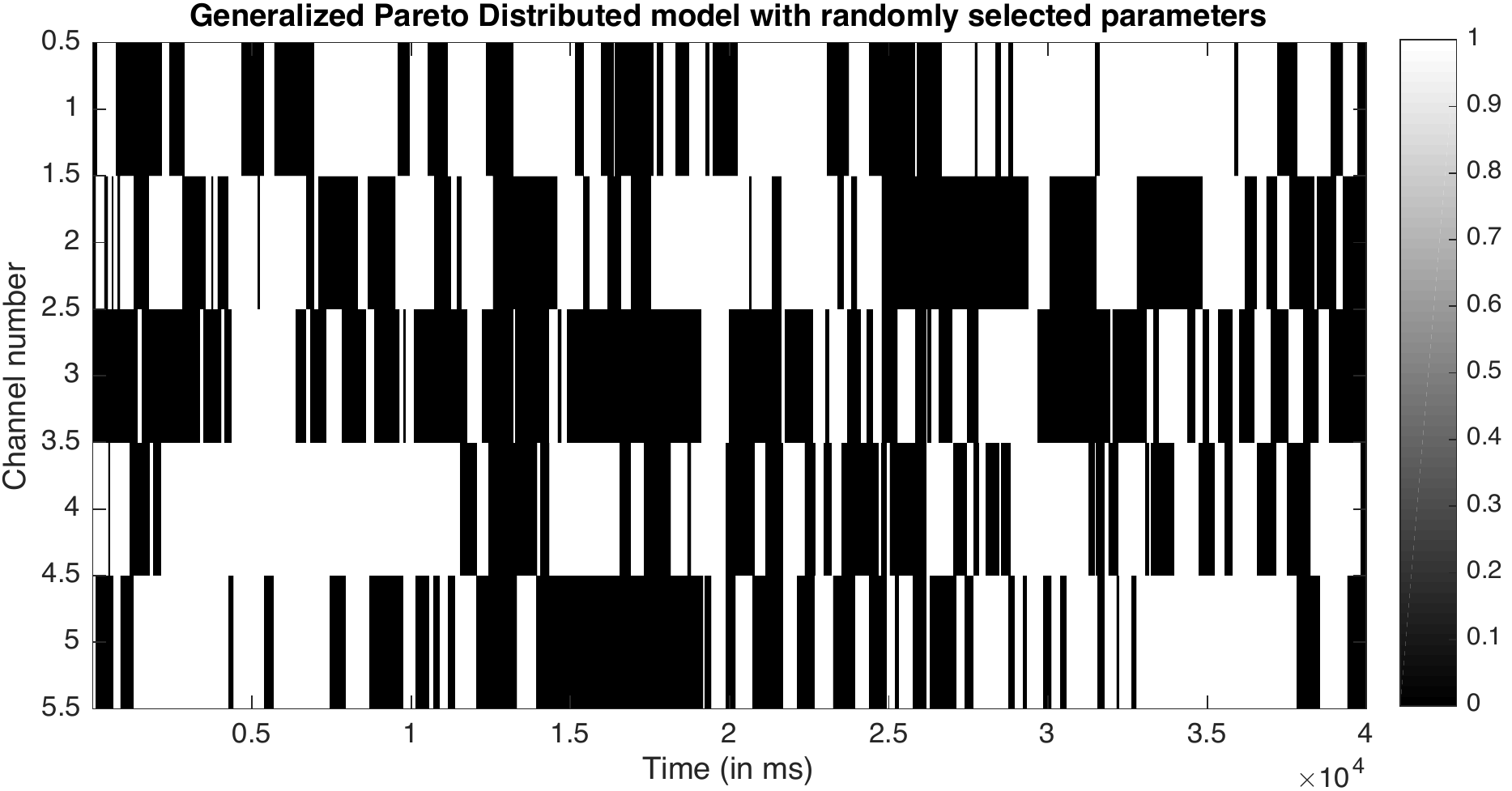}
        \caption{Generalized Pareto distributed primary traffic}
        \label{fig:GPDmodel}
    \end{figure}
    Figure \ref{fig:GPDmodel} shows the evolution of primary traffic modeled as GPD over 40 seconds for 5 independent channels. We can see that there are large blocks of time for which some of the channels are in the OFF state. From the figure, it is apparent that we need not spend time sensing the channel in every frame. In fact, one can save significant amount of sensing time and energy if one could learn the average on/off behaviour of the channels. We aim to exploit this in our proposed approach. To maximally utilize these gaps in the spectrum, we propose a two-stage reinforcement learning technique. Before describing our proposed method, we first briefly summarize key ideas and popular methods in reinforcement learning in the succeeding section. 
	
	\section{Reinforcement Learning} \label{sec:Background}
     Learning can be broadly categorized as supervised, unsupervised and reinforcement learning. In Reinforcement Learning(RL), the agent learns to act appropriately by interacting with the environment \cite{sutton1998reinforcement}. With each action, a reward is obtained which is the only mode of feedback for the learning agent. RL is attractive as it does not require a dataset for training; it learns from data and hence, can be used in real-time applications; this is why RL has also been applied in cognitive radio to find which channel to sense. In general, an RL problem is modelled as a Markov Decision Process in which there are $S$ states and $A$ possible actions that can be taken from each state. 
	
	A multi-armed bandit(MAB) problem is the simplest form(single state) of an RL problem in which the agent learns to choose the most optimal option among available options so as to maximize the cumulative reward by sampling one option at a time. A learning agent chooses one of the $N$ actions available to it and updates the value of the action according to the reward it obtains. The next action  chosen is a function of the learnt values. In the cognitive radio setting, the different channels to be sensed by a SU act as the arms of the bandit. On pulling an arm, that is, picking a channel to sense, we obtain a reward which can be the SU throughput achieved or an indicator which gives us information on whether a collision with the primary user had occurred or not. 
	
	A standard phenomenon in the bandit setting is the exploration exploitation trade-off. The algorithm needs to optimally balance utilizing the action that has performed the best so far, and looking for new options that can be potentially better.  Various algorithms have been proposed to address this trade-off: the classic Thompson sampling algorithm for Bernoulli arms \cite{thompson1933likelihood}, $\epsilon$-greedy algorithm, Upper confidence bound algorithms\cite{auer2002finite}, EXP3 for the case of adversarial bandits \cite{auer2002nonstochastic}, etc. Two of the RL algorithms which have been used in the cognitive radio context are briefly explained below. We will use these two algorithms for comparison against our proposed approach in section \ref{sec:SimRes}.
	
	\subsection{Algorithms} \label{algorithms}
	   \subsubsection{Q Learning (Algorithm 1)}
	        The Q Learning Algorithm is a widely used reinforcement learning algorithm in wireless communication introduced in 1989 by Watkins \cite{Watkins1992} . It works on a Markov model with multiple states and actions. The algorithm selects the optimal action in each state by learning Q-values associated with each action. It also incorporates an exploration strategy like $\epsilon$-greedy  or softmax in its framework. The Q Learning algorithm has been shown to converge to the optimal action for  each state. \\
	        
	        \begin{algorithm}    
                \caption{Q Learning}    \label{alg:Q Learning}
                \begin{algorithmic}[1]
                    \State \textbf{Initialization:} $Q(s,a) = 0, \: \forall \: a \: \in \: A, A$ is set of all possible actions and$ \: \forall \: s \: \in \: S, S$ is set of all states
                    \For t = 1, 2, 3, \ldots
                        \State Select the action based on the exploration strategy and Q values in state $s$
                        \State Take action selected and obtain reward $r$, $s'$ is new state reached
                        \State Update the Q value for arm selected in state $s$ as 
                        \begin{equation*}    
                        Q_{t + 1  }(s,a) \gets (1 - \alpha)Q_t(s,a) + \alpha*(r + \gamma* \max \limits _ {b \in A(s')} Q_t(s',b))
                        \end{equation*} where $\alpha$ is the learning rate, $\gamma$ is the discount factor
                    \EndFor
                \end{algorithmic}
            \end{algorithm}
	        
	        The stateless Q Learning variant \cite{claus1998dynamics} assumes a single state with multiple actions. The update equation for stateless Q Learning is obtained by setting the discounting parameter, $\gamma$, to zero. Q-learning is also employed for channel selection in cognitive radio by \cite{Li2009}.
	        
	        \subsubsection{Thompson Sampling (Algorithm 2)}
	        Thompson Sampling comes from a family of randomized probability matching algorithms.
            It is proposed for Bernoulli arms; therefore, the rewards are binary in nature. A prior Beta distribution - $Beta(\alpha,\beta)$ - is maintained which tracks the probability of success of each arm, which is the parameter of the Bernoulli distribution. Here $\alpha$ and $\beta$ correspond to the number of successes and failures observed respectively. The Beta distribution peaks at the actual Bernoulli parameter when one has observed sufficient number of samples. This algorithm was first introduced in \cite{Thompson1933} and was further analyzed by \cite{agrawal2012analysis}. The work in \cite{gwon2013optimizing} and \cite{Zhang2013a} use ideas from Thompson sampling to propose an algorithm for channel selection in CR. In \cite{gwon2013optimizing}, a Weibull likelihood and a inverse gamma conjugate prior is assumed for performing the Bayesian update instead of the Bernoulli- Beta distribution used in TS. In \cite{Zhang2013a}, the authors propose TS for multislot sensing by listing the channels in the descending order of their preference.
            \begin{algorithm} 
                \caption{Thompson Sampling \cite{agrawal2012analysis}} \label{alg:TS}
                \begin{algorithmic}[1]
                    \State \textbf{Initialization:} $\alpha_0^i = 2, \beta_0^i = 2 \: \forall \: i = 1,\ldots,K$
                    \For $t = 1, 2, 3, \ldots$
                        \State Sample $\theta^i$ from Beta($\alpha_{t-1}^i,\beta_{t-1}^{i}$) for all arms
                        \State Select arm with highest value of $\theta^i$
                        \State Pull that arm $I_t$ and receive reward $x_{I_t,t}$
                        \State $\alpha_t^{I_t} \gets \alpha_{t-1}^{I_t} + x_{I_t,t}$
                        \State $\beta_t^{I_t} \gets \beta_{t-1}^{I_t} + (1-x_{I_t,t})$
                    \EndFor
                \end{algorithmic}
            \end{algorithm}
            
	 These RL algorithms can be used to suggest which channel to select at the start of every frame, when used in the cognitive radio context. However, we can see that, when a channel is unoccupied, it is highly likely to be unoccupied in the next instant as well. Our approach additionally exploits this behavior of the primary traffic.

	\section{Proposed Approach} \label{sec:PropApproach}
    
    \subsection{Motivation for the Proposed Approach} \label{sec:MotPropApproach}
	
	Most of the work in CR literature focuses on finding the optimal channel to sense. However, they do not address how often to sense the channel after selection. It is apparent from Figure \ref{fig:GPDmodel} that SU can forgo sensing the selected channel for some duration depending on the channel traffic.  Hence, we propose an algorithm that incorporates two stages of learning. Learning to pick the right channel through an MAB formulation and learning an appropriate strategy for sensing that channel. We believe that our work is the first to suggest a two-stage learning algorithm so that time and energy spent in sensing can be minimized.
	
	In the first stage of learning, a metric known as value is maintained for all the arms which quantitatively indicates how beneficial it is to pick that channel. In the context of cognitive radio, this is an indicator of whether a channel is likely to be idle or vacant. When an arm is played, i.e., when a channel is chosen for sensing, the energy detector that indicates whether the channel is busy or available. The value of the arm is penalized if it found to be busy by the energy detector. When a vacant channel is found, we also obtain a reward from the environment based on whether the frame was transmitted successfully. This reward is also used to update the value of the arm. At any instant, the SU can decide to exploit - pick the channel with the maximum value, or explore - pick among other channels so as to learn more about their behaviour. We illustrate our algorithm using an optimistic variant of the Thompson sampling algorithm, although any of the other bandit or RL algorithms can be employed. 
    
    To determine the strategy to skip sensing, one possible approach is to have the secondary user transmit on a channel until it encounters a collision from the primary data. However, this implies that sending on a channel is interrupted only when a collision occurs. Depending on the primary user traffic, this can lead to significant interference to the PU and one needs a more conservative approach. Hence, we propose the following approach. Ideally, we should skip a few frames and stop transmitting before a collision occurs. To do so, we need to estimate the underlying traffic distribution of the primary user. A prior distribution for a parameter of the estimated primary traffic is maintained; it is updated when a data sample is observed. In practical situations, secondary users do not have exact knowledge about the primary traffic; hence, in our algorithm, SU assumes that the primary traffic follows exponential on/off model and maintains a prior for the exponential parameter, $\theta$. The conjugate prior for an exponential distribution is a gamma distribution whose parameters we update to estimate $\theta$. The Generalized Pareto Distribution has a heavier tail when compared to the exponential model; hence, assuming an exponential distribution is justified as it prepares the SU for shorter idle times than the actual traffic and hence leads to lesser probability of interference with the PU. 
    
    The OFF time of the channel is assumed to be a sample from an exponential distribution with parameter $\theta$, i.e.,  \begin{equation*}
        T_{OFF} \sim Exp(\theta) = \theta e^{-\theta t_{OFF}}
    \end{equation*}
The mean of the distribution is $1/\theta$ and this corresponds to the mean OFF time. Therefore, to determine the number of steps for which sensing can be skipped($t_{skip}$), we consider the inverse of the sample obtained from the prior distribution. Physically, $1/\theta$ signifies the mean time duration for which the channel stays idle. 
    
    Let the SU assume that the primary traffic OFF time, $t_{OFF}$, is exponential with parameter $\theta$ for a specific channel; let the conjugate gamma prior be parameterized by $\alpha$ and $\beta$, i.e., $\theta \sim \mathcal{G}(\alpha,\beta)$.
	\begin{equation}
	    p(\theta) = \dfrac{\beta^\alpha}{\Gamma(\alpha)} \theta^{\alpha - 1} e^{ - \beta \theta}
	\end{equation}
    In the cognitive radio context we get a data sample $x$ which denotes the duration for which data is transmitted without experiencing a collision from the time that channel was selected. It is essentially a sample quantized to the frame size $T$ as we assume that we do not have information of any collisions that occur is less than one frame duration. The posterior distribution of $\theta$ is given by
	\begin{align} 
	    p(\theta | x) & \propto p(\theta) l(\theta|x) \nonumber\\
	                  & \propto \dfrac{\beta^\alpha}{\Gamma(\alpha)} \theta^{\alpha - 1} e^{ - \beta \theta}  \theta e^{-\theta x} \nonumber\\
	                  & \propto \theta^{\alpha +1 -1 } e^{- \left(\beta + x\right) \theta} \nonumber\\
	   p(\theta|x)    & \sim \mathcal{G}(\alpha +1 , \beta + x) \nonumber \end{align}
	   When we have $n$ samples, the posterior is given by \begin{equation} \label{eqn:px}
	   p(\theta|x_1,...x_n) \sim \mathcal{G}(\alpha +n , \beta + \sum_{i=1}^n x_i)    
	\end{equation}
	Note that in Equation (\ref{eqn:px}), $x_i$s play the role of realizations of the OFF time of the primary traffic. As we observe more and more samples of $x_i$, $p(\theta|x_1, ..x_n)$ becomes more and more concentrated around its mean. However, we make the following observation. $x$ is not a sample of the entire OFF time; it is a sample of the OFF time observed after the secondary user selects the channel. Let $Y$ and $V$ denote the random variables for the exponential OFF time and the time at which an empty channel is selected by the SU respectively. In the absence of prior information, we can assume that $V$ is a uniform distribution, i.e., a SU is likely to occupy an idle channel anytime during the OFF period. As $V$ is assumed to be uniform, i.e., $ V \sim U[0,y]$ where $y$ is a realization of $Y$, the actual OFF time. Hence, we have
	\begin{align*}
	    \mathbb{E}[Y-V] &= \mathbb{E}[Y]-\mathbb{E}[V]\\
	    &= \frac{1}{\theta} - \mathbb{E}_Y[\frac{y}{2}]\\
	    &= \frac{1}{2 \theta}
	\end{align*}
	On an average, the mean OFF time observed after SU selects the channel is half the actual OFF time. Therefore, we incorporate a correcting factor in the posterior update.
	\begin{align} \label{eqn:posterior}
	    p(\theta|x)       & \sim \mathcal{G}(\alpha +1 , \beta + 2x)
	\end{align}
    The gamma distribution is hence updated by the above equation. The number of frames to be skipped is then obtained as a function of inverse of the sample from this posterior. As Equation (\ref{eqn:posterior}) refers to the posterior of the PU OFF time, the inverse of a sample from $p(\theta|x)$ is indicative of the entire OFF time. Note that, just as the SU could have begun sensing anywhere during the PU OFF time, SU transmission could also start anywhere in that PU OFF period; the entire OFF period may not be available for SU transmission. Hence, the average time for which sensing should be skipped is typically half the inverse of a sample from $p(\theta|x)$. The step-wise description of the algorithm is given in Section \ref{sec:AlgDes} and the intuition behind its working is described in Section \ref{sec:working}.

    \subsection{Algorithm Description} \label{sec:AlgDes}
    In the classic multi-armed bandit problem, the Thompson sampling algorithm suggests one arm to be played out of the available options as illustrated in Algorithm \ref{alg:TS}. However, when applied to channel selection in cognitive radio using multi-slot sensing, we require a list of channels to sense. In \cite{zhang2013channel}, Thompson sampling is implemented to return a ranked list of channels so as to perform multi-slot sensing. We further modify this algorithm using a recent work \cite{may2012optimistic} to obtain an optimistic variant. The Optimistic Thompson Sampling (OTS) algorithm modified to return a ranked list of channels instead of a single channel is given by Algorithm \ref{alg:OTS}. Channel sensing is performed in the order indicated by the list until a vacant channel is found. Following this, the optimal number of frames to skip for the chosen channel is chosen.
    
    The algorithm operates in two states: the \textit{SENSE} state and the \textit{SKIP} state. 
    \begin{itemize}
        \item \textit{SENSE} state: In this state, the algorithm always asks for a ranked list of preferred channels from the OTS algorithm (Algorithm \ref{alg:OTS}). Then, it follows the multi-slot sensing policy to sense the channels until a vacant channel is found. Only on finding a vacant channel, will it transmit the data. Also, the number of frames to skip, $t_{skip}$, is calculated. 
        \item \textit{SKIP} state: In this state the algorithm, primary channel sensing is not performed. The data to be transmitted is sent on the channel which was previously found to be vacant. $N_{skip}$, a counter for number of skips performed, is also incremented. 

    \end{itemize}  
    The detailed algorithm is as given below:\\
    \textbf{Step 1}: \textit{Initialization} - Set the parameters of the gamma prior for all channels to $\alpha_i = 1$ and $\beta_i = frameLength$ . Also, set the algorithm in the \textit{SENSE} state and $N_{skip}$ to $0$.\\
	\textbf{Step 2}: \textit{Channel Selection} - If in the \textit{SENSE} state then call \texttt{GetRankedList()} of Algorithm \ref{alg:OTS} to obtain a ranked list of preferred channels. Sense the channels sequentially to find a vacant channel to transmit, say $c_t$ at time $t$. A sample $\hat{\theta}$ is drawn from $\mathcal{G}(\alpha_{c_t},\beta_{c_t})$ and $t_{skip}$ is calculated as 
	\begin{equation*}
	t_{skip} = \frac{max\left(\dfrac{1}{\hat{\theta}}, \dfrac{1}{\alpha_{c_t}/\beta_{c_t}}\right)}{2}
	\end{equation*} 
	Note that $\alpha_{c_t}/\beta_{c_t}$ is the mean of the gamma distribution. Else, if the algorithm is in the \textit{SKIP} state, then transmit on channel used in the previous time step and increment $N_{skip}$. \\ 
    \textbf{Step 3}: \textit{Obtain feedback} - After frame transmission, a reward is obtained in the form of a collision indicator which is also used to update the OTS algorithm. (The collision indicator is nothing but the ACK or NACK from the device to which data was transmitted by the SU. Let $z$ denote the collision indicator.) \\
    \textbf{Step 4}: \textit{Prior and OTS Updation} - Three cases can occur here as given below:
    \begin{enumerate} 
            \item Successful transmission and $N_{skip} < t_{skip}$: Call \texttt{UpdateObservation}($c_t,z=0$) then continue in \textit{SKIP} state.
            \item Successful transmission and $N_{skip} = t_{skip}$: Reset $N_{skip}$ and call \texttt{UpdateObservation}($c_t,z=0$). Go to \textit{SENSE} state.The parameters of the gamma prior are not updated until $c_{t+1}$ is picked. Only if the next channel chosen by OTS is not the same as the current channel ($c_{t+1} \neq c_t$), $\alpha_{c_t}$ and $\beta_{c_t}$ are updated as given below 
            \begin{align*}
                \alpha_{c_t} &\gets  \alpha_{c_t} + 1 \\
                \beta_{c_t} &\gets \beta_{c_t} + 2*\hat{t}_{skip}*frameLength
            \end{align*}
            where $\hat{t}_{skip}$ indicates the number of frames skipped successfully in the same channel consecutively. 
            \item Unsuccessful transmission: Reset $N_{skip}$ and call \texttt{UpdateObservation}($c_t,z=1$). Go to \textit{SENSE} state. Update $\alpha_{c_t}$ and $\beta_{c_t}$ of gamma prior of channel $c_t$ as given below.
            \begin{align*}
                \alpha_{c_t} &\gets  \alpha_{c_t} + 1 \\
                \beta_{c_t} &\gets \beta_{c_t} + 2*\hat{t}_{skip}*frameLength
            \end{align*}    
    \end{enumerate}
    \textbf{Step 5}: Go to \textbf{Step 2} \\

    \begin{algorithm}[H]
            \caption{Optimistic Thompson Sampling (OTS) functions}   \label{alg:OTS}
            \begin{algorithmic}[1]
            \State \textbf{Parameters} $S_i = 1, F_i = 1 \: \forall \: i \in \{1,\ldots,N\}$
            \Function{\texttt{GetRankedList()}}{}
                \State $d_i \sim Beta(S_i,F_i)$
                \State $\hat{d}_i = max(d_i, \mathbb{E}[Beta(S_i,F_i)])$
                \State Sort $\hat{d}_i$ in descending order
                \State Return the index of sorted order
            \EndFunction
            \Function{\texttt{UpdateObservation}}{$c_t$,collision}
                \If{collision}
                    \State $F_{c_t} \gets F_{c_t} + 1$
                \Else
                    \State $S_{c_t} \gets S_{c_t} + 1$
                \EndIf
            \EndFunction
            \end{algorithmic}
    \end{algorithm}

    \subsection{Working of  the algorithm} \label{sec:working}
    We first explain the \textit{SENSE} state of the algorithm. The OTS algorithm maintains a Beta distribution with success and failure parameters as $S_i$ and $F_i$ for each of the $N$ channels which are updated after every frame transmission. To obtain a priority list, an optimistic sample is drawn from each of the $N$ Beta distributions i.e., the maximum of the mean and the sample is picked as the optimistic sample. This is done so that the samples obtained are atleast as much as the mean of the corresponding Beta distributions. The samples that are drawn are then arranged in descending order and their indexes are returned. This is the order in which sensing should be done.
    
    Now moving on to the \textit{SKIP} state of the algorithm, we aim to estimate the idle time of the primary user on each of the channels using a Bayesian approach. The duration for which we can forgo sensing a channel is found by drawing a sample from the conjugate posterior distribution of $\theta$ and taking the inverse. The motivation behind drawing a sample from the gamma distribution rather than using the mean of the posterior is that drawing a sample from a distribution holds the possibility of exploring both lower and higher value for time duration to skip sensing which will help to converge to the best possible skip duration. As we observe higher number of samples, the variance of the gamma distribution, $ \alpha / \beta^2$, decreases anyway. This implies that with more number of samples, picking a sample from the gamma distribution is very close to picking its mean value $\alpha / \beta$. In other words, though we start with picking a sample from the posterior distribution instead of its mean, as we observe more samples, the picked sample $\hat{\theta}$ becomes close to the mean. (Also note that $\alpha / \beta$ is the MMSE estimate of $\theta$.) 
    
    Furthermore, inspired by the idea behind the optimistic variant of Thompson sampling, we wish to be more optimistic in our estimation initially and hence favour the higher values for the number of frames to be skipped, i.e., we wish to pick a sample, $t_{OFF}$, that corresponds to atmost the mean of the gamma distribution in the worst case. Therefore, $t_{skip}=t_{OFF}/2$ corresponds to half of either the sample from the gamma distribution or the mean of the distribution, whichever is lower.  
    \begin{equation} \label{eqn:tskip}
        t_{skip} = \frac{max\left(\dfrac{1}{\hat{\theta}}, \dfrac{1}{\alpha_{c_t}/\beta_{c_t}}\right)}{2}
    \end{equation}
    Note that the factor of 2 in the denominator is because the entire OFF period may not be available to the SU for transmission. Please see Section \ref{sec:MotPropApproach} for detailed explanation.
    
    The $\alpha$ parameter accounts for the number of OFF period samples observed. It is incremented by 1 every time a collision occurs. If there is no collision at the end of $t_{skip}$, then the OFF period is ongoing. As $\alpha$ corresponds to an entire OFF period, one cannot just update $\alpha$ at the end of $t_{skip}$. Hence, $\alpha$ is not updated if the same channel is picked again for transmission as it would indicate that the same OFF period is ongoing. It is updated only when a different channel is picked at the end of $t_{skip}$, i.e., $c_t \neq c_{t+1}$ or when a collision occurs in the current channel. The $\beta$ parameter is updated with twice the duration upto which data was successfully transmitted on the same channel. As mentioned earlier, a factor of 2 is incorporated so that the gamma distribution is indicative of the entire OFF period. 
    
    As mentioned in Section \ref{sec:MotPropApproach}, the number of frames (OFF duration of primary traffic) that can be skipped is inversely proportional to $\theta$. When no collision occurs, we update $\beta$ with a higher value; this reduces the mean of the posterior, which implies longer OFF times. When collision occurs, $\beta$ is updated with a comparatively lower value. On performing this update with each sample, we arrive at an appropriate duration to skip as the posterior mean converges to the mean OFF time as number of samples increases. Note that the Thompson sampling algorithm is also updated in parallel to keep track of the channel occupancy every time a frame is transmitted. We increment the success parameter $S_{c_t}$ if the transmission is successful; if a collision is encountered, the failure parameter $F_{c_t}$ is incremented.  
    
    In our proposed algorithm, even when the frames are skipped without encountering a collision, we go back to the sensing phase after $t_{skip}$ frames. This is done to ensure that we skip conservatively instead of waiting for a collision. Instead, if we choose to transmit on the same channel even after $t_{skip}$ frames, we may obtain a better throughput at the cost of higher interference in heavy traffic scenarios. However, this will increase the number of collisions, while we want to keep the number of collisions to be similar to what is seen when only RL algorithms are used.

\ifCLASSOPTIONonecolumn
        
\else
        \begin{figure*}[!h]
            \begin{subfigure}{.33\textwidth}
                \includegraphics[width=1.\linewidth]{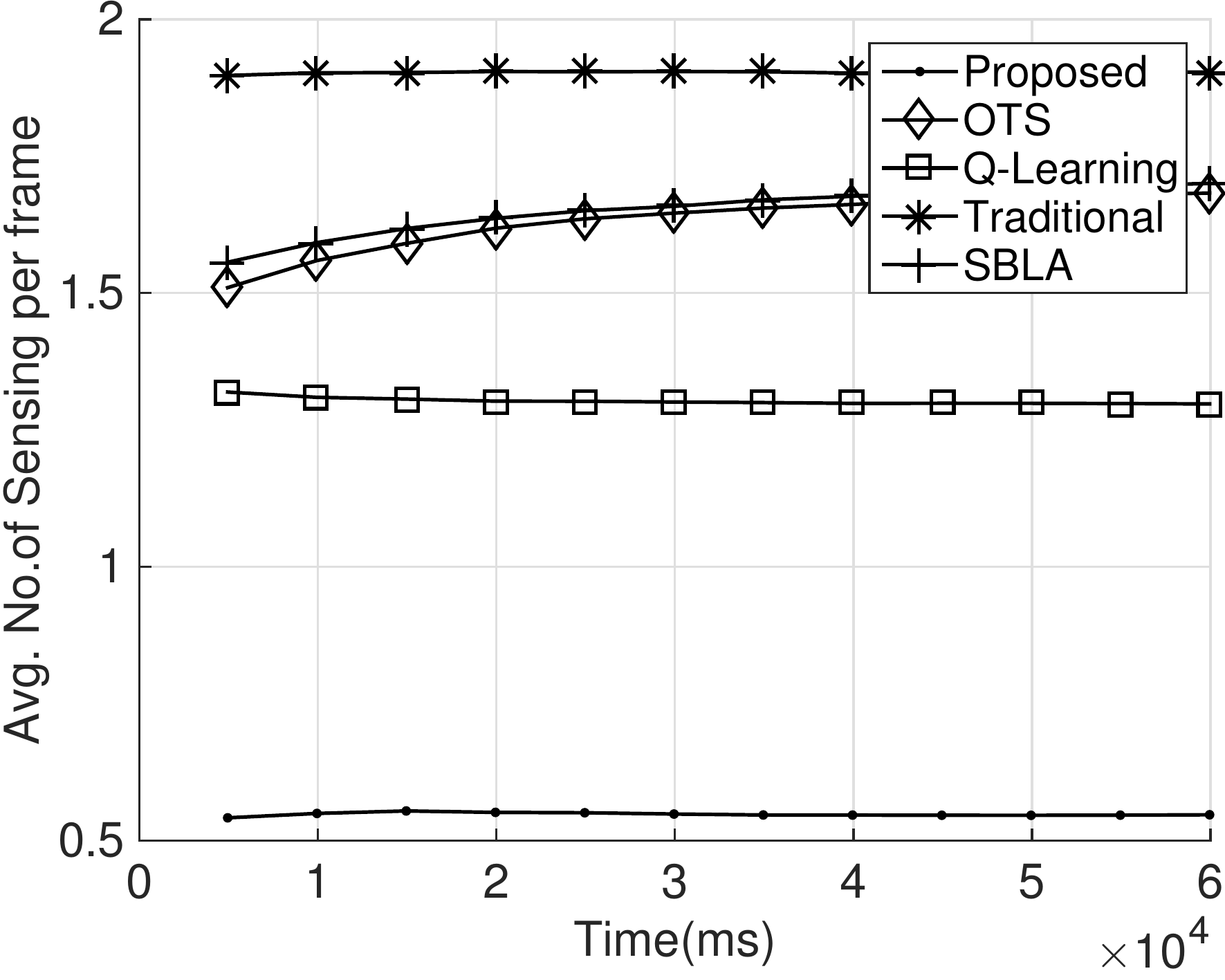}
                \caption{Avg. No.of Sensing per frame}
                \label{fig:GPD_C05_F050_S03_SE}
            \end{subfigure}%
            \begin{subfigure}{.33\textwidth}
                \includegraphics[width=1.\linewidth]{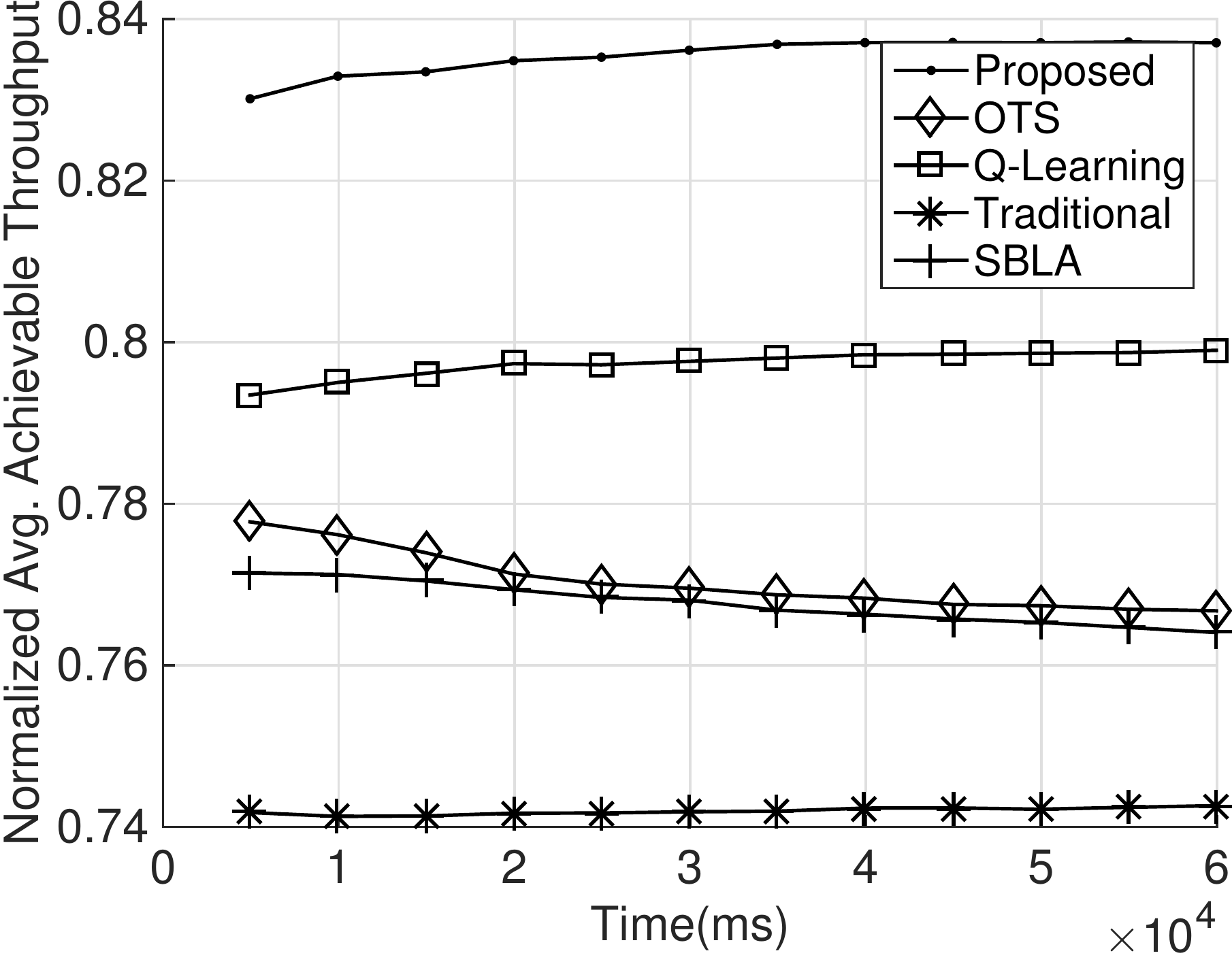}
                \caption{Avg. Achievable Throughput}
                \label{fig:GPD_C05_F050_S03_TP}
            \end{subfigure}%
            \begin{subfigure}{.33\textwidth}
                \includegraphics[width=1.\linewidth]{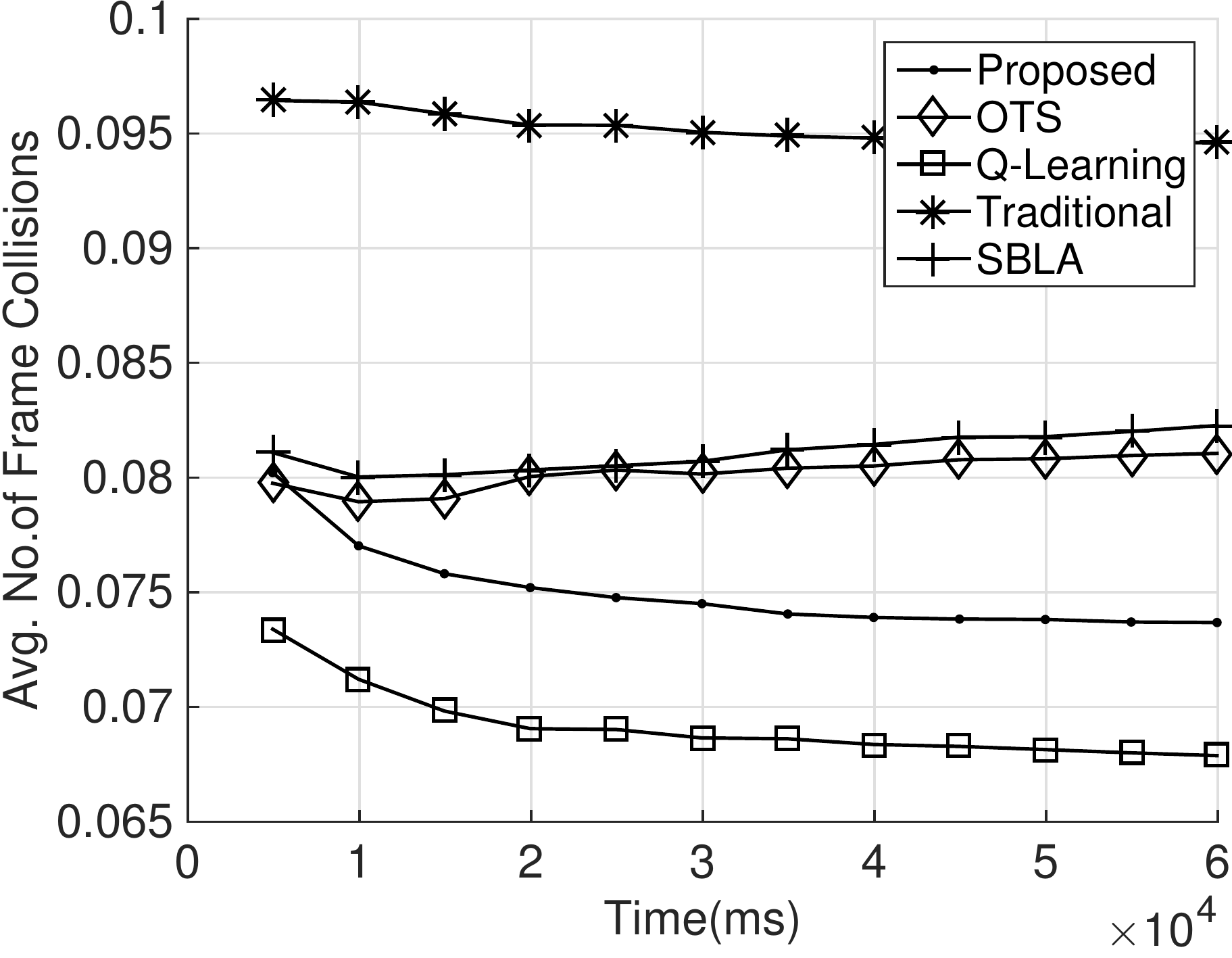}
                \caption{Avg. Frame Collision}
                \label{fig:GPD_C05_F050_S03_FC}
            \end{subfigure}%
            \caption{Results for GPD Model for frame size $50$ms for $5$ channels.}
            \label{fig:GPD_C05_F050_S03}
        \end{figure*}
        
        \begin{figure*}[!h]
            \begin{subfigure}{.33\textwidth}
                \includegraphics[width=1.\linewidth]{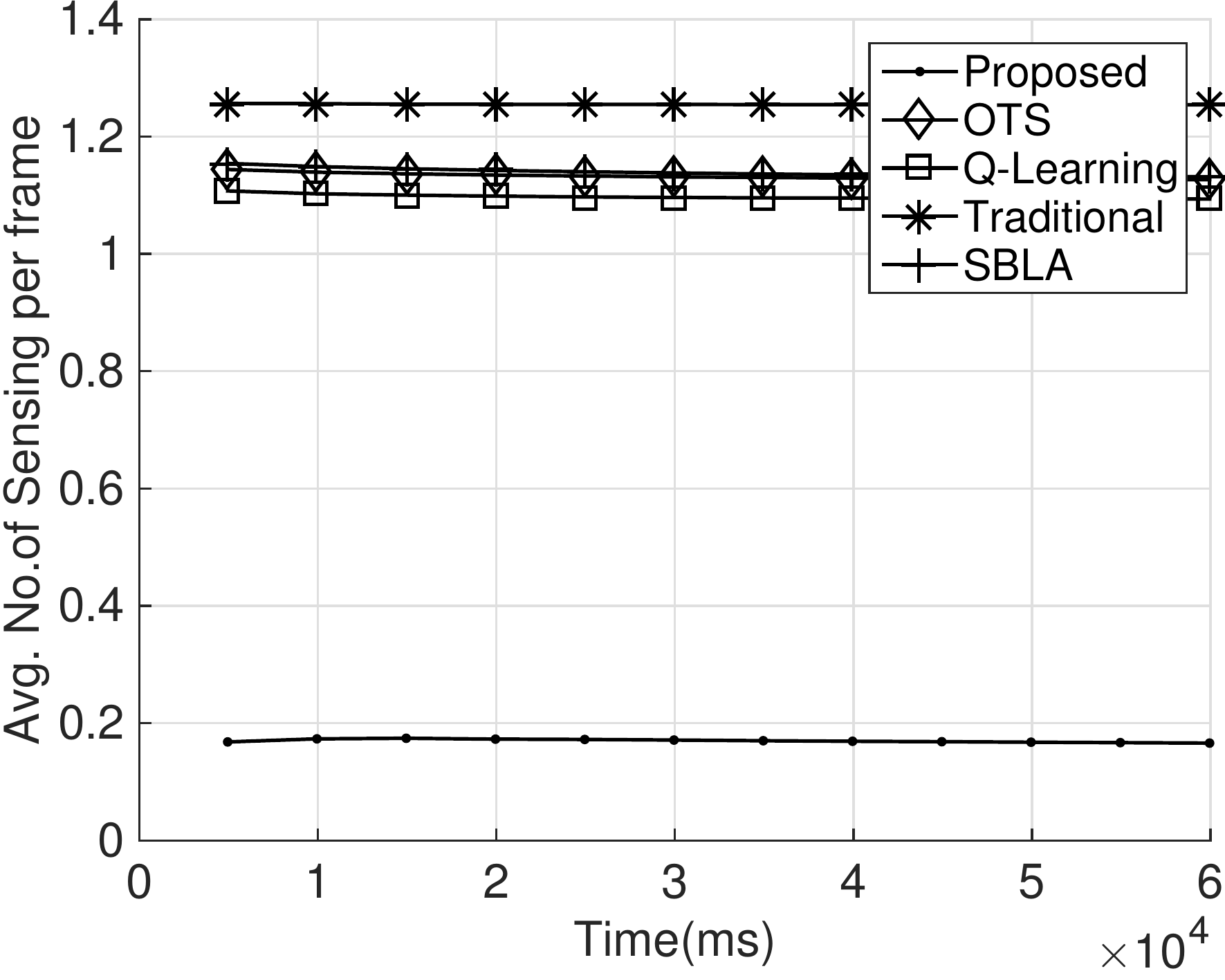}
                \caption{Avg. No.of Sensing per frame}
                \label{fig:DTMC_Beta_L-I_C05_F050_S03_SE}
            \end{subfigure}%
            \begin{subfigure}{.33\textwidth}
                \includegraphics[width=1.\linewidth]{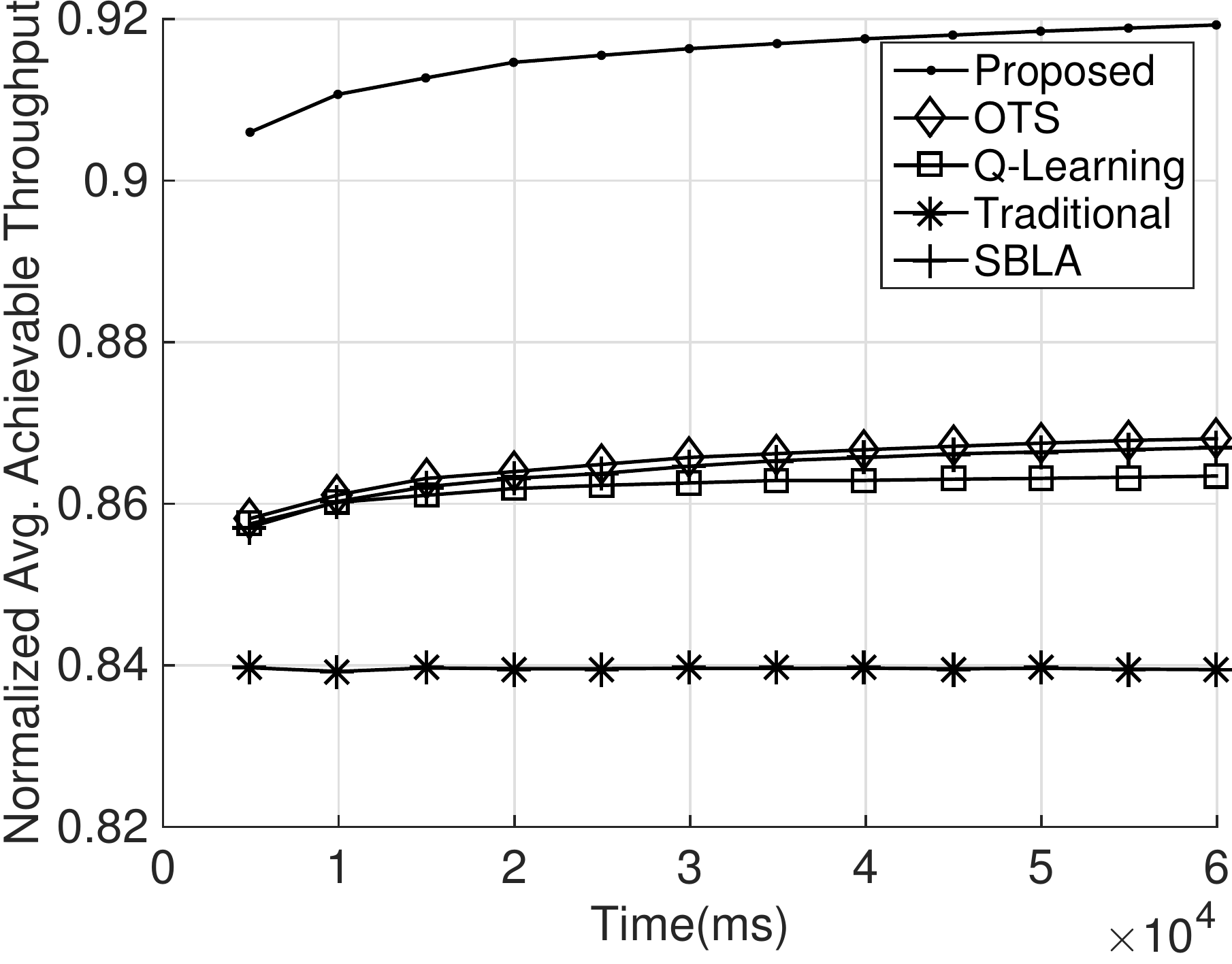}
                \caption{Avg. Achievable Throughput}
                \label{fig:DTMC_Beta_L-I_C05_F050_S03_TP}
            \end{subfigure}%
            \begin{subfigure}{.33\textwidth}
                \includegraphics[width=1.\linewidth]{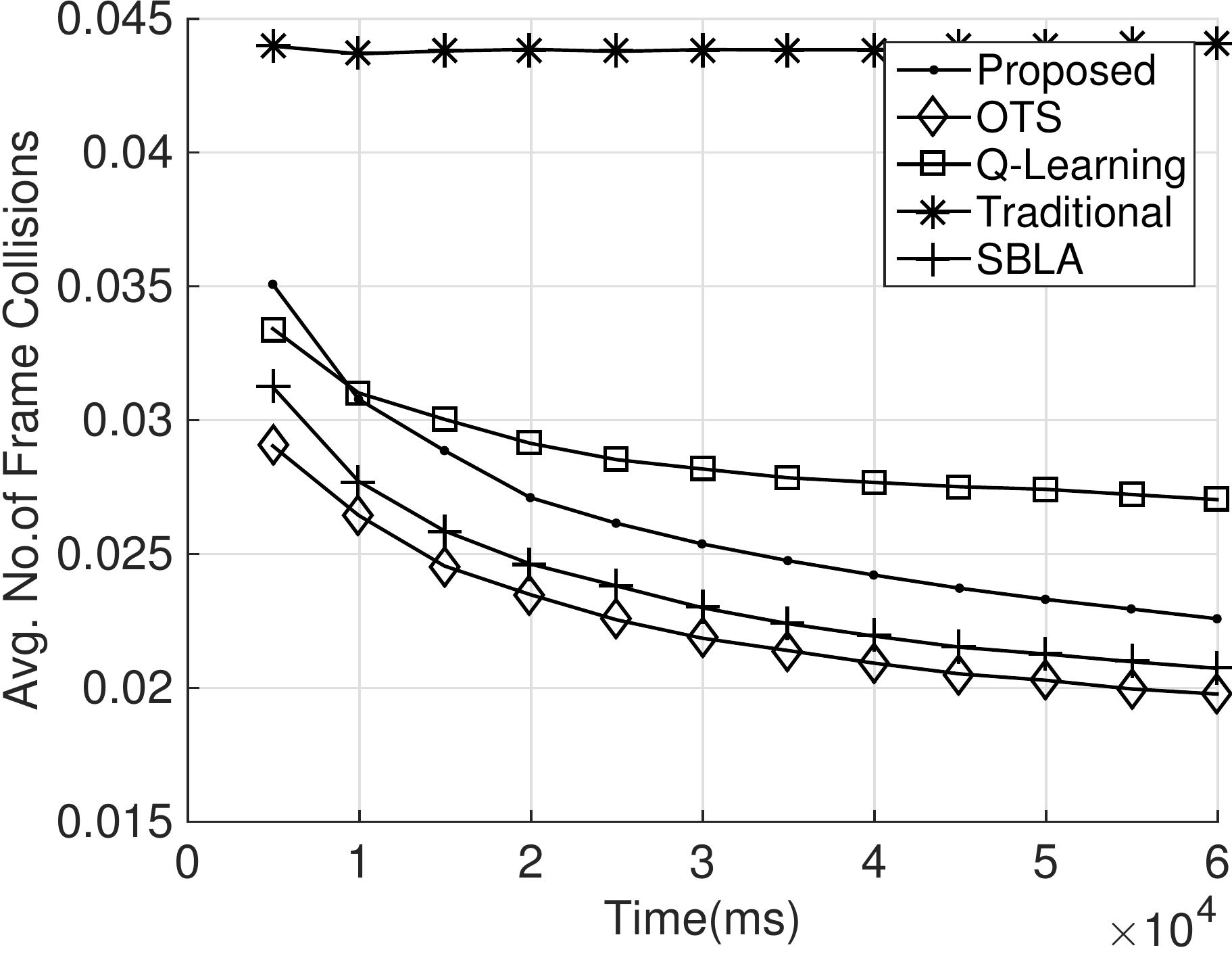}
                \caption{Avg. Frame Collision}
                \label{fig:DTMC_Beta_L-I_C05_F050_S03_FC}
            \end{subfigure}%
            \caption{Results for DTMC Low Traffic Model (DTMC L-I) for frame size $50$ms for $5$ channels.}
            \label{fig:DTMC_Beta_L-I_C05_F050_S03}
        \end{figure*}
        
        \begin{figure*}[!h]
            \begin{subfigure}{.33\textwidth}
                \includegraphics[width=1.\linewidth]{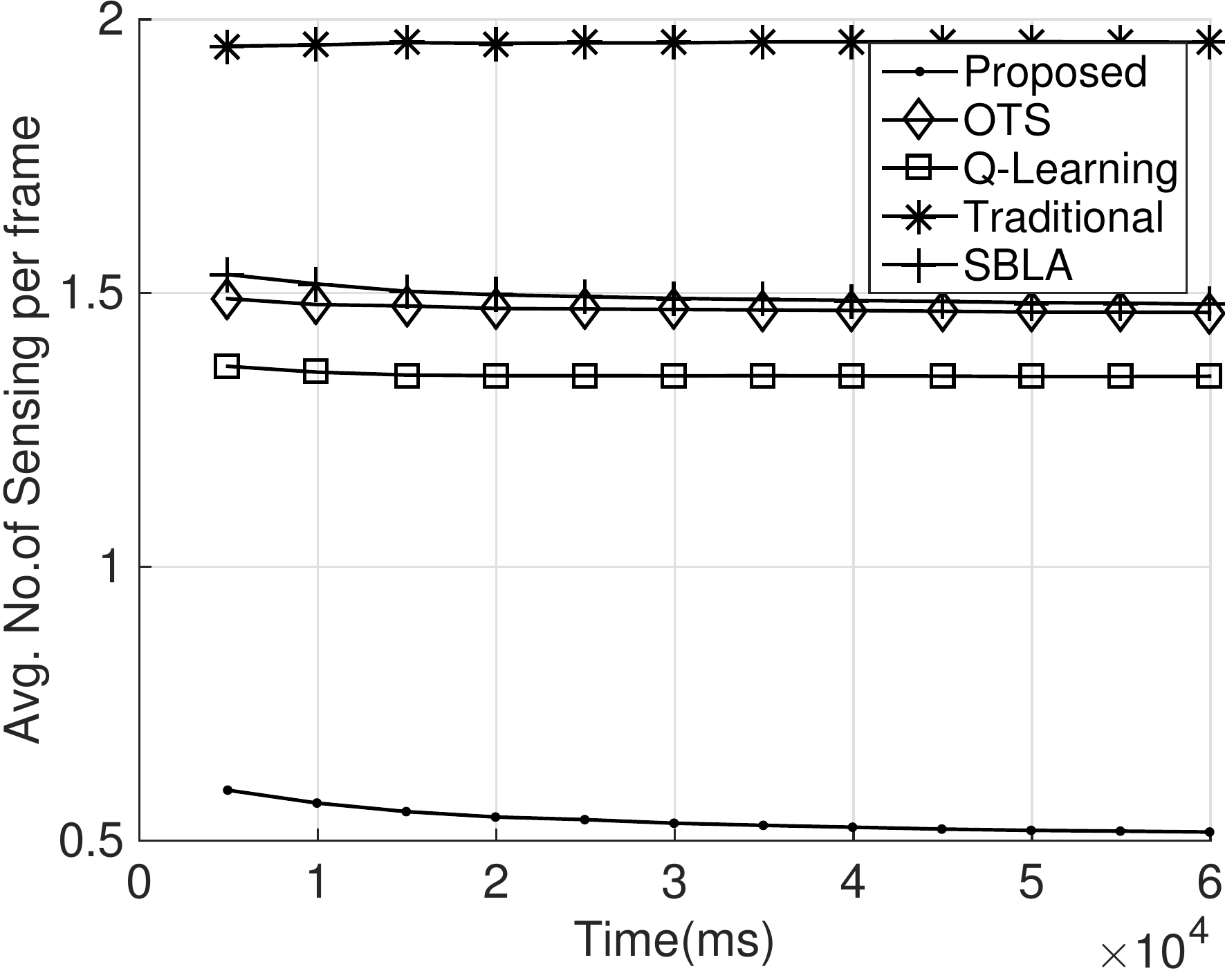}
                \caption{Avg. No.of Sensing per frame}
                \label{fig:DTMC_Beta_M-I_C05_F050_S03_SE}
            \end{subfigure}%
            \begin{subfigure}{.33\textwidth}
                \includegraphics[width=1.\linewidth]{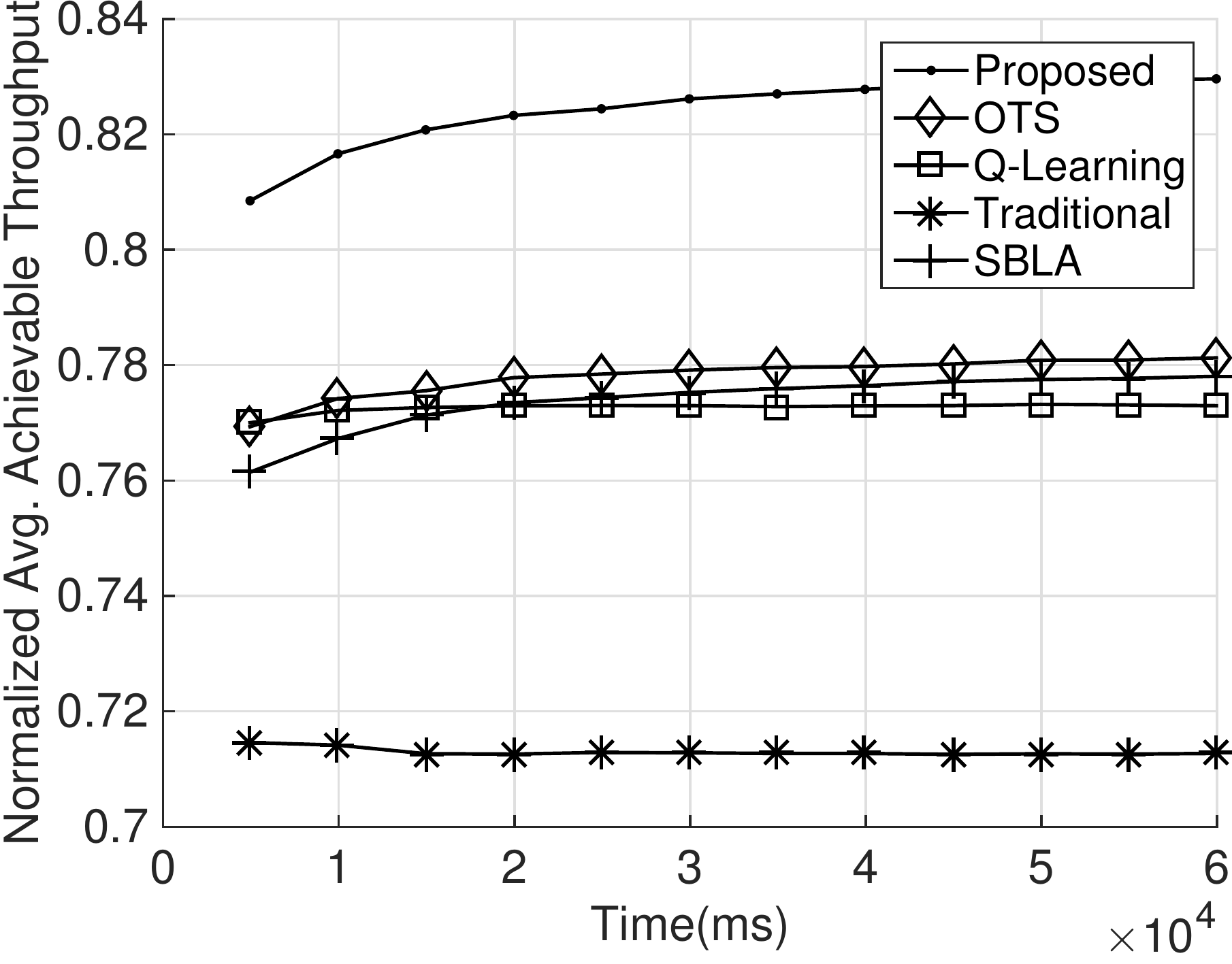}
                \caption{Avg. Achievable Throughput}
                \label{fig:DTMC_Beta_M-I_C05_F050_S03_TP}
            \end{subfigure}%
            \begin{subfigure}{.33\textwidth}
                \includegraphics[width=1.\linewidth]{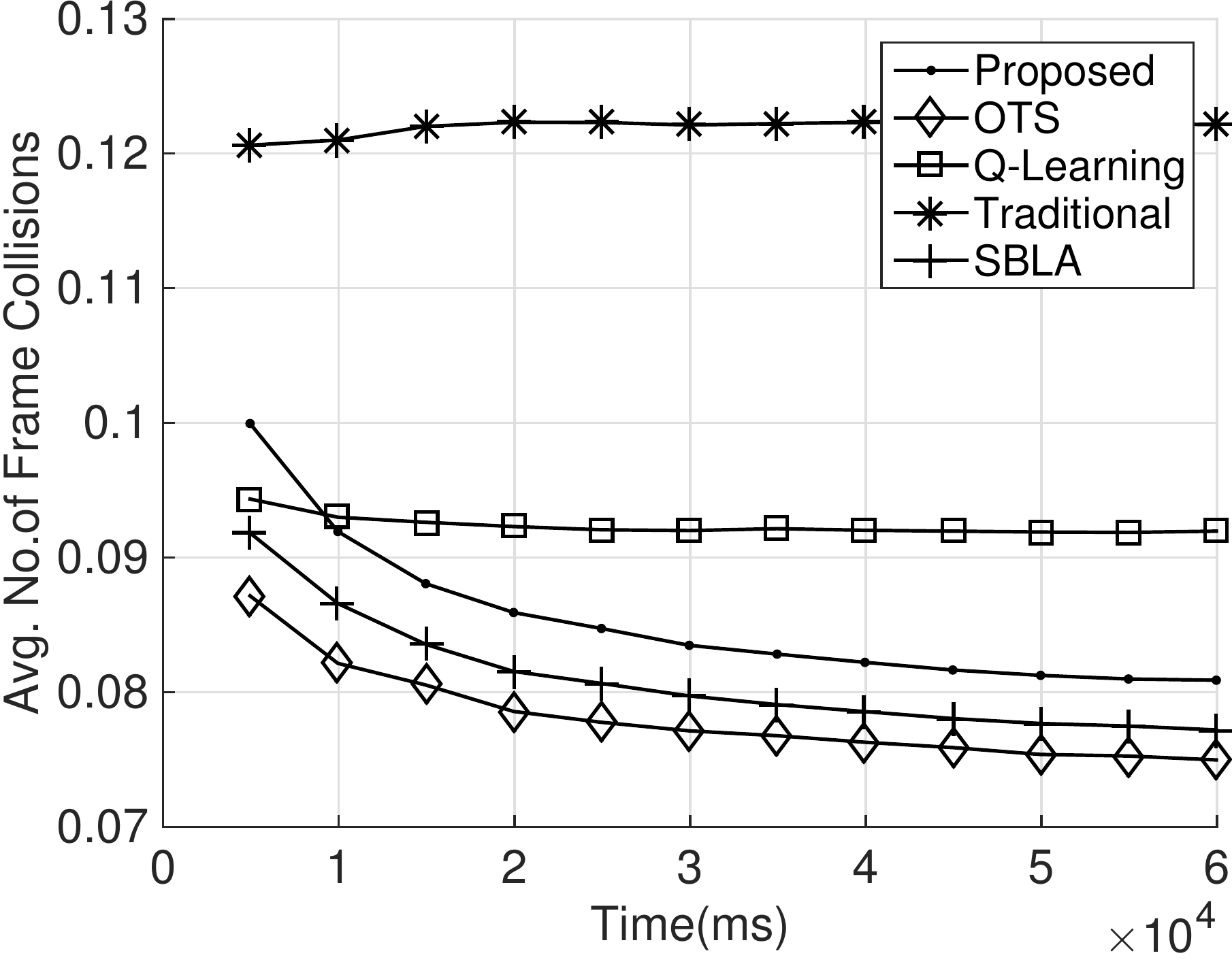}
                \caption{Avg. Frame Collision}
                \label{fig:DTMC_Beta_M-I_C05_F050_S03_FC}
            \end{subfigure}%
            \caption{Results for DTMC Medium Traffic Model (DTMC M-I) for frame size $50$ms for $5$ channels.}
            \label{fig:DTMC_Beta_M-I_C05_F050_S03}
        \end{figure*}
        
        \begin{figure*}[!h]
            \begin{subfigure}{.33\textwidth}
                \includegraphics[width=1.\linewidth]{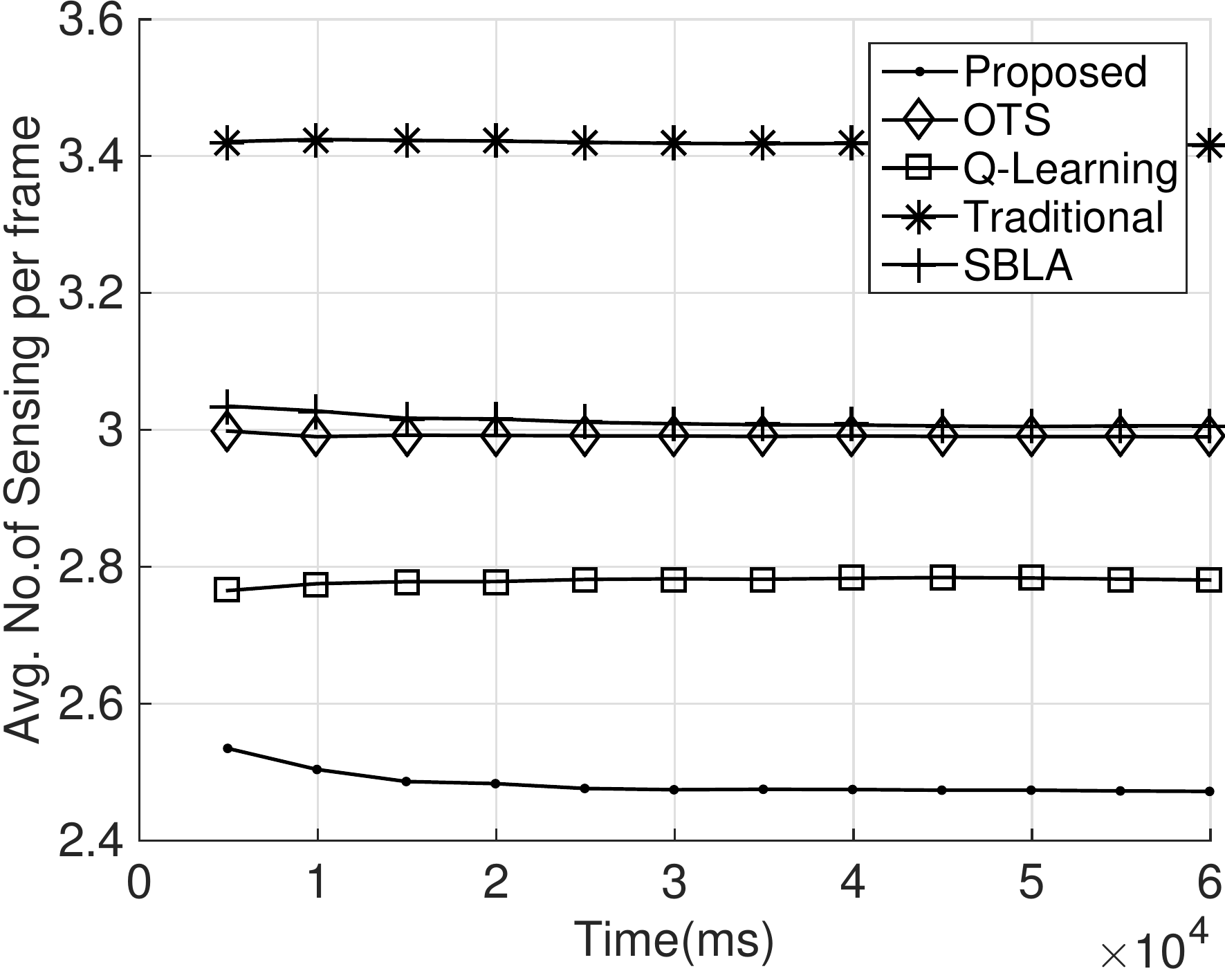}
                \caption{Avg. No.of Sensing per frame}
                \label{fig:DTMC_Beta_H-I_C05_F050_S03_SE}
            \end{subfigure}%
            \begin{subfigure}{.33\textwidth}
                \includegraphics[width=1.\linewidth]{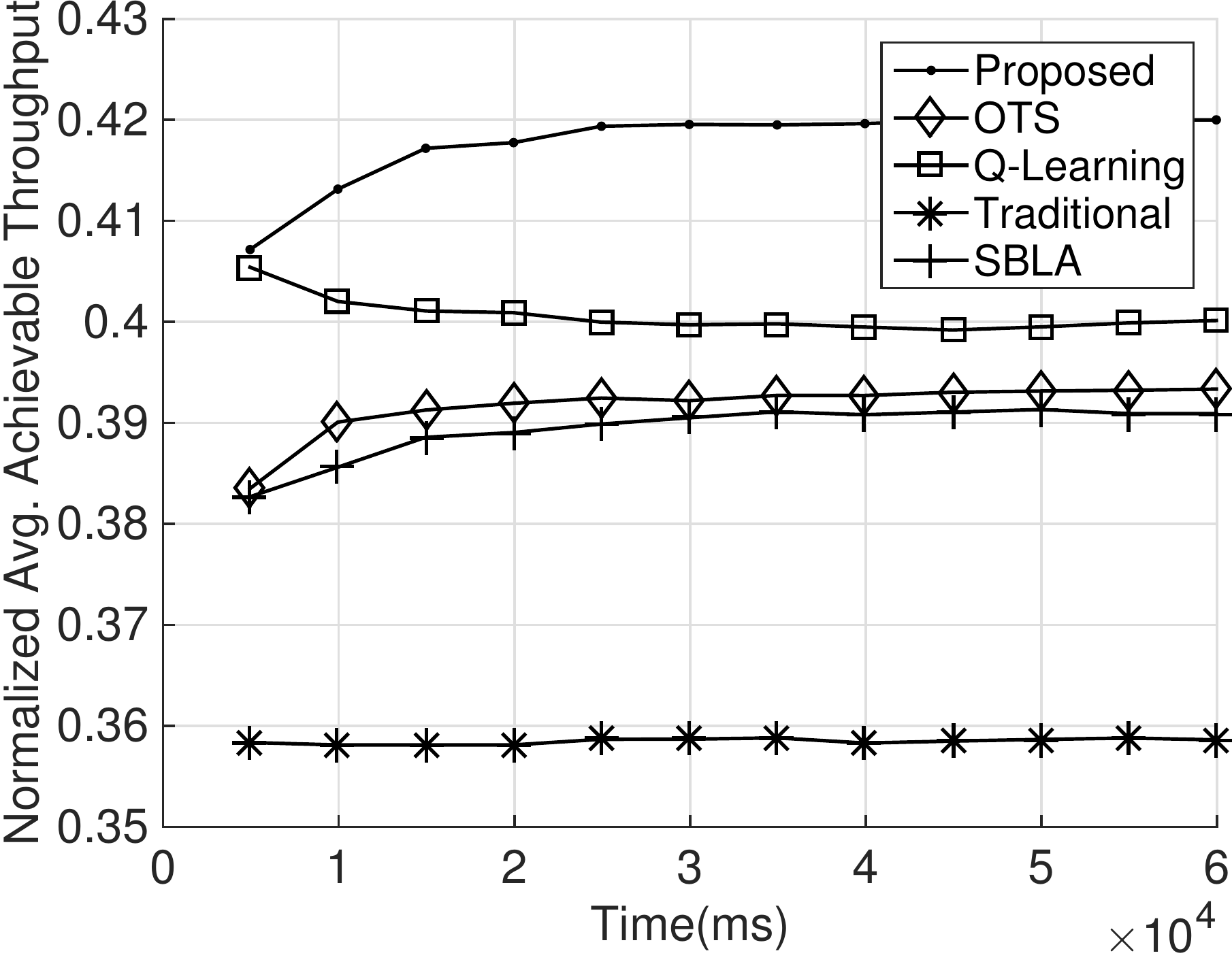}
                \caption{Avg. Achievable Throughput}
                \label{fig:DTMC_Beta_H-I_C05_F050_S03_TP}
            \end{subfigure}%
            \begin{subfigure}{.33\textwidth}
                \includegraphics[width=1.\linewidth]{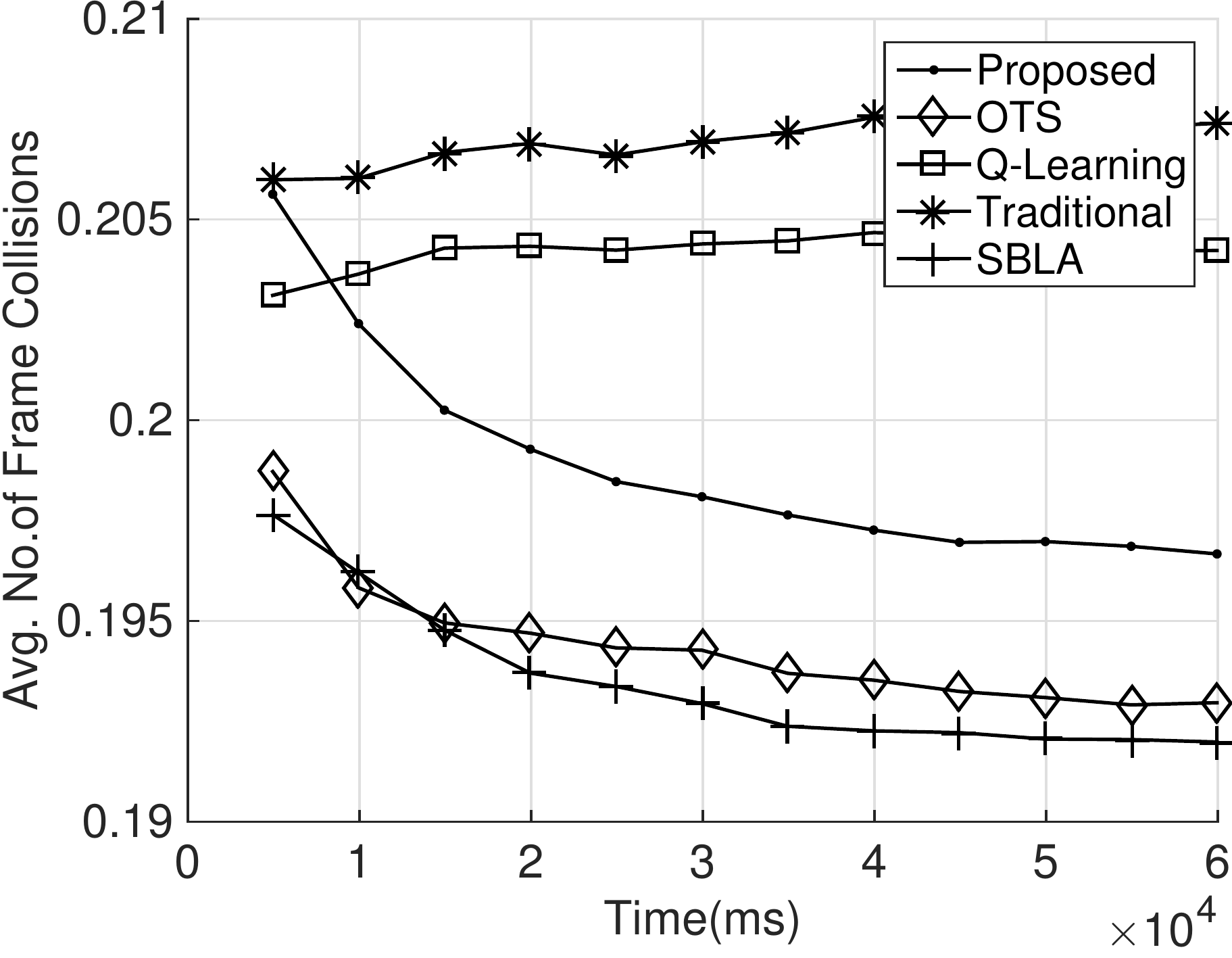}
                \caption{Avg. Frame Collision}
                \label{fig:DTMC_Beta_H-I_C05_F050_S03_FC}
            \end{subfigure}%
            \caption{Results for DTMC High Traffic Model (DTMC H-I) for frame size $50$ms for $5$ channels.}
            \label{fig:DTMC_Beta_H-I_C05_F050_S03}
        \end{figure*}
        
        \begin{figure*}[!h]
            \begin{subfigure}{.33\textwidth}
                \includegraphics[width=1.\linewidth]{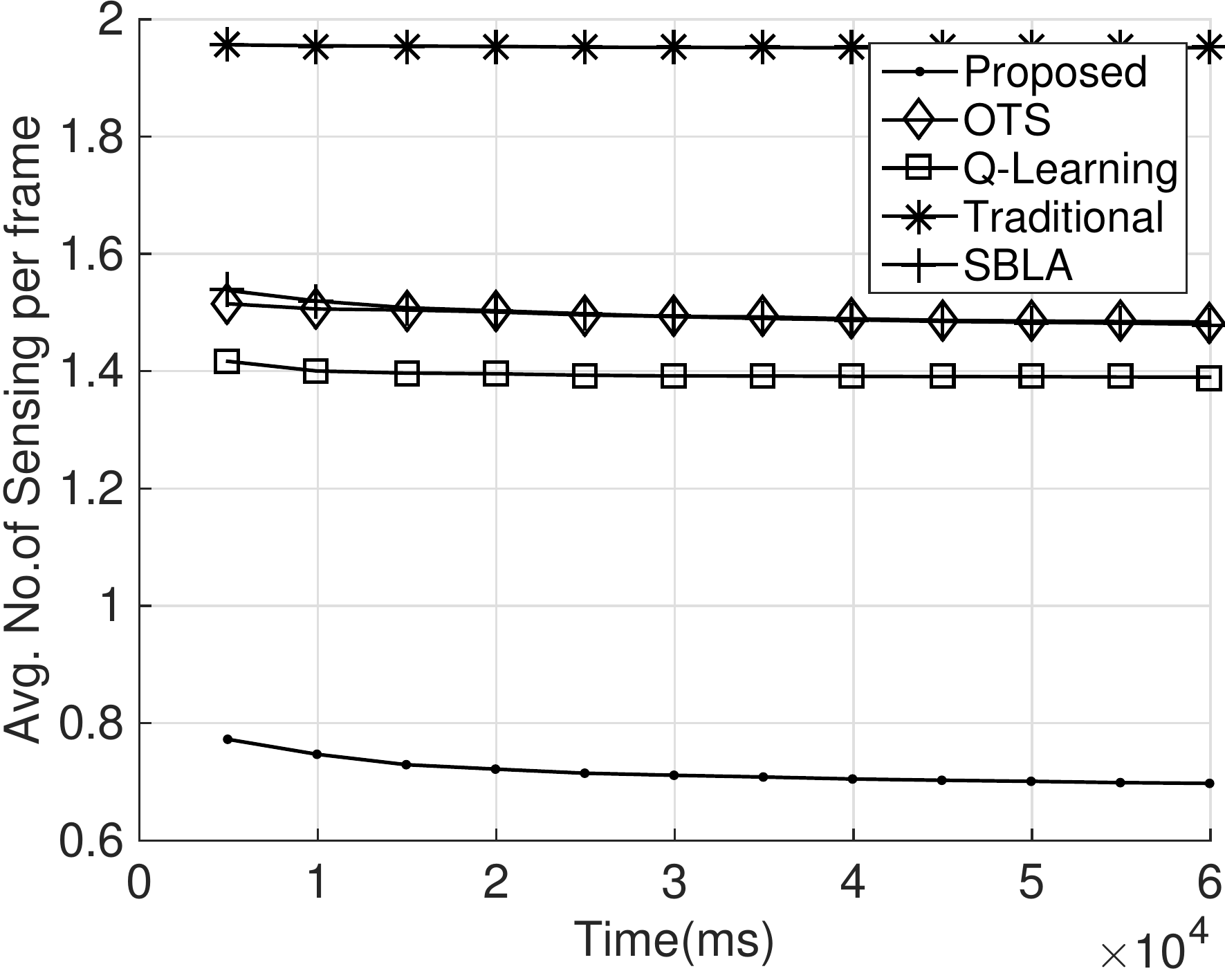}
                \caption{Avg. No.of Sensing per frame}
                \label{fig:Exp_C05_F050_S03_SE}
            \end{subfigure}%
            \begin{subfigure}{.33\textwidth}
                \includegraphics[width=1.\linewidth]{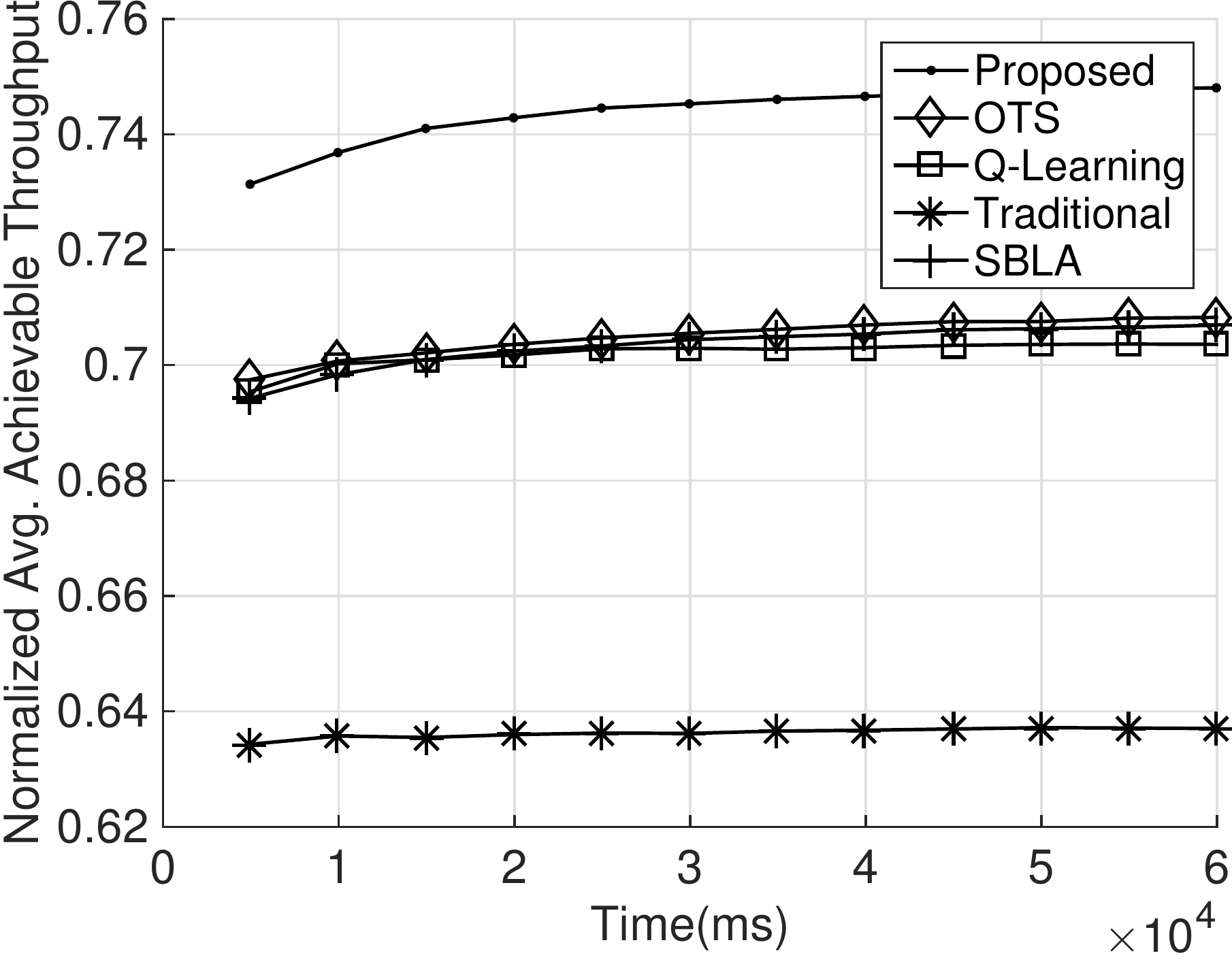}
                \caption{Avg. Achievable Throughput}
                \label{fig:Exp_C05_F050_S03_TP}
            \end{subfigure}%
            \begin{subfigure}{.33\textwidth}
                \includegraphics[width=1.\linewidth]{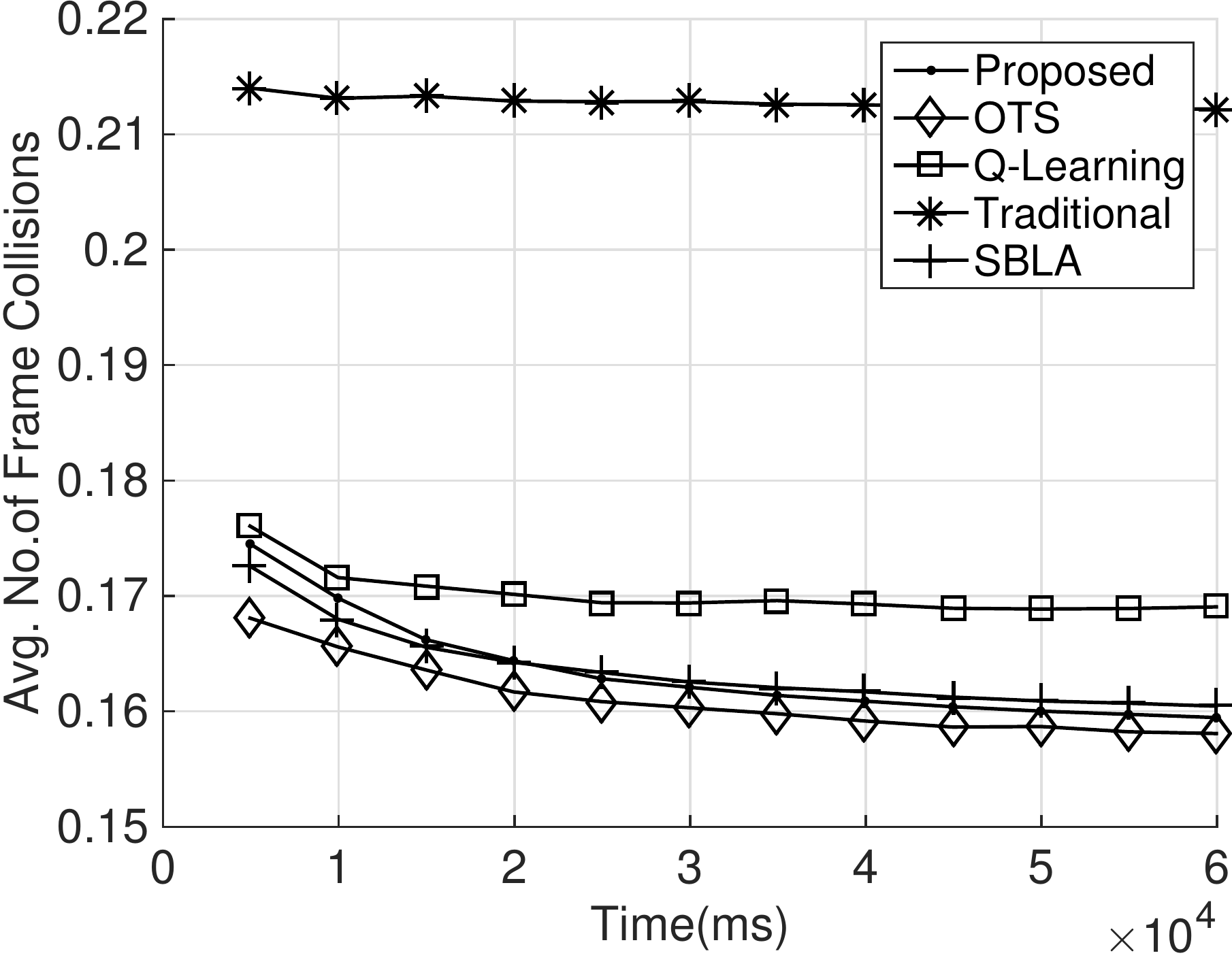}
                \caption{Avg. Frame Collision}
                \label{fig:Exp_C05_F050_S03_FC}
            \end{subfigure}%
            \caption{Results for Exponential Traffic Model for frame size $50$ms for $5$ channels.}
            \label{fig:Exp_C05_F050_S03}
        \end{figure*}
\fi
	\section{Simulation Results} \label{sec:SimRes}
    \subsection{Simulation Setup}
	In this section, we present simulation results to evaluate the performance of our proposed approach with three other learning based algorithms that have been employed in CR literature for channel selection. Most of the papers that use learning techniques in CR networks assume either a Discrete Time Markov model or a channel model based on a Bernoulli Distribution\cite{Zhu2016}. Hence we also consider a primary user traffic model based on Discrete Time Markov Chain (DTMC) model. However it has been pointed out in \cite{lopez2012spectrum} that a continuous time model based on GPD can capture traffic characteristics better. Therefore, we also consider a continuous time primary user traffic model based on GPD model. 
	For DTMC traffic model, we provide results for low, medium and high traffic models. For GPD and exponential models, the parameters are set to mimic  medium traffic in channels. The traffic on each channel is modeled independently by assuming different parameters for different channels in the ranges specified in Table \ref{table:param}. 
	We employ an energy detector to sense the presence of any primary traffic in available channels \cite{atapattu2011energy} during the sensing phase. The number of samples for the energy detector is calculated using the formula given in \cite{liang2008sensing} for a given value of probability of false alarm ($P_f$) and probability of detection($P_d$). For the values of $P_f$ and $P_d$ provided in Table \ref{table:param}, the number of samples required are $1180$. Note that while \cite{digham2007energy,atapattu2011energy,ghasemi2007optimization} implement an energy detector, most of the work that uses RL algorithms assume perfect sensing \cite{Liu2010}, i.e., $P_f = 0$ and $P_d = 1$. We evaluate the performance of the proposed algorithm in terms of three metrics:
 	\begin{enumerate}
 	    \item Normalized Average Achievable Throughput: This metric corresponds to the average observed throughput normalized to maximum throughput achievable. 
 	    \item Average Number of frame collisions: Average number of frame collisions incurred with primary user traffic.
 	    \item Average Number of sensing per frame: Number of sensing operations done at the beginning of each frame until a vacant channel is found.
 	\end{enumerate} 

\begin{table}[h]
    \centering
    \begin{tabular}{ |c|c| } 
         \hline
         Parameter & Value \\ 
         \hline
         DTMC Model Low Traffice (L-I) & $\alpha\in(0, 1]$, $\beta\in[1,5]$   \\
         \hline
         Medium Traffic (M-I)  & $ \alpha \in (0,1]$, $ \beta \in (0,1]$ \\
         \hline
         High Traffic (H-I)    & $ \alpha \in [1,5]$, $ \beta \in [1,5]$ \\
         \hline
         Continuous Traffic Model-GPD & $\sigma=500$,$k\in[0,0.5]$,$\theta\in[50,100]$ \\ 
        \hline
         Exponential Traffic Model & $\theta^{-1} = (0,500]$\\
        \hline
        $P_d$ & 0.95\\
         \hline
        $P_f$ & 0.05\\
         \hline
         Frame duration & $50ms$\\
         \hline
         Sensing duration & $3ms$ \\   
         \hline
         SNR of SU at SU receiver & $20dB$ \\
        \hline
        SNR of PU at SU receiver & $-10dB$ \\
        \hline
        Channel Error & $5\%$\\
        \hline
    \end{tabular}
    \caption{Parameters used for simulation}
    \label{table:param}    
\end{table} 	

The throughput is calculated as in \cite{liang2008sensing,Yang2015a} using the formula given below.
	\begin{align}
	    TP_{achievable} &= \frac{T - k \tau}{T} \cdot C
 	\end{align}  
 	 
 	\[ 
 	    C = 
 	        \begin{cases}
                log_2\left( 1 + SNR_{sec} \right) & ;\text{on successful transmission}\\
                0 & ;\text{on frame loss}
            \end{cases}
    \]

Here, $T$ is the frame duration and $\tau$ is the sensing duration. $k$ denotes the number of channels sensed before finding a vacant channel and can take values between $1$ and $N$, which is the total number of channels. The SNR of the secondary user ($SNR_{sec}$) at the SU receiver is taken to be $20dB$. To report the results, we normalize the achieved throughput to maximum throughput that can be achieved. This metric can vary between $[0,1]$ with $1$ indicating maximum usage of spectrum for data transmission. Normalized throughput is calculated as
\begin{align}
    TP_{normlaized} = \frac{1}{C_{max}} \cdot TP_{achievable}
\end{align}
where $C_{max} = \log_2( 1 + SNR_{sec} )$.

We have also taken into account a possibility of channel error. To the best of our knowledge, existing literature in CR applying learning techniques does not consider channel error. Channel error causes the frame to be lost even when the channel was not actually occupied by the primary user. In all the simulation results provided below, the probability of channel error is taken to be $0.05$. This was done to simulate a more realistic environment. 
All the results reported are plotted for values averaged over $1000$ independent runs for a time duration of $60$s. The proposed algorithms are compared with Q-Learning (QL) \cite{Li2009}, Switchable Bayesian Learning Automata (SBLA) \cite{Zhang2013a}, Optimistic Thompson Sampling (OTS) \cite{May2012} and a traditional method (RAND) \cite{Yang2015a}. The traditional method is the method where we pick the channel to sense randomly till we find a vacant channel or till all channels are sensed.
    
        For all the metrics i.e., throughput, number of sensing and frame collisions, we plot the cumulative values normalized to the number of frames the SU attempts to transmit till time $t$, say $F_t$. That is, for a metric $X$, we plot
        \begin{align}
            y_t = \frac{1}{F_t} \cdot \sum \limits_{n=1}^{t} X_{n}
        \end{align}
        This metric is akin to the sample-averaged reward definition popularly used to evaluate reinforcement learning algorithms. The aim of this work is to maximize the achievable throughput of secondary user while minimizing number of sensing required by secondary user and keeping the collisions with PU to be comparable to that of the existing RL algorithms in literature which sense the the channel in every frame.
        
\ifCLASSOPTIONonecolumn

\else
        \begin{figure*}[h]
            \begin{subfigure}{.33\textwidth}
                \includegraphics[width=1.\linewidth]{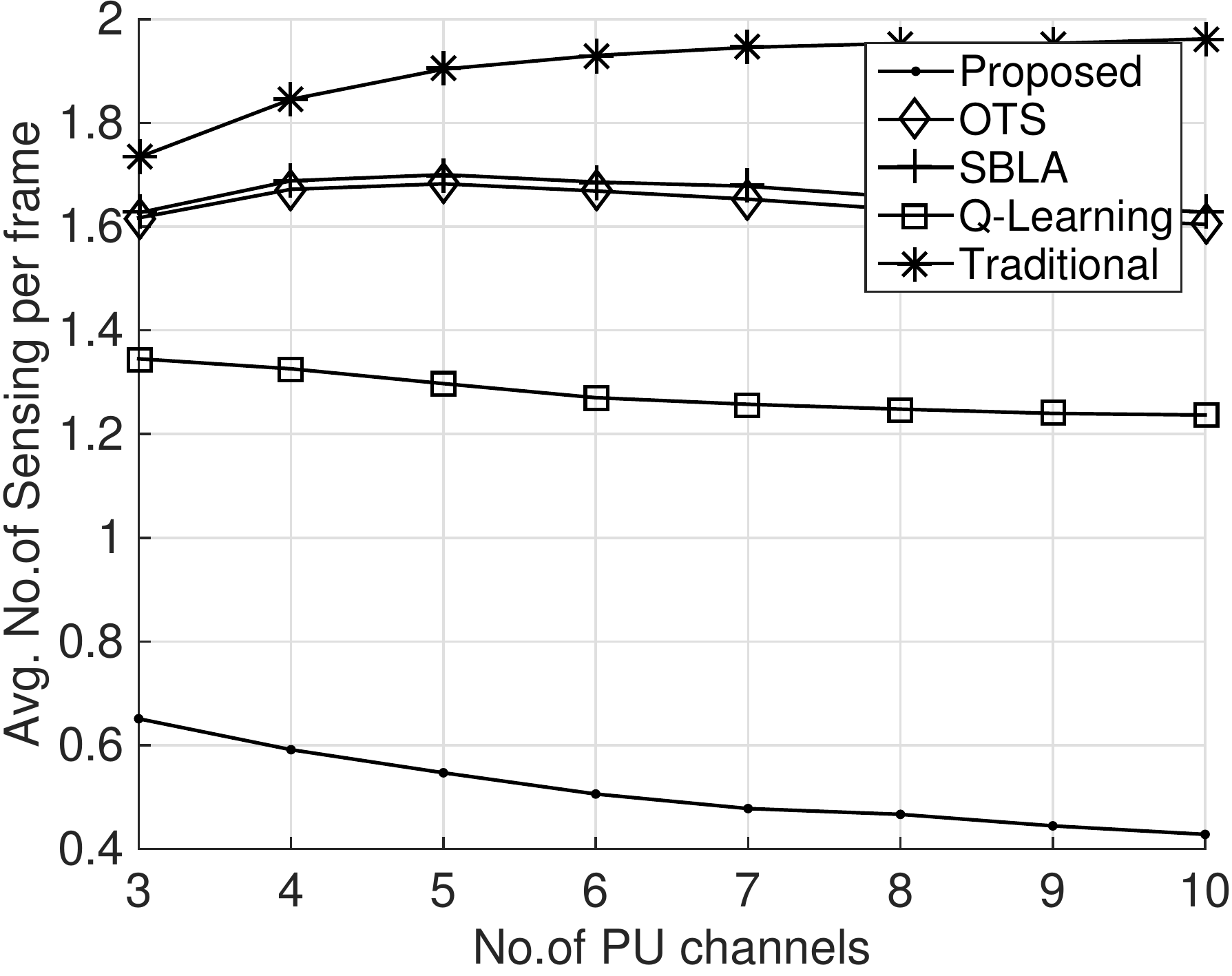}
                \caption{Avg. No.of Sensing per frame}
                \label{fig:CH_GPD_F050_S03_SE}
            \end{subfigure}%
            \begin{subfigure}{.33\textwidth}
                \includegraphics[width=1.\linewidth]{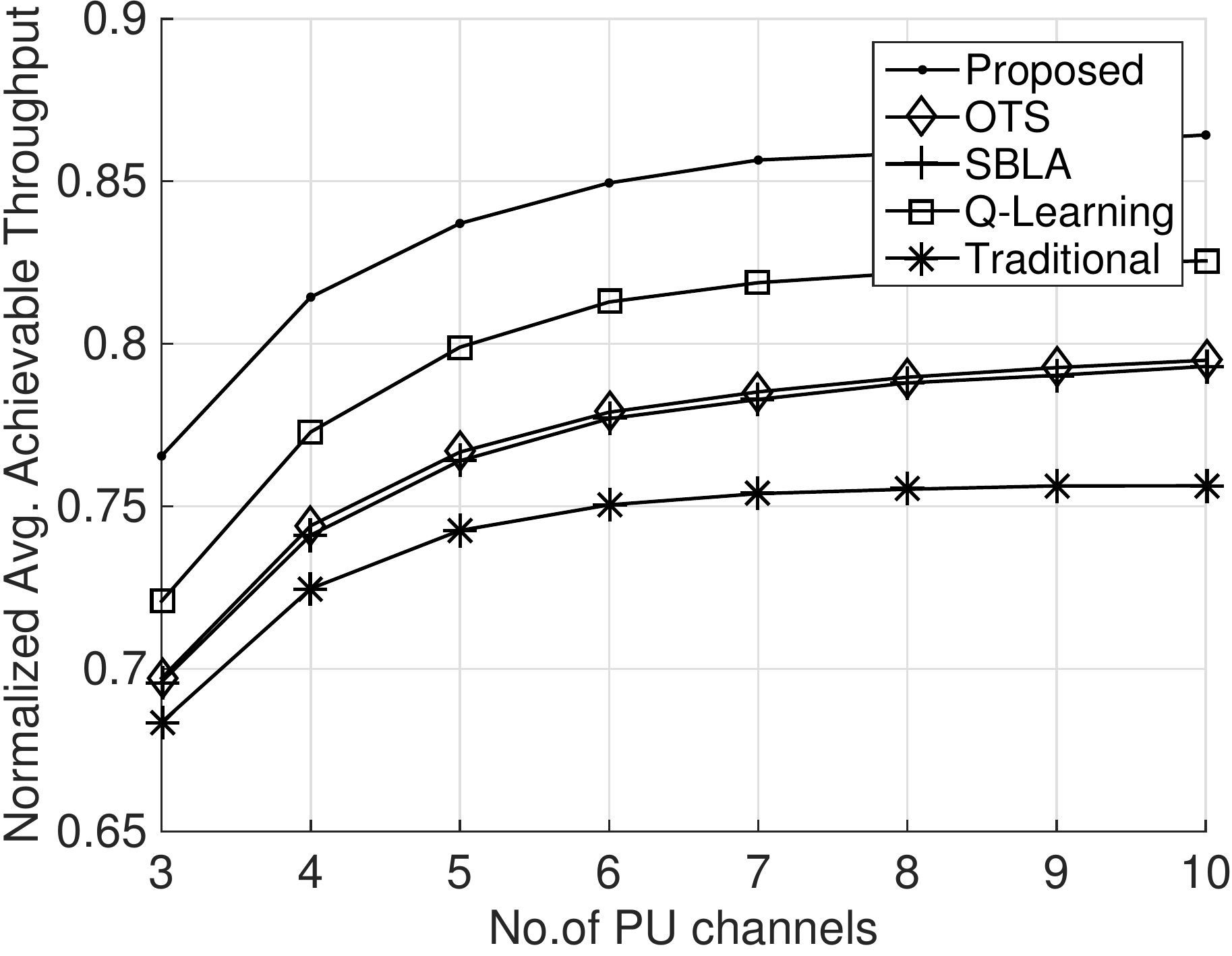}
                \caption{Avg. Achievable Throughput}
                \label{fig:CH_GPD_F050_S03_TP}
            \end{subfigure}%
            \begin{subfigure}{.33\textwidth}
                \includegraphics[width=1.\linewidth]{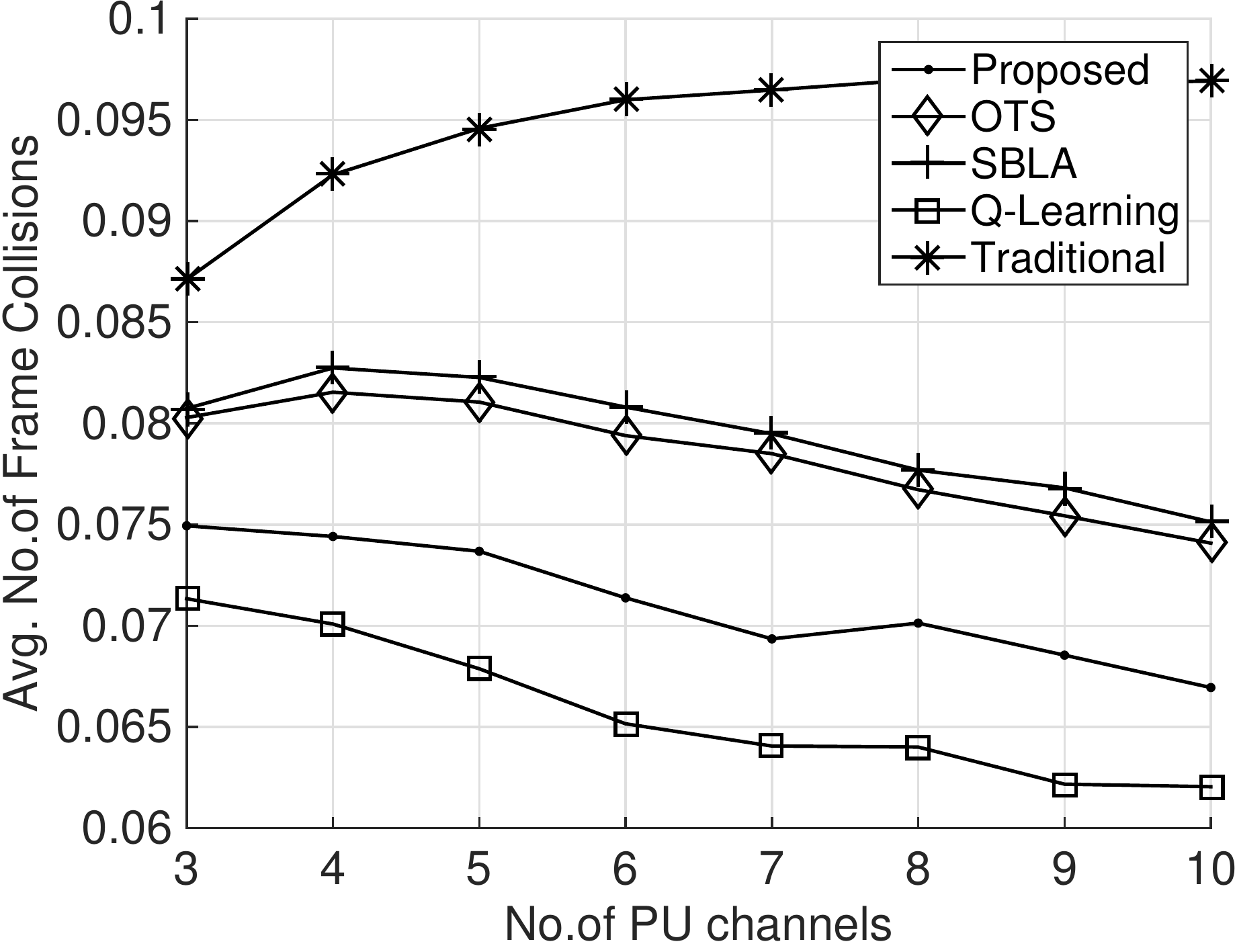}
                \caption{Avg. Frame Collision}
                \label{fig:CH_GPD_F050_S03_FC}
            \end{subfigure}%
            \caption{Results for GPD Traffic Model with varying number of channels for frame size $50$ms.}
            \label{fig:CH_GPD_F050_S03}
        \end{figure*}

        \begin{figure*}[h]
            \begin{subfigure}{.33\textwidth}
                \includegraphics[width=1.\linewidth]{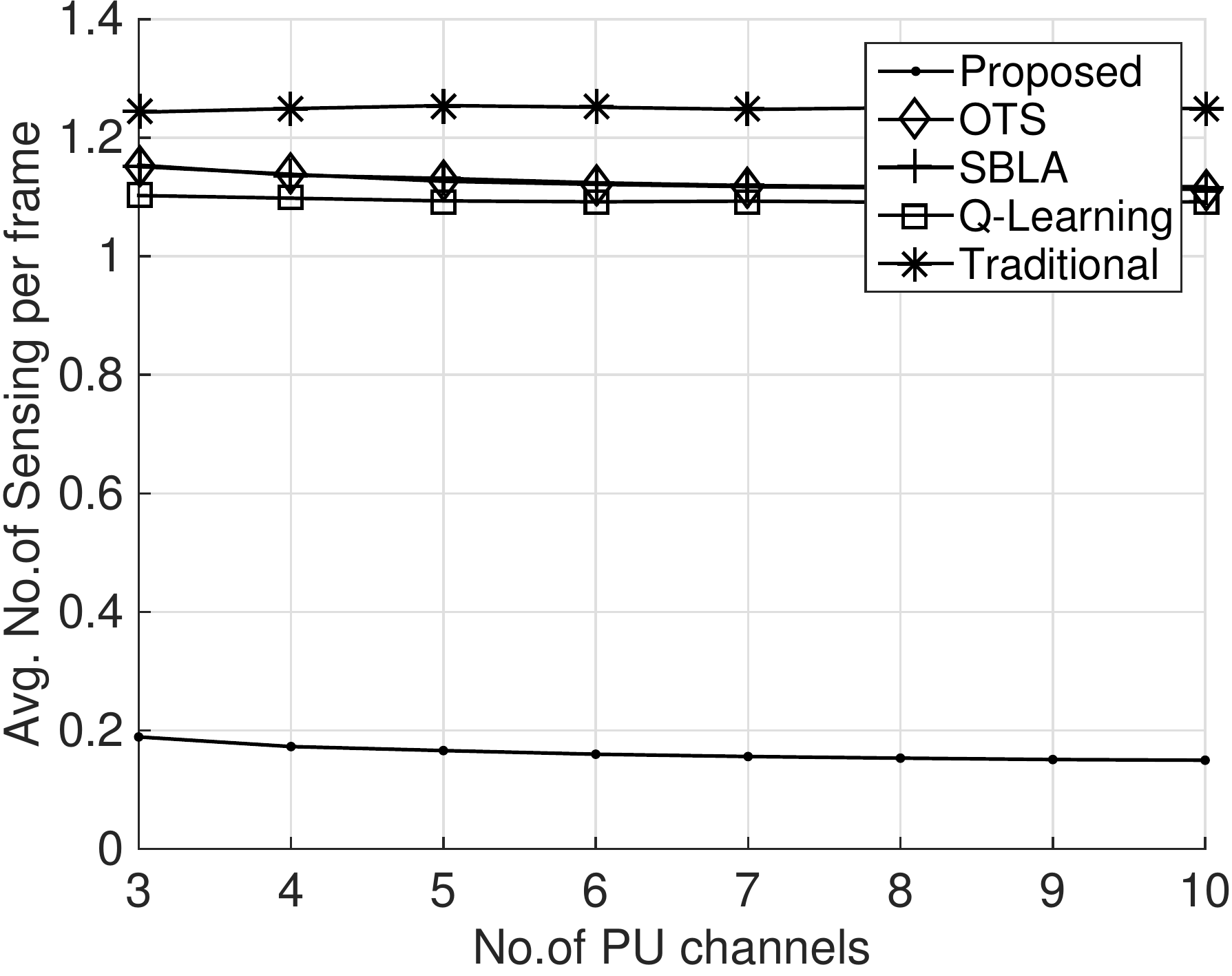}
                \caption{Avg. No.of Sensing per frame}
                \label{fig:CH_DTMC_Beta_L-I_F050_S03_SE}
            \end{subfigure}%
            \begin{subfigure}{.33\textwidth}
                \includegraphics[width=1.\linewidth]{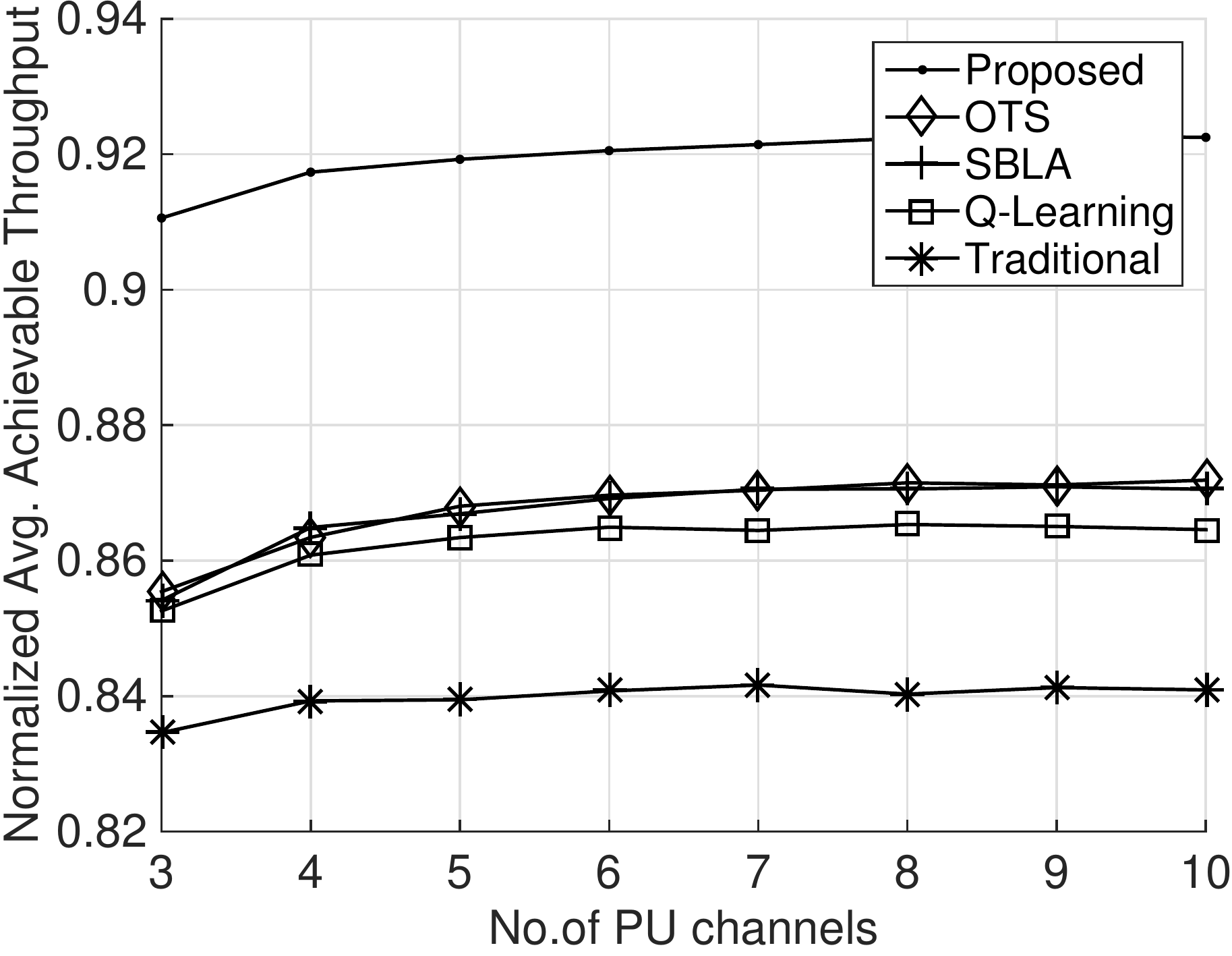}
                \caption{Avg. Achievable Throughput}
                \label{fig:CH_DTMC_Beta_L-I_F050_S03_TP}
            \end{subfigure}%
            \begin{subfigure}{.33\textwidth}
                \includegraphics[width=1.\linewidth]{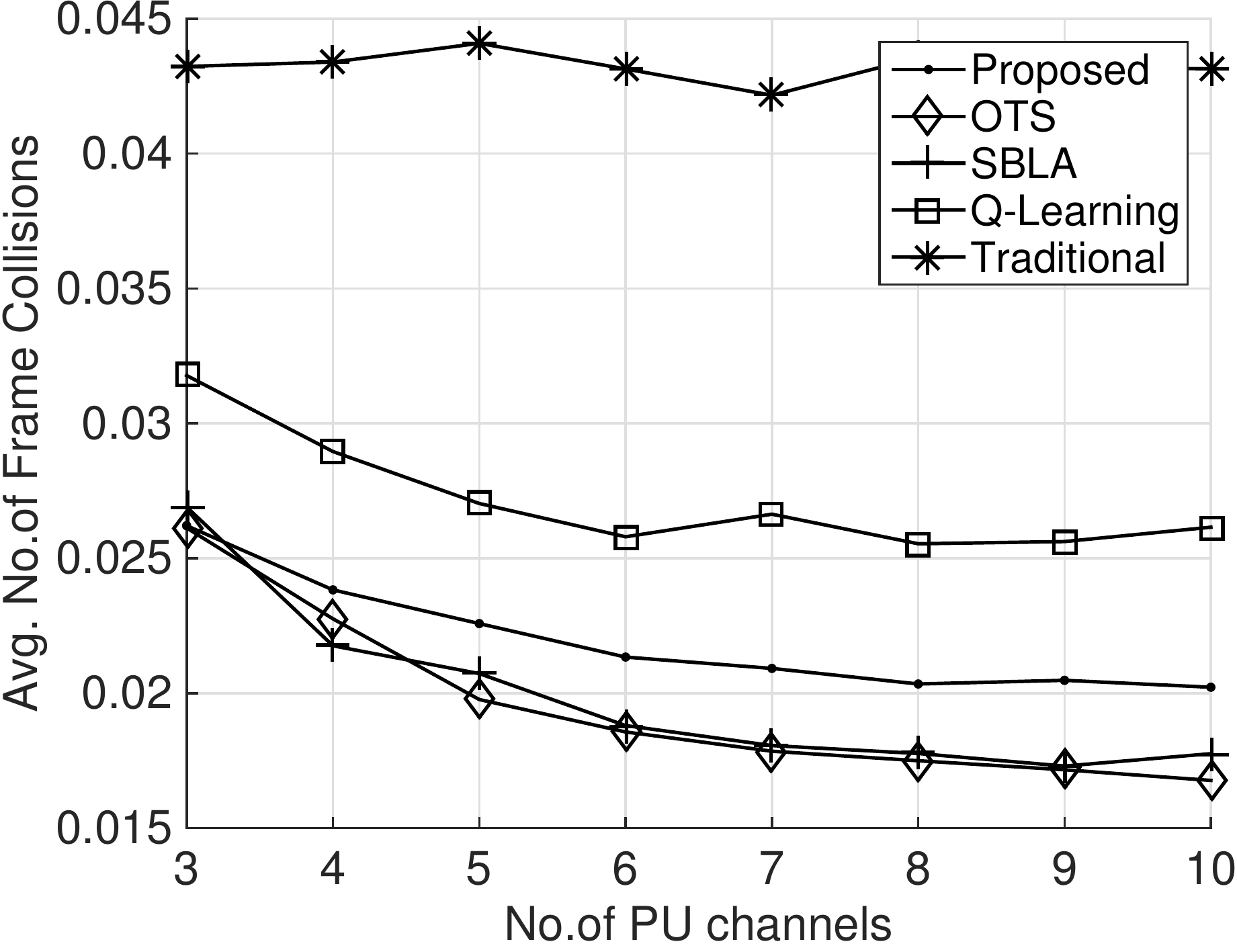}
                \caption{Avg. Frame Collision}
                \label{fig:CH_DTMC_Beta_L-I_F050_S03_FC}
            \end{subfigure}%
            \caption{Results for DTMC Low Traffic Model with varying number of channels for frame size $50$ms.}
            \label{fig:CH_DTMC_Beta_L-I_F050_S03}
        \end{figure*}
        
        \begin{figure*}[h]
            \begin{subfigure}{.33\textwidth}
                \includegraphics[width=1.\linewidth]{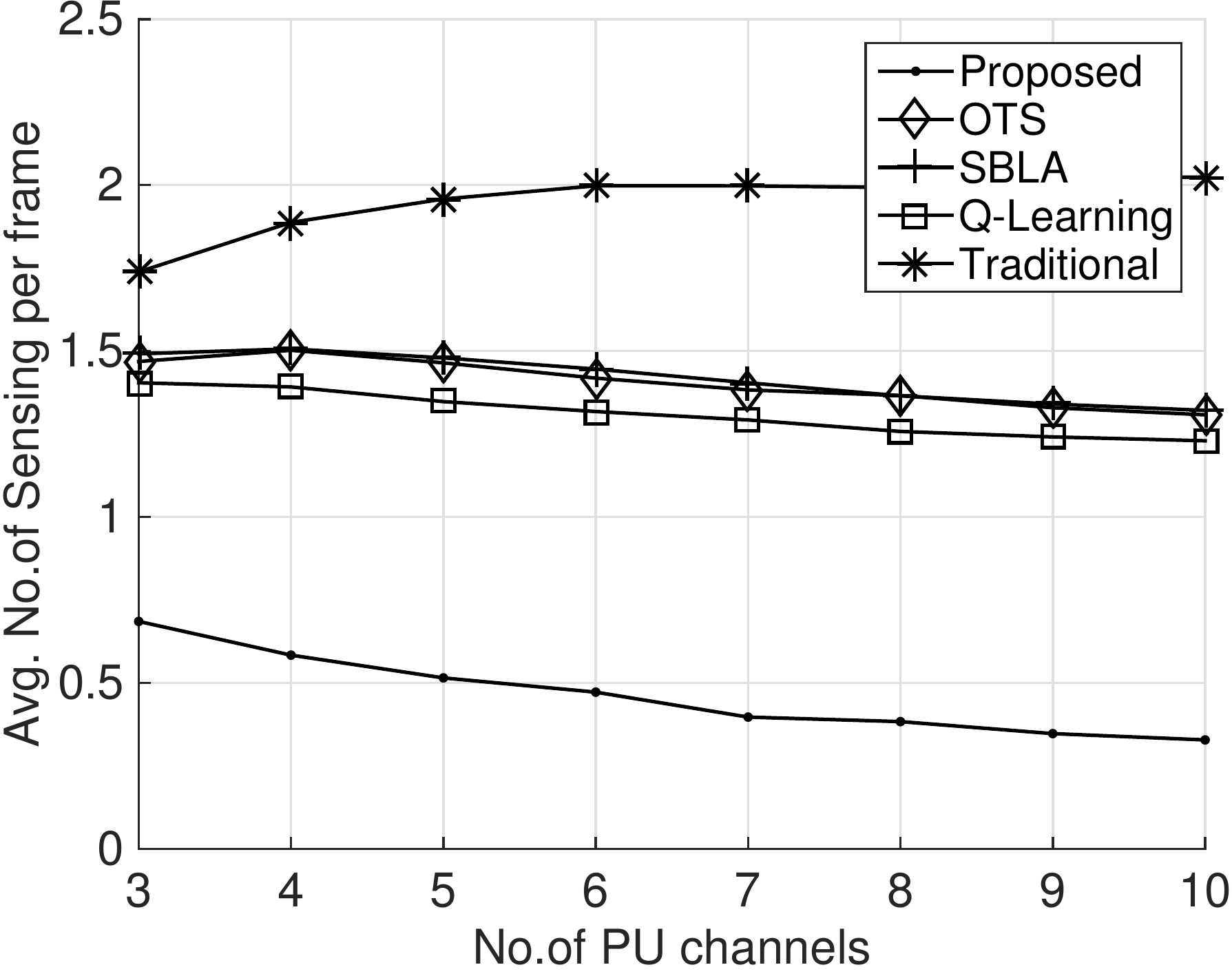}
                \caption{Avg. No.of Sensing per frame}
                \label{fig:CH_DTMC_Beta_M-I_F050_S03_SE}
            \end{subfigure}%
            \begin{subfigure}{.33\textwidth}
                \includegraphics[width=1.\linewidth]{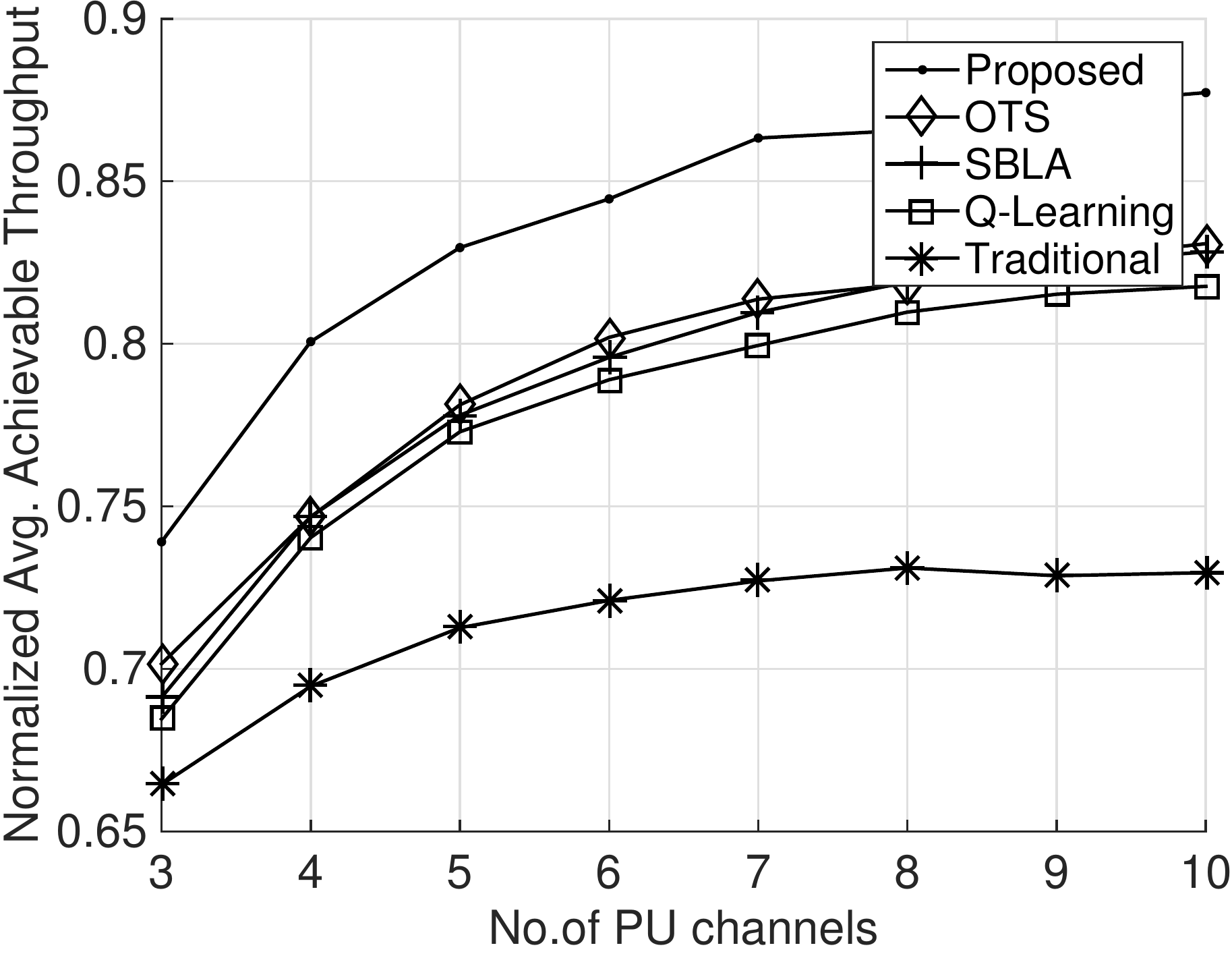}
                \caption{Avg. Achievable Throughput}
                \label{fig:CH_DTMC_Beta_M-I_F050_S03_TP}
            \end{subfigure}%
            \begin{subfigure}{.33\textwidth}
                \includegraphics[width=1.\linewidth]{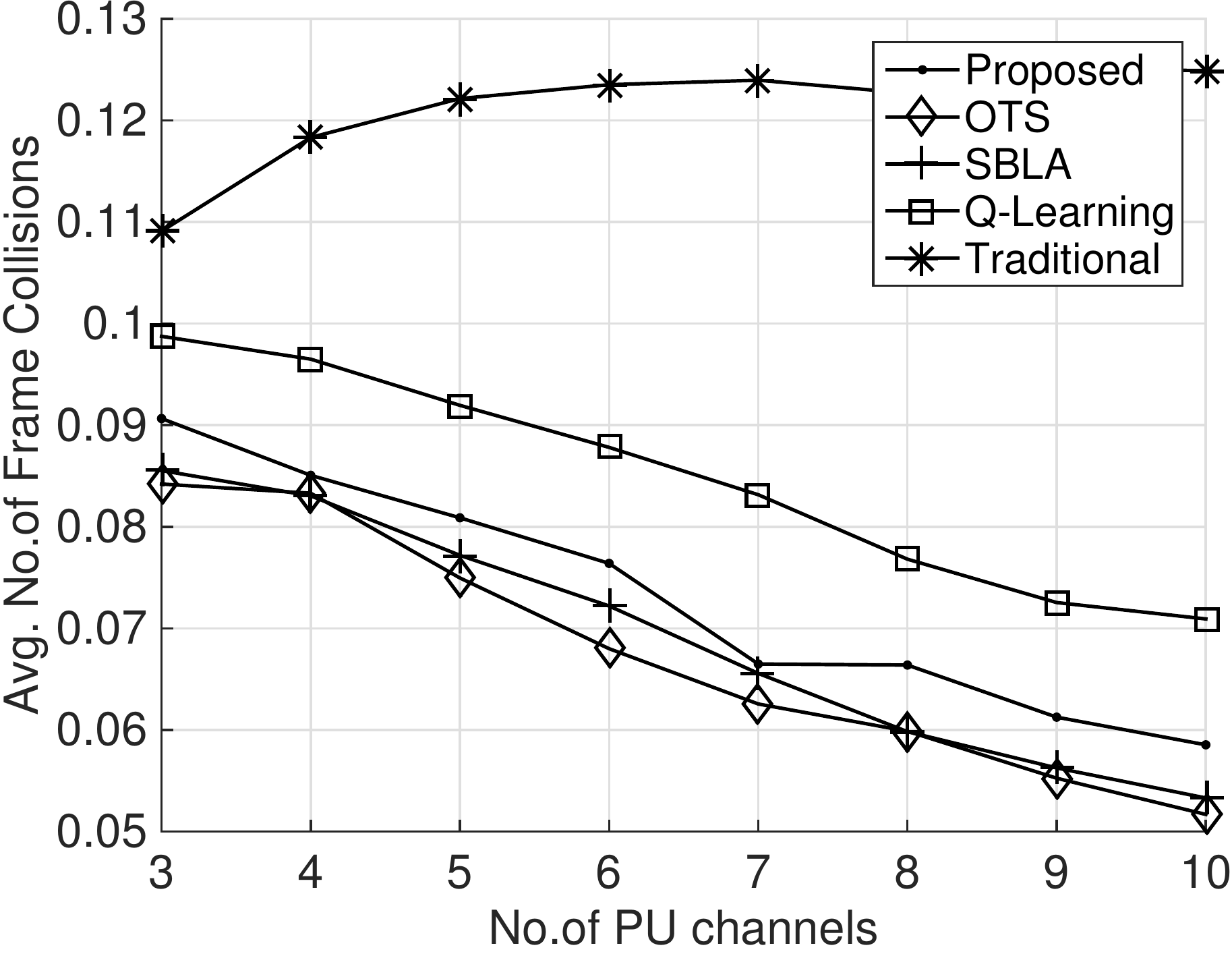}
                \caption{Avg. Frame Collision}
                \label{fig:CH_DTMC_Beta_M-I_F050_S03_FC}
            \end{subfigure}%
            \caption{Results for DTMC Medium Traffic Model with varying number of channels for frame size $50$ms.}
            \label{fig:CH_DTMC_Beta_M-I_F050_S03}
        \end{figure*}
        

        \begin{figure*}[h]
            \begin{subfigure}{.33\textwidth}
                \includegraphics[width=1.\linewidth]{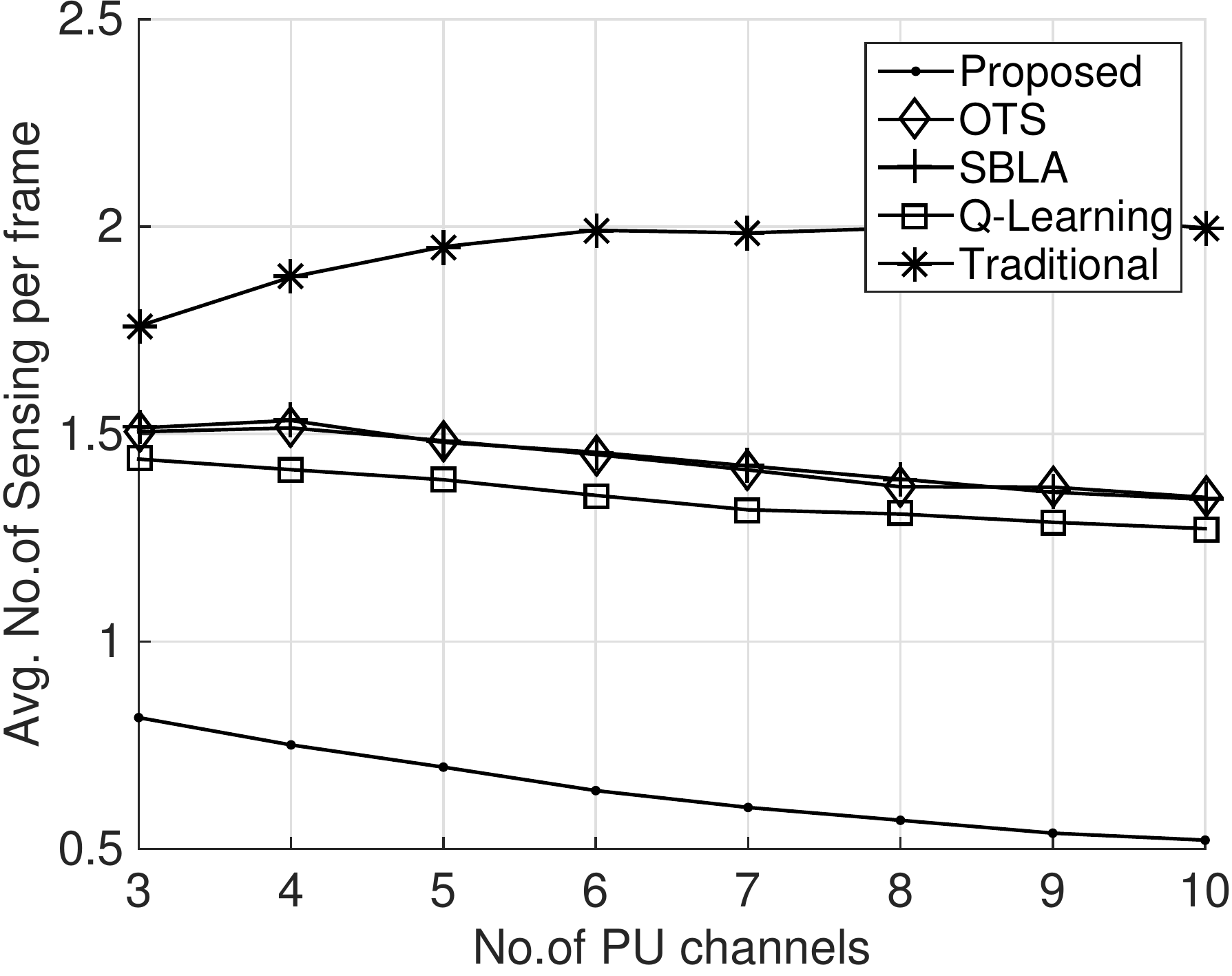}
                \caption{Avg. No.of Sensing per frame}
                \label{fig:CH_Exp_F050_S03_SE}
            \end{subfigure}%
            \begin{subfigure}{.33\textwidth}
                \includegraphics[width=1.\linewidth]{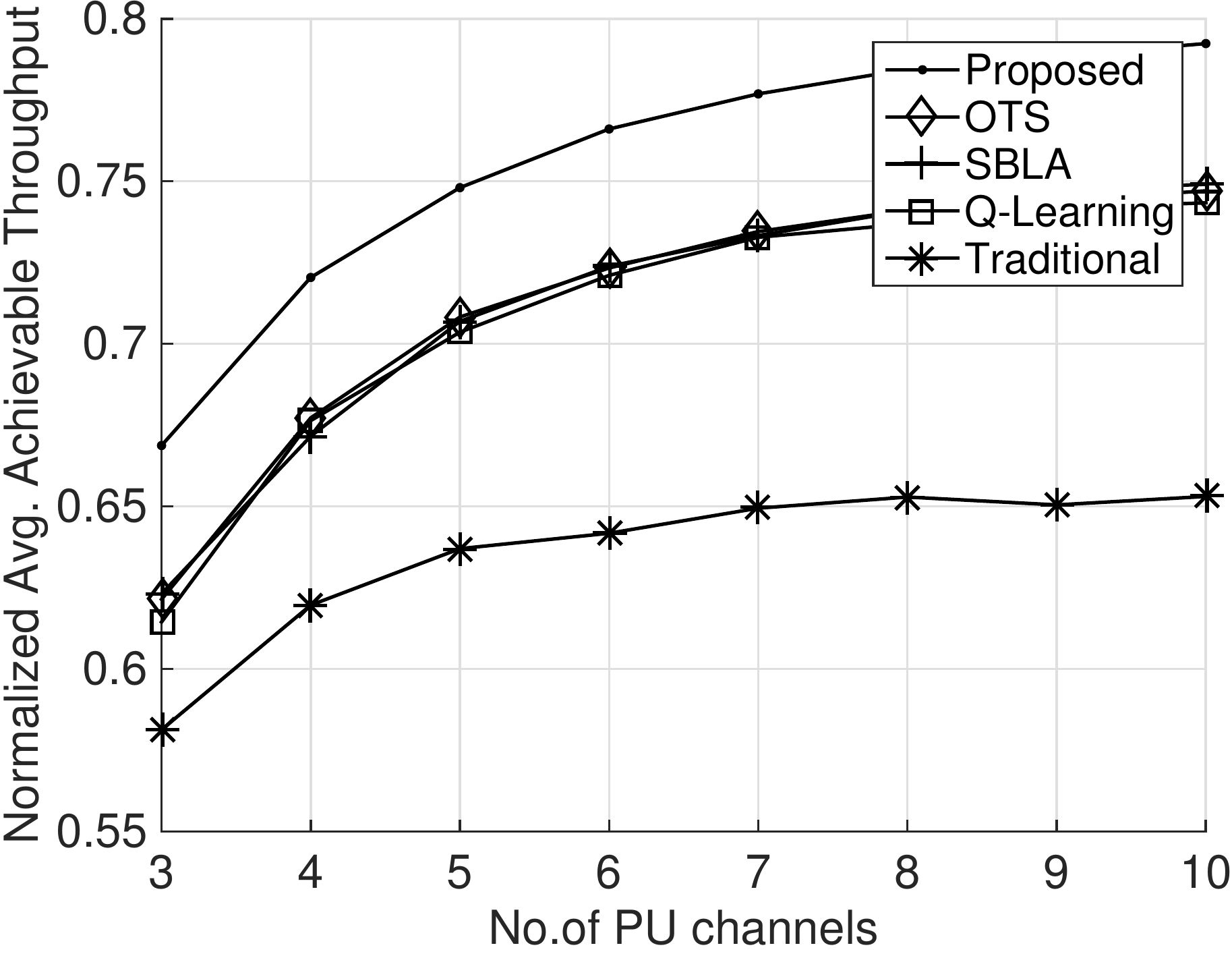}
                \caption{Avg. Achievable Throughput}
                \label{fig:CH_Exp_F050_S03_TP}
            \end{subfigure}%
            \begin{subfigure}{.33\textwidth}
                \includegraphics[width=1.\linewidth]{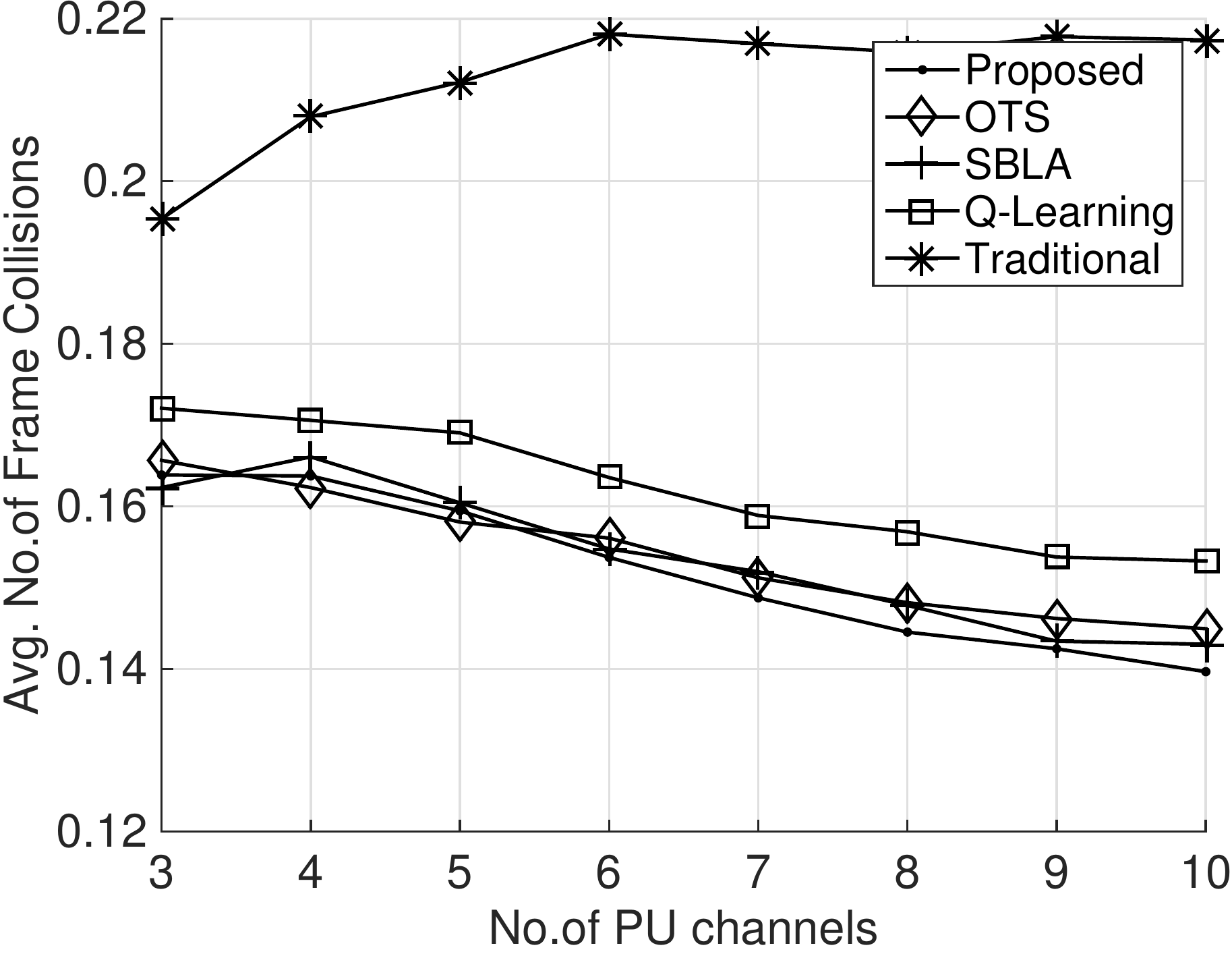}
                \caption{Avg. Frame Collision}
                \label{fig:CH_Exp_F050_S03_FC}
            \end{subfigure}%
            \caption{Results for Exponential Traffic Model with varying number of channels for frame size $50$ms.}
            \label{fig:CH_Exp_F050_S03}
        \end{figure*}
\fi

    \subsection{Learning over time}
        In this section, we present the results of how each of the methods perform over time. We consider a 5 channel system ($N=5$) with a channel error of $5\%$. 
        
        Figure \ref{fig:GPD_C05_F050_S03} shows the results for GPD traffic model. From the figures, we can see that the proposed method is able to provide a $10\%$ improvement in throughput and is able to reduce the number of sensing significantly. While the comparing algorithms require to do $1.5$ sensing to transmit one frame on an average, the proposed method requires only $1/3rd$ of that. The proposed approach causes only a $0.5\%$ increase in frame collision as compared to the the best performing method (Q-Learning), which is negligible considering the improvements in other metrics.

\ifCLASSOPTIONonecolumn
        \begin{figure*}[h]
            \begin{subfigure}{.33\textwidth}
                \includegraphics[width=1.\linewidth]{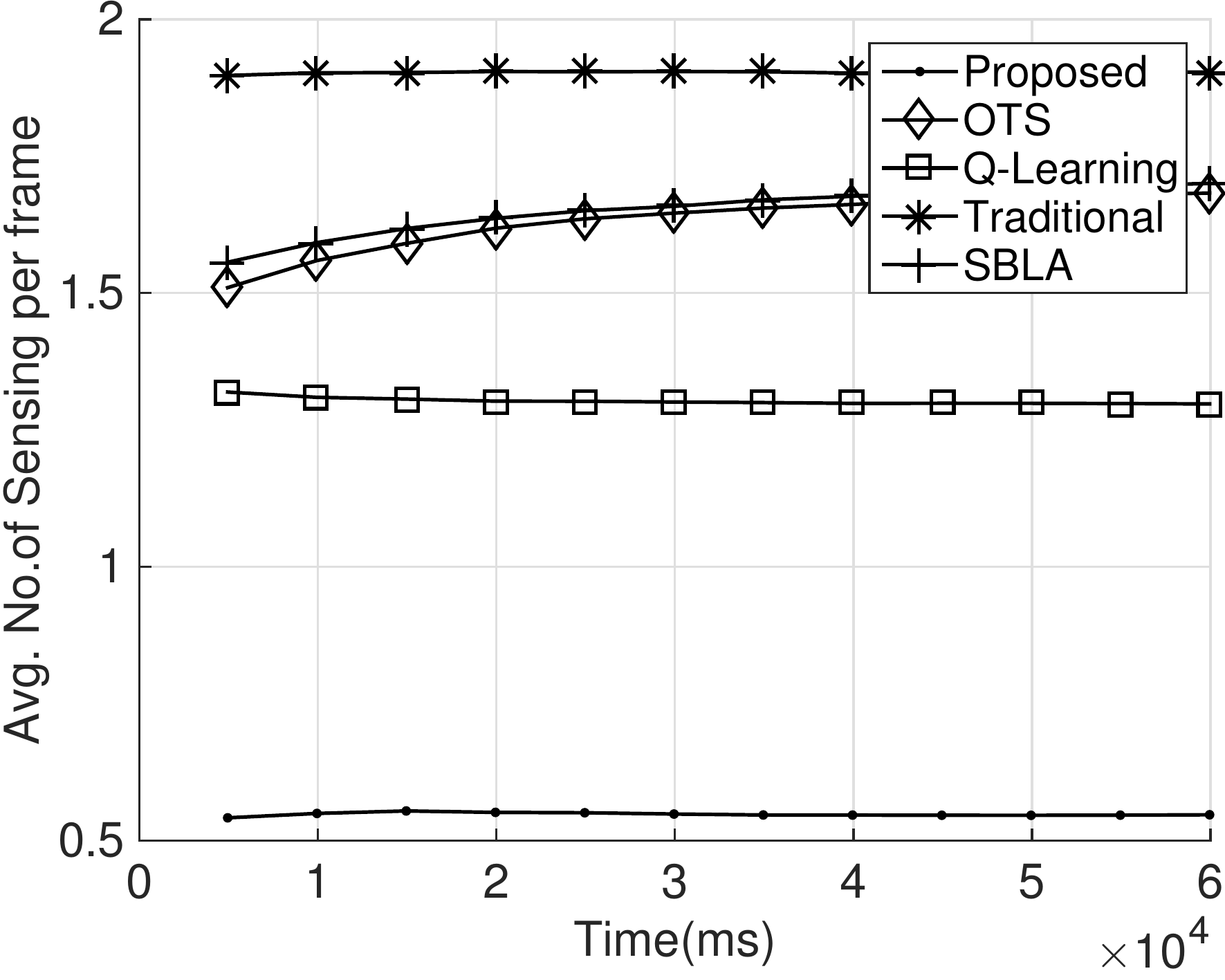}
                \caption{Avg. No.of Sensing per frame}
                \label{fig:GPD_C05_F050_S03_SE}
            \end{subfigure}%
            \begin{subfigure}{.33\textwidth}
                \includegraphics[width=1.\linewidth]{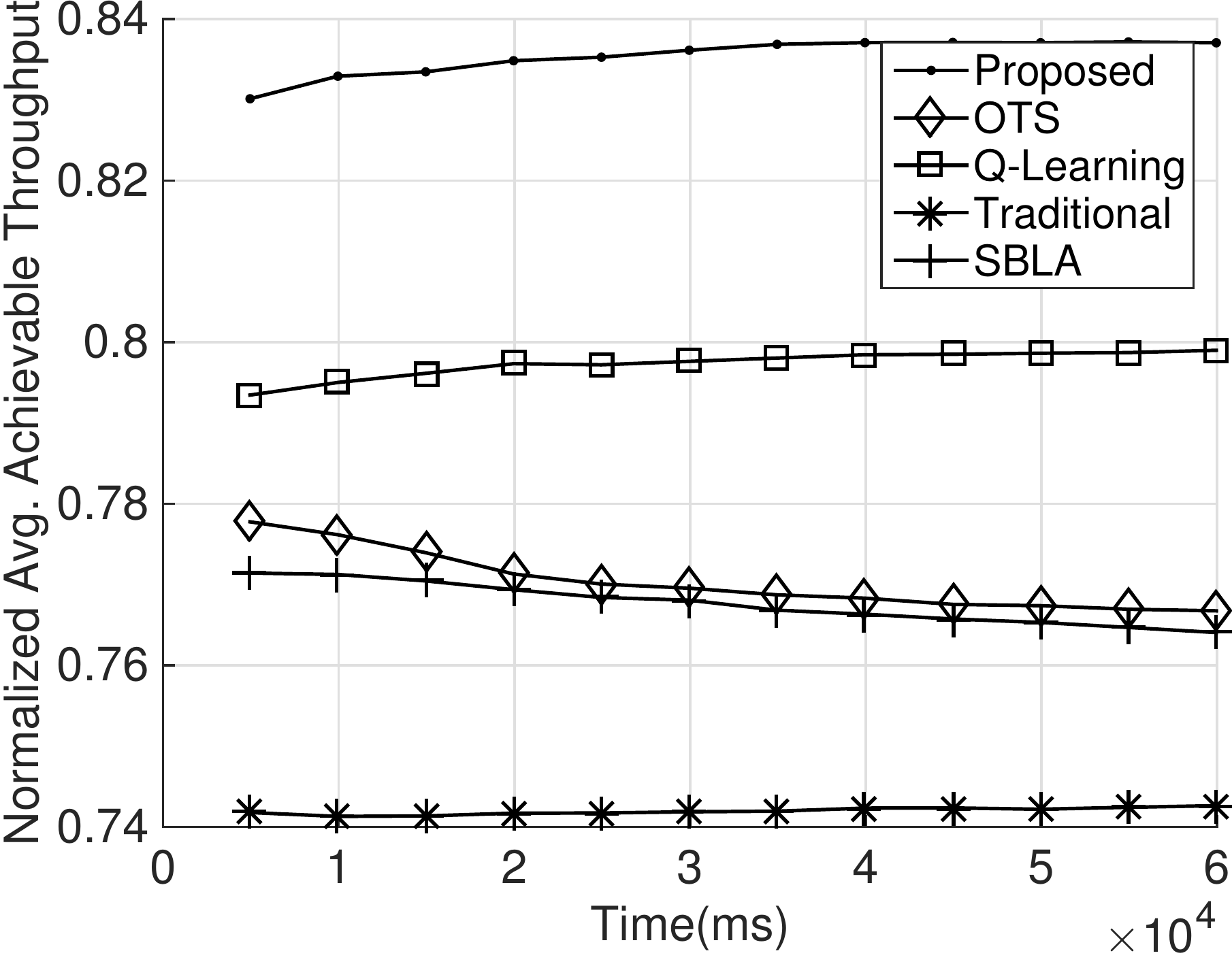}
                \caption{Avg. Achievable Throughput}
                \label{fig:GPD_C05_F050_S03_TP}
            \end{subfigure}%
            \begin{subfigure}{.33\textwidth}
                \includegraphics[width=1.\linewidth]{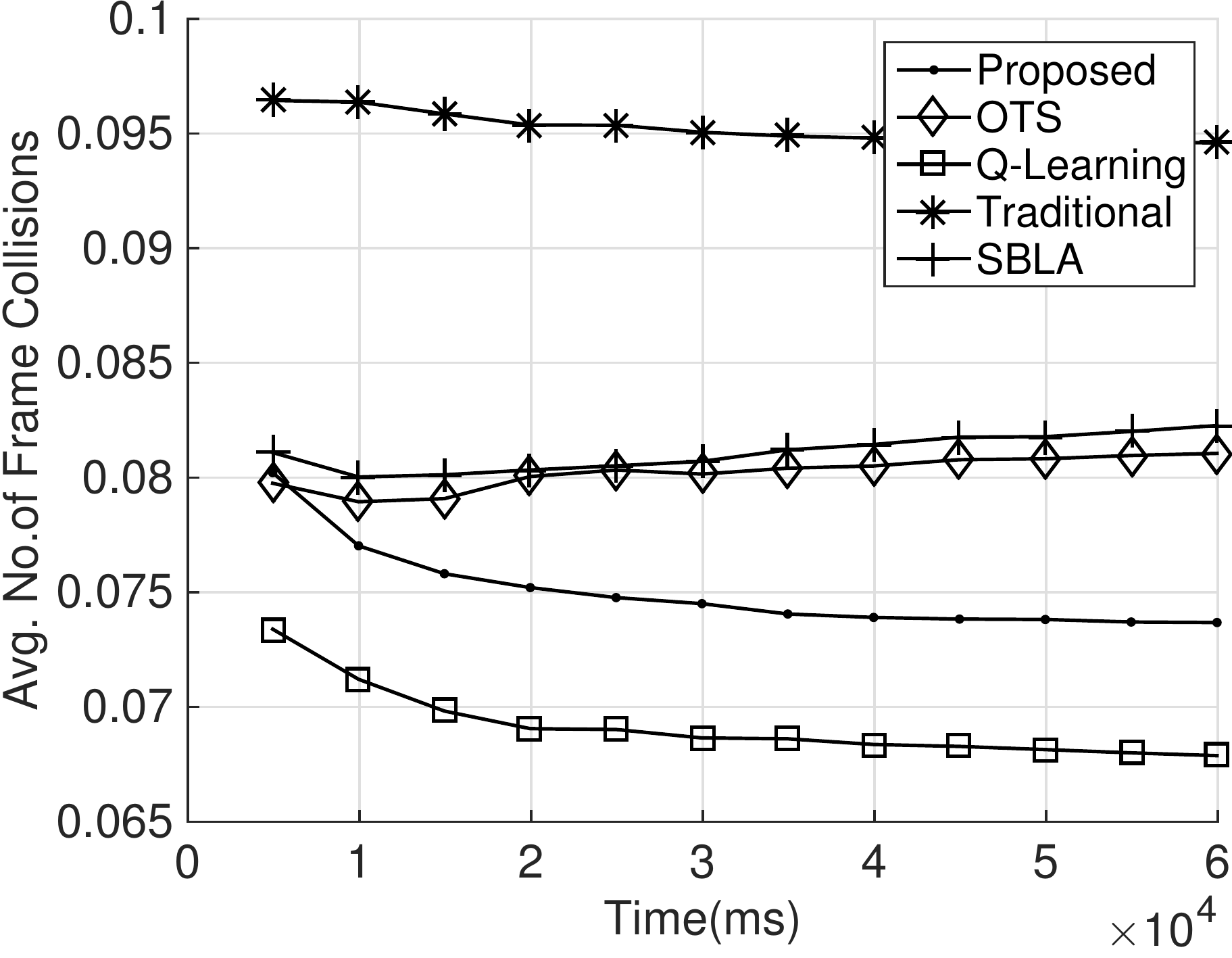}
                \caption{Avg. Frame Collision}
                \label{fig:GPD_C05_F050_S03_FC}
            \end{subfigure}%
            \caption{Results for GPD Model for frame size $50$ms for $5$ channels.}
            \label{fig:GPD_C05_F050_S03}
        \end{figure*}
\else

\fi
        
        Figures \ref{fig:DTMC_Beta_L-I_C05_F050_S03}-\ref{fig:DTMC_Beta_H-I_C05_F050_S03} show the results for the DTMC model for three different intensities of traffic. In all the traffic intensities that were considered, we can see that the proposed method is able to attain a gain of approximately $8\%$ in throughput while creating only around $0.5\%$ increase in frame collision when compared to the best performing alternative while significantly reducing the number of sensing required. In case of high traffic intensity, we can see that the proposed method is able to achieve considerable improvement in terms of number of sensing required and achievable throughput with only marginal increase in frame collision to PU. Overall, we can conclude that, the proposed method shows improvement in the popular DTMC traffic model in all traffic intensities when compared to other methods.
        
\ifCLASSOPTIONonecolumn
        \begin{figure*}[h]
            \begin{subfigure}{.33\textwidth}
                \includegraphics[width=1.\linewidth]{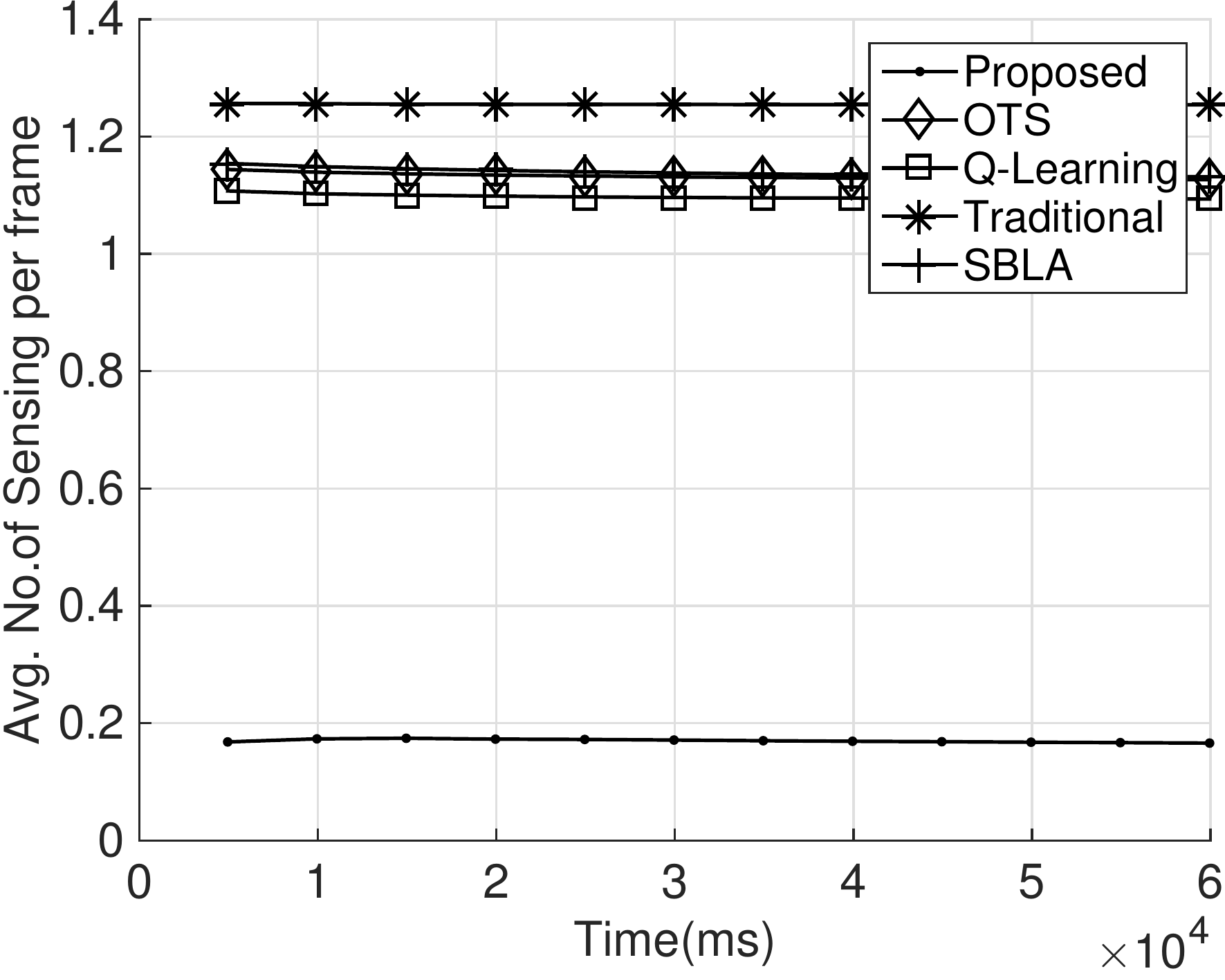}
                \caption{Avg. No.of Sensing per frame}
                \label{fig:DTMC_Beta_L-I_C05_F050_S03_SE}
            \end{subfigure}%
            \begin{subfigure}{.33\textwidth}
                \includegraphics[width=1.\linewidth]{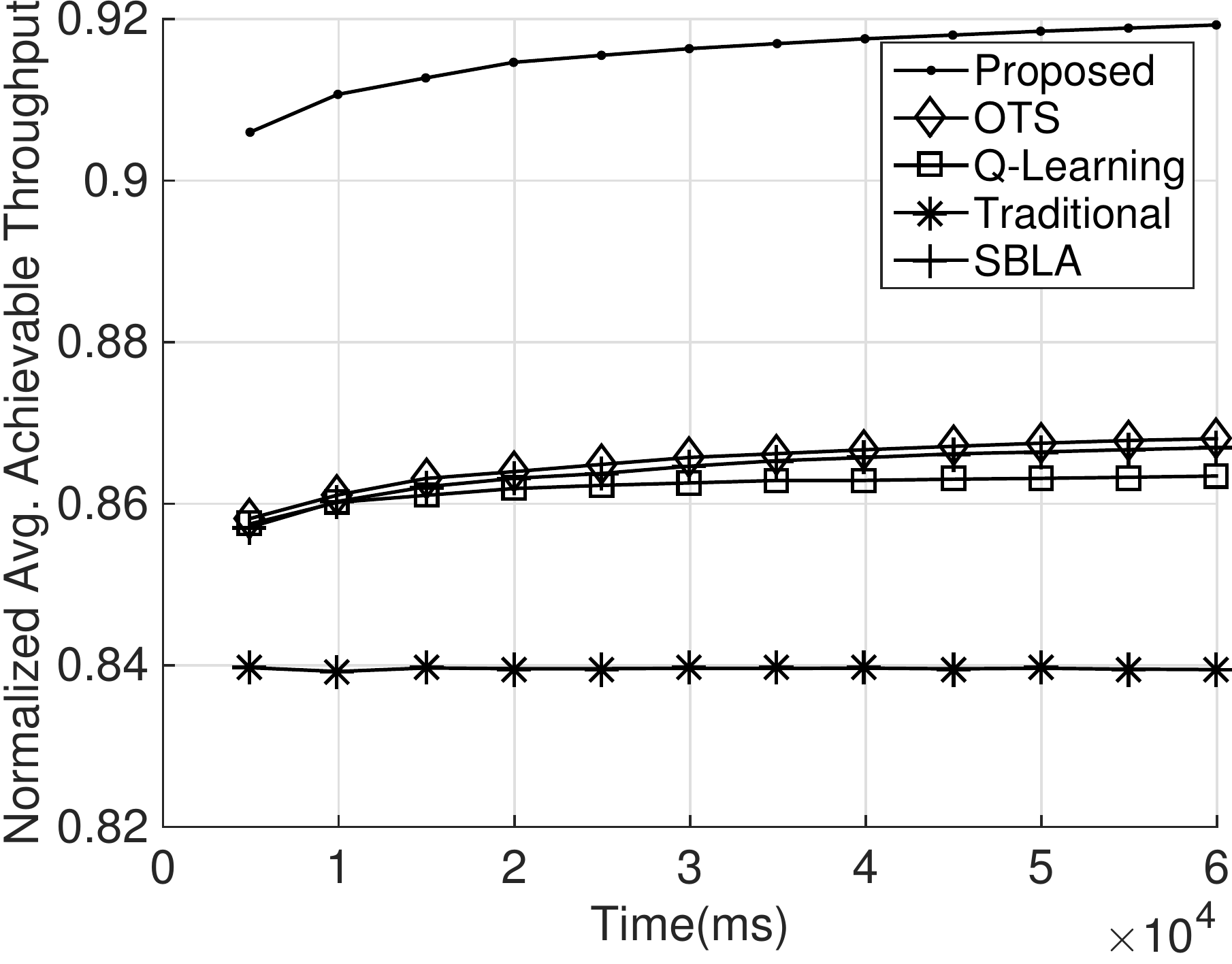}
                \caption{Avg. Achievable Throughput}
                \label{fig:DTMC_Beta_L-I_C05_F050_S03_TP}
            \end{subfigure}%
            \begin{subfigure}{.33\textwidth}
                \includegraphics[width=1.\linewidth]{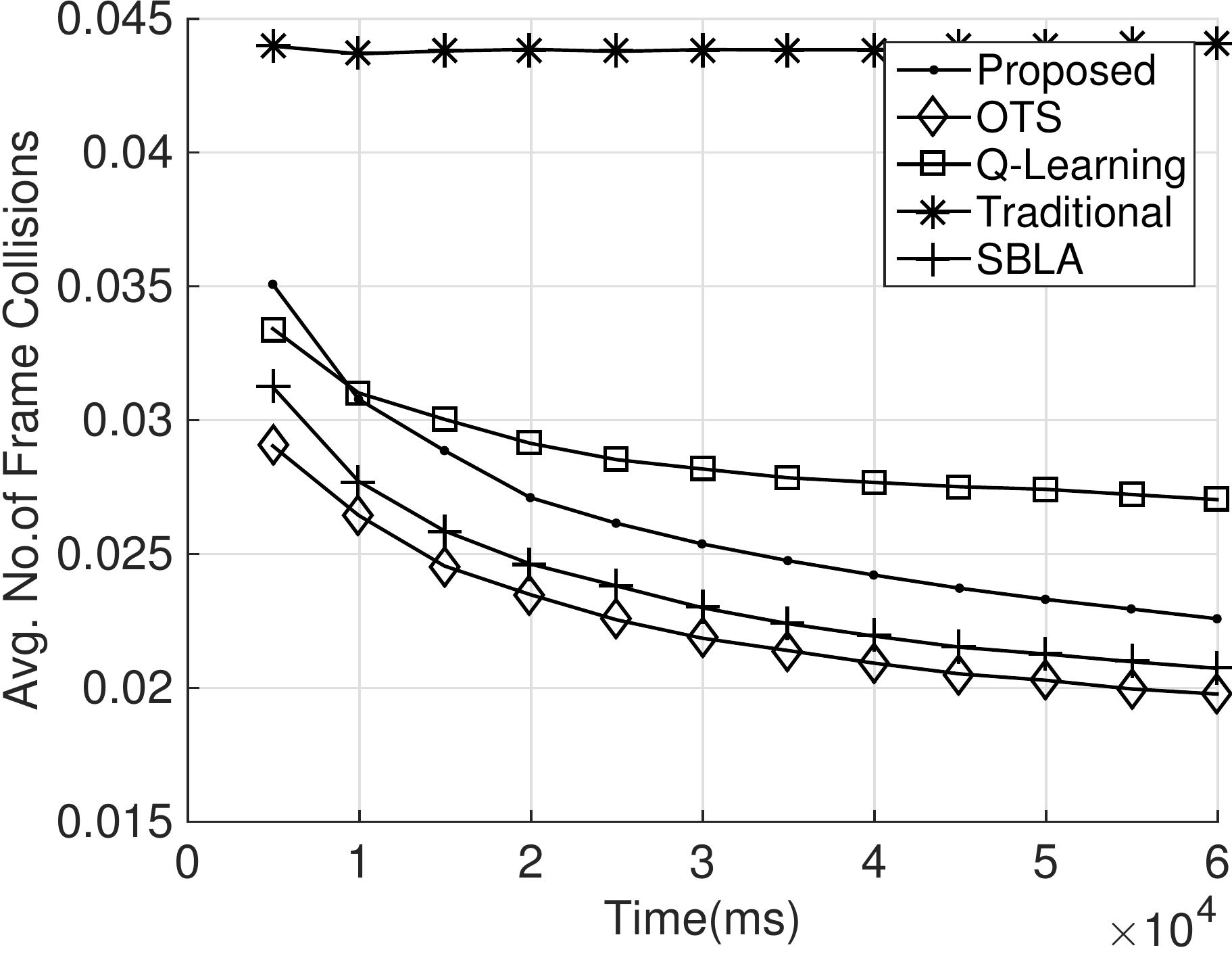}
                \caption{Avg. Frame Collision}
                \label{fig:DTMC_Beta_L-I_C05_F050_S03_FC}
            \end{subfigure}%
            \caption{Results for DTMC Low Traffic Model (DTMC L-I) for frame size $50$ms for $5$ channels.}
            \label{fig:DTMC_Beta_L-I_C05_F050_S03}
        \end{figure*}
        
        \begin{figure*}[h]
            \begin{subfigure}{.33\textwidth}
                \includegraphics[width=1.\linewidth]{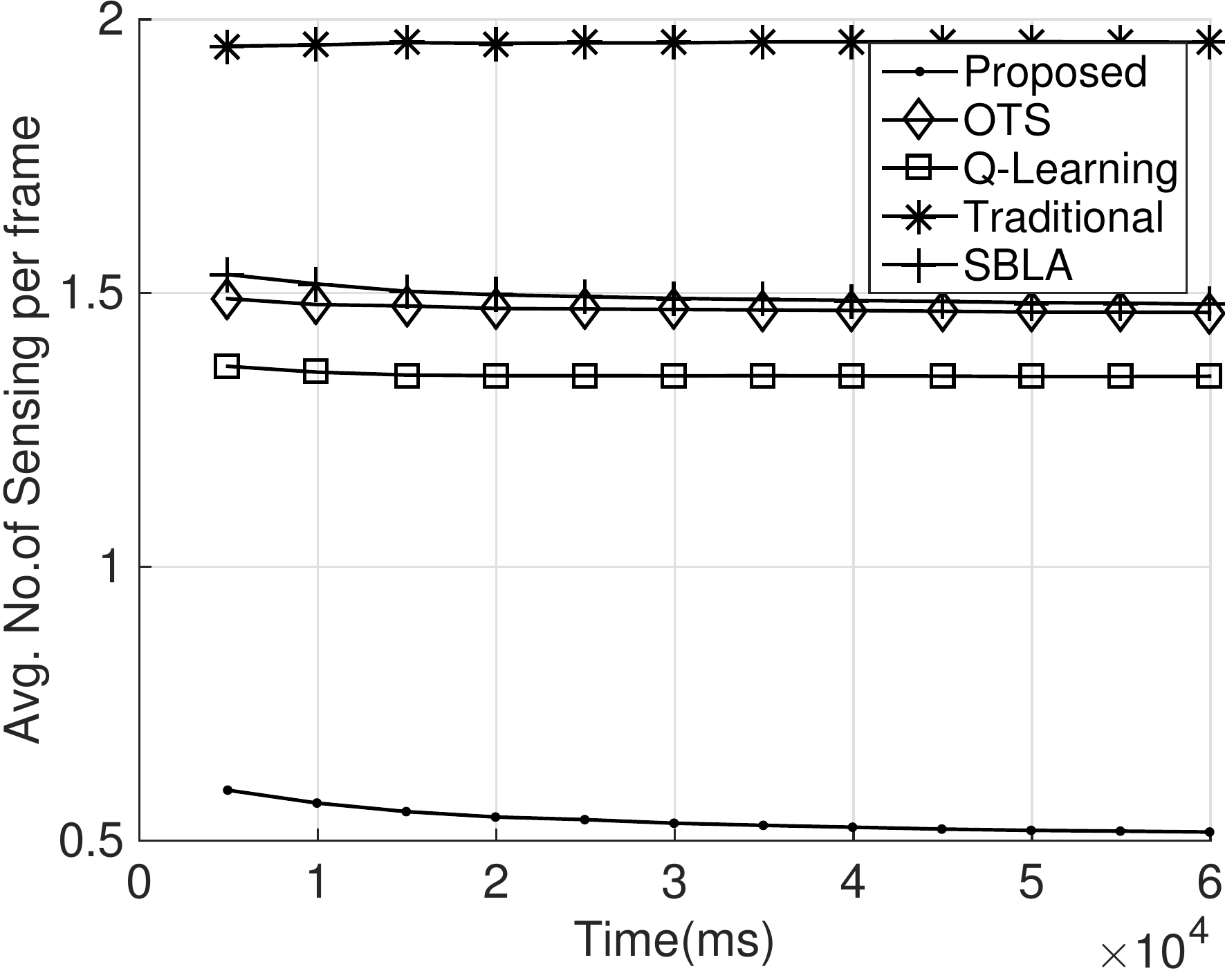}
                \caption{Avg. No.of Sensing per frame}
                \label{fig:DTMC_Beta_M-I_C05_F050_S03_SE}
            \end{subfigure}%
            \begin{subfigure}{.33\textwidth}
                \includegraphics[width=1.\linewidth]{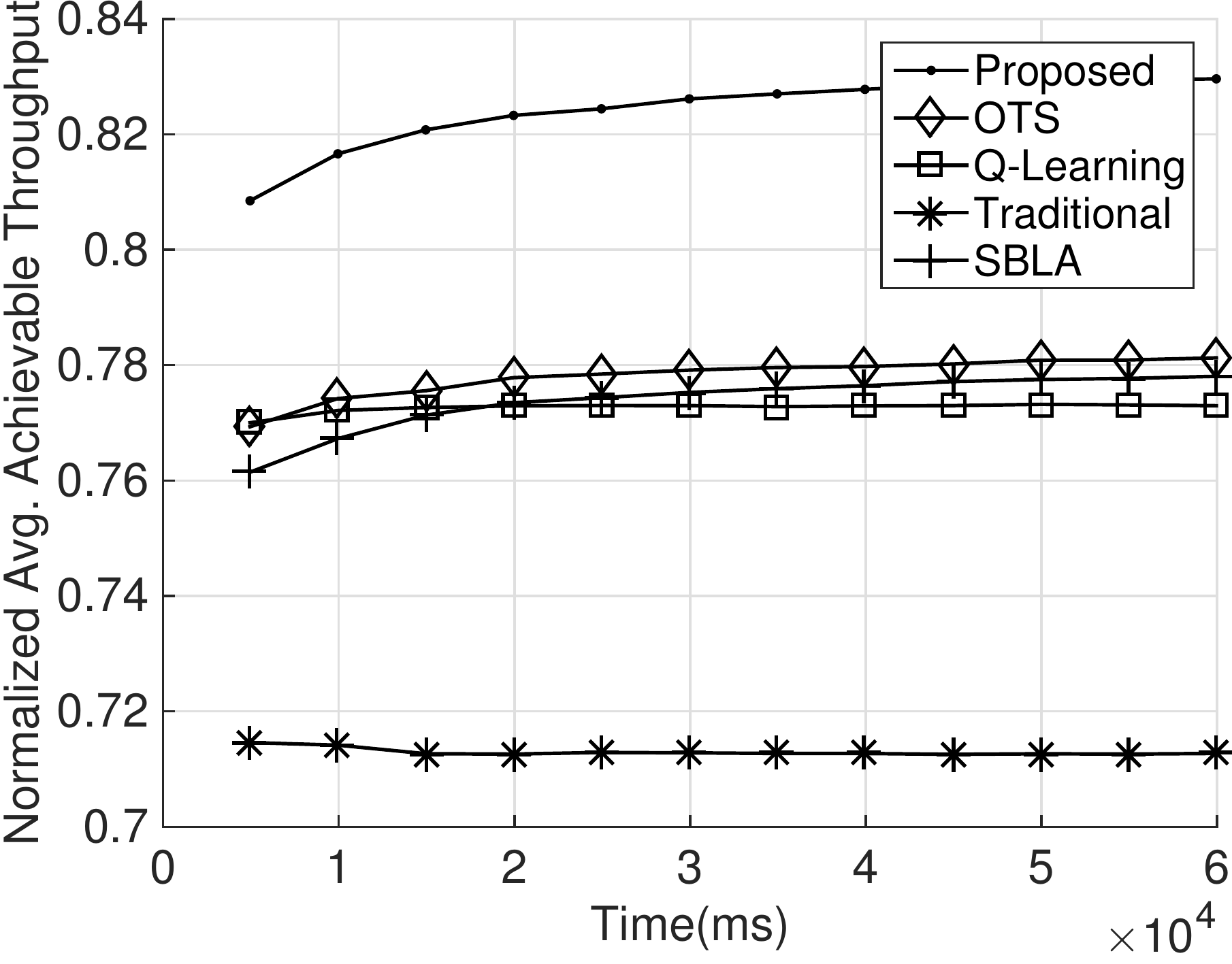}
                \caption{Avg. Achievable Throughput}
                \label{fig:DTMC_Beta_M-I_C05_F050_S03_TP}
            \end{subfigure}%
            \begin{subfigure}{.33\textwidth}
                \includegraphics[width=1.\linewidth]{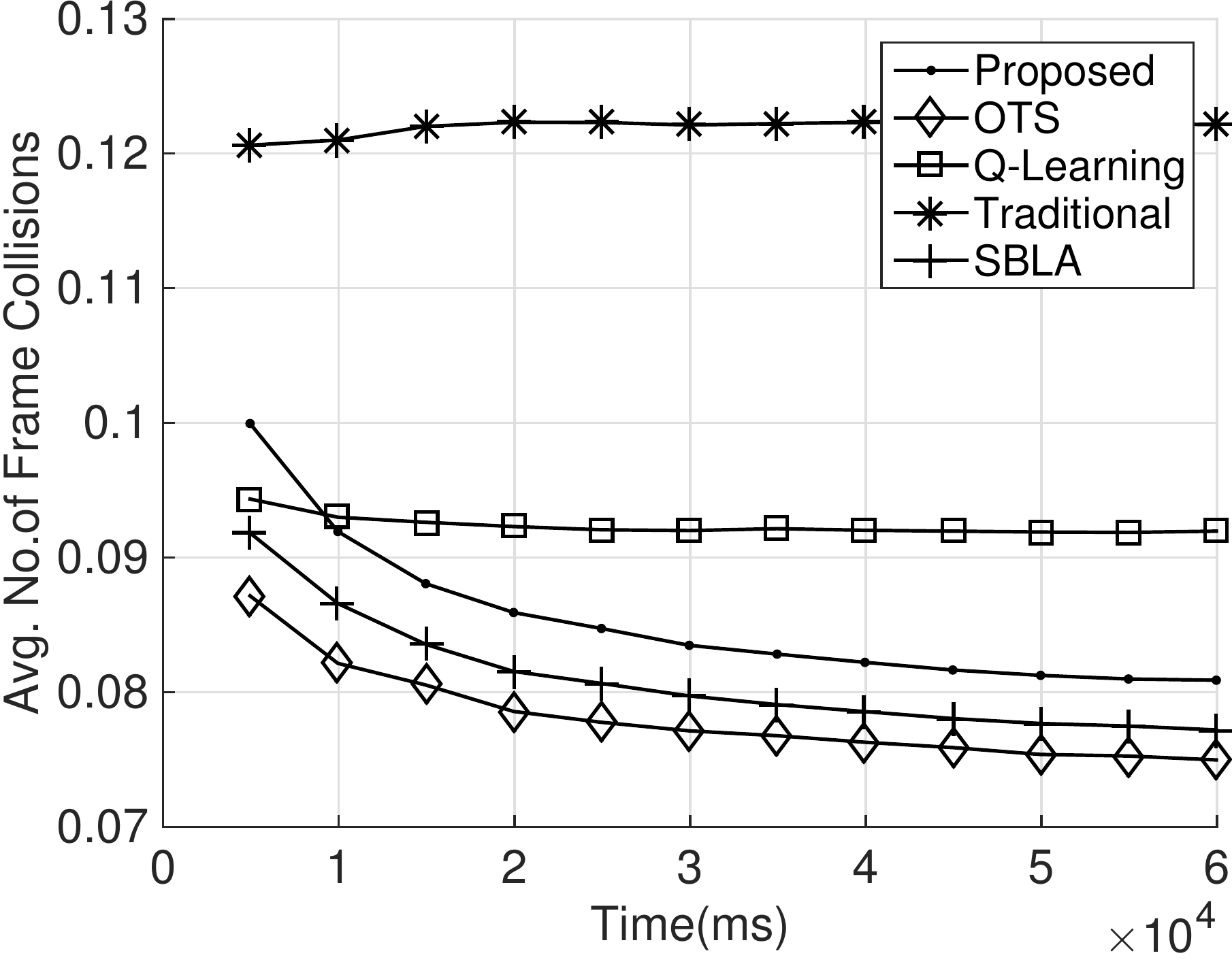}
                \caption{Avg. Frame Collision}
                \label{fig:DTMC_Beta_M-I_C05_F050_S03_FC}
            \end{subfigure}%
            \caption{Results for DTMC Medium Traffic Model (DTMC M-I) for frame size $50$ms for $5$ channels.}
            \label{fig:DTMC_Beta_M-I_C05_F050_S03}
        \end{figure*}
        
        \begin{figure*}[h]
            \begin{subfigure}{.33\textwidth}
                \includegraphics[width=1.\linewidth]{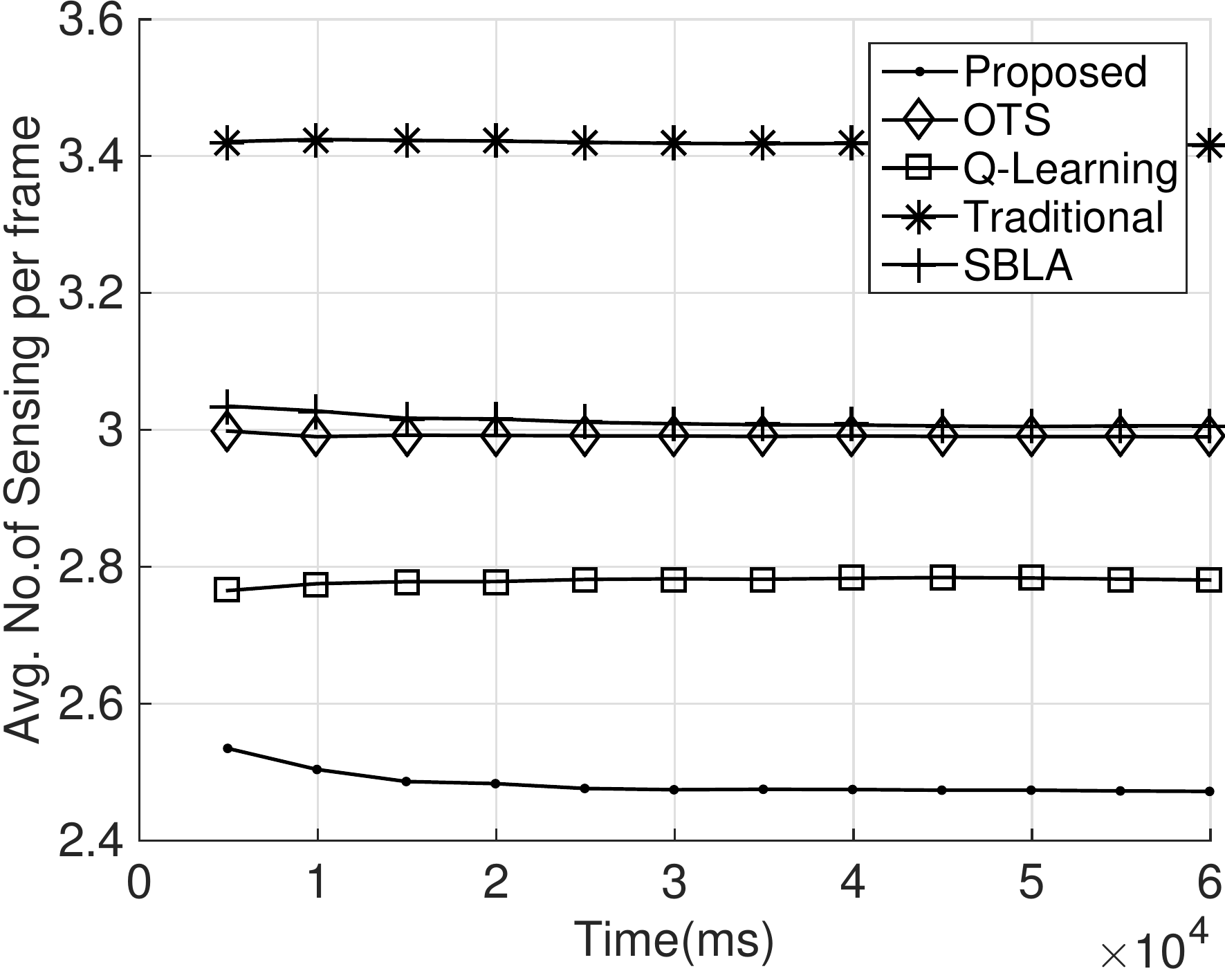}
                \caption{Avg. No.of Sensing per frame}
                \label{fig:DTMC_Beta_H-I_C05_F050_S03_SE}
            \end{subfigure}%
            \begin{subfigure}{.33\textwidth}
                \includegraphics[width=1.\linewidth]{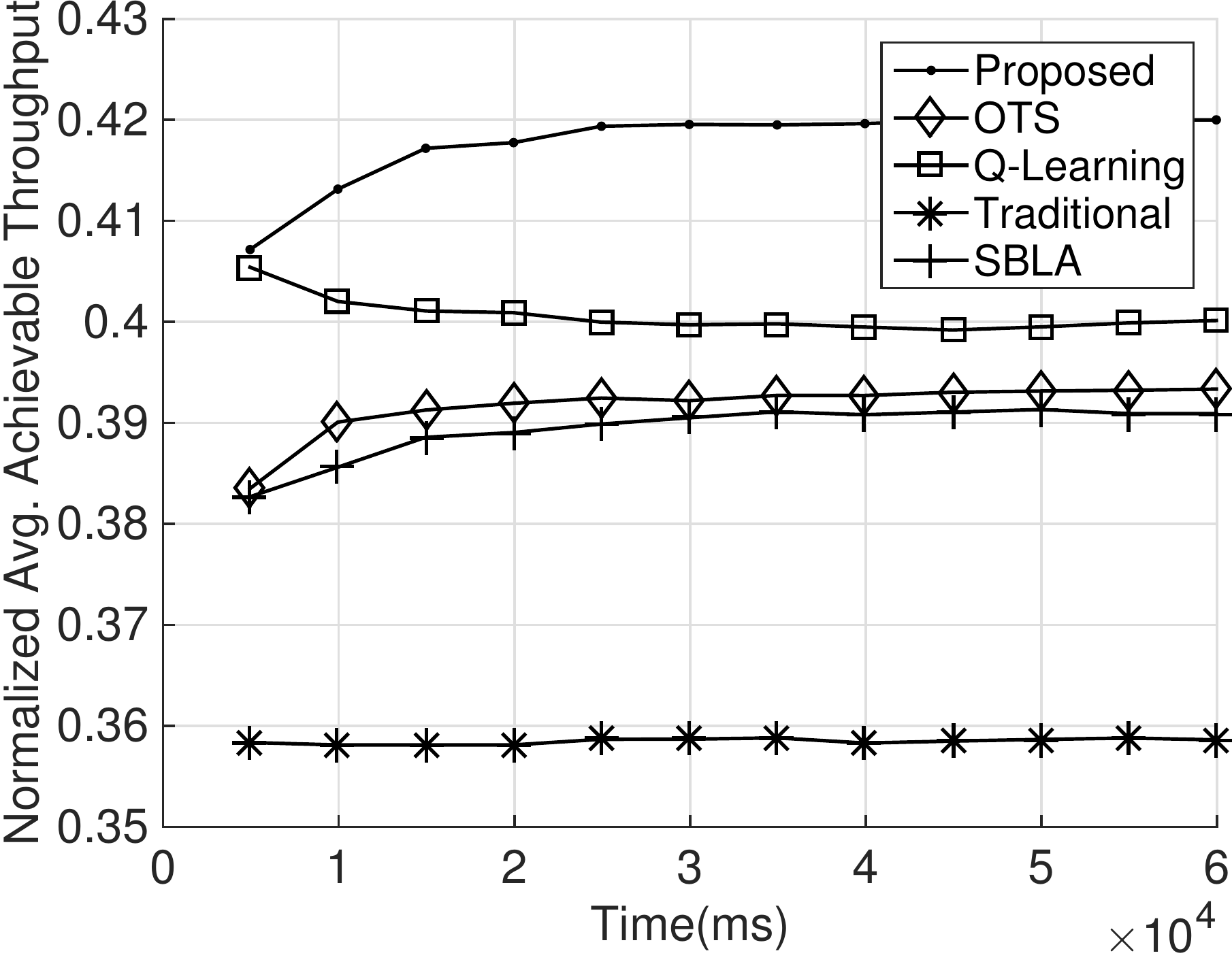}
                \caption{Avg. Achievable Throughput}
                \label{fig:DTMC_Beta_H-I_C05_F050_S03_TP}
            \end{subfigure}%
            \begin{subfigure}{.33\textwidth}
                \includegraphics[width=1.\linewidth]{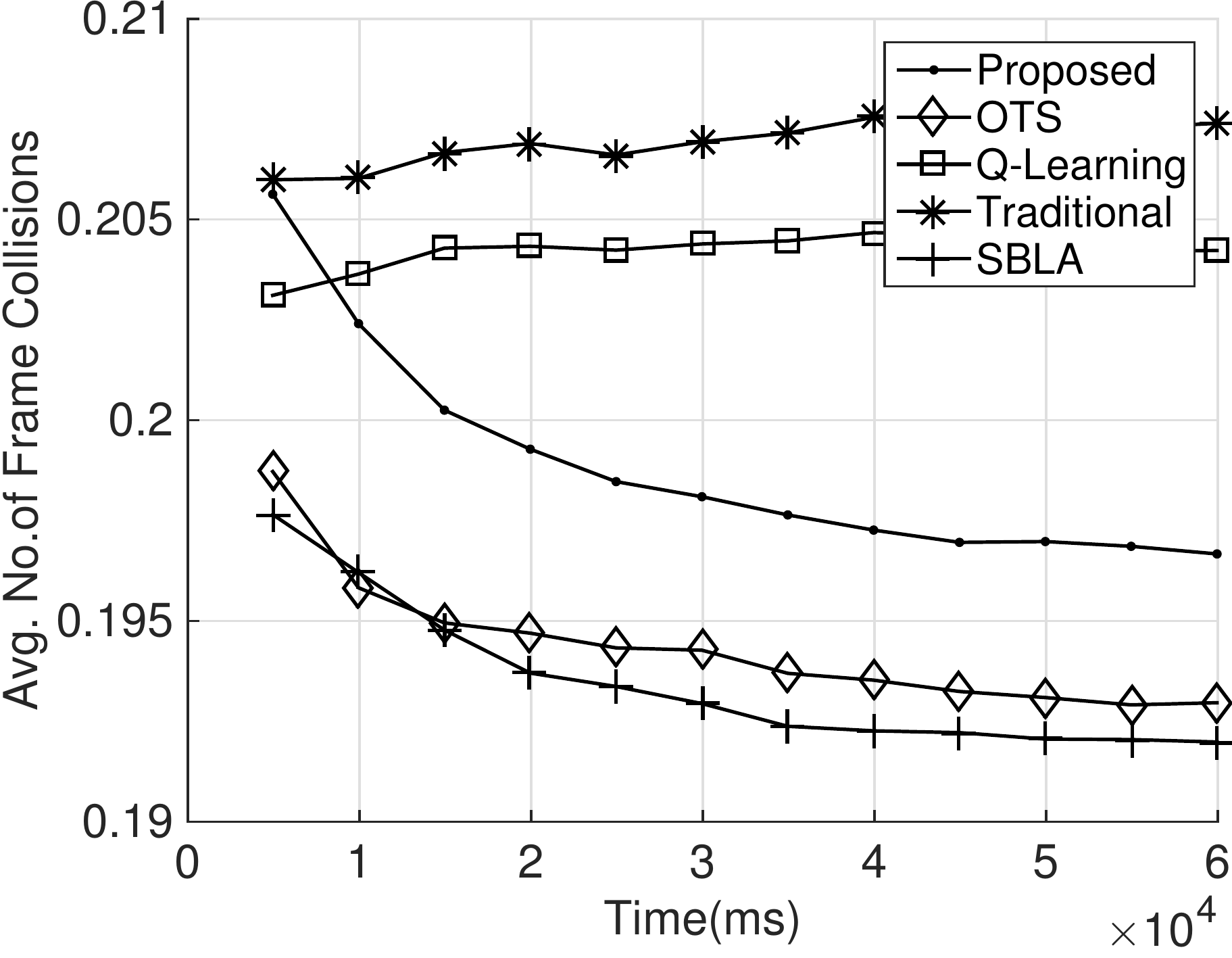}
                \caption{Avg. Frame Collision}
                \label{fig:DTMC_Beta_H-I_C05_F050_S03_FC}
            \end{subfigure}%
            \caption{Results for DTMC High Traffic Model (DTMC H-I) for frame size $50$ms for $5$ channels.}
            \label{fig:DTMC_Beta_H-I_C05_F050_S03}
        \end{figure*}
\else
\fi   

\ifCLASSOPTIONonecolumn
        \begin{figure*}[h]
            \begin{subfigure}{.33\textwidth}
                \includegraphics[width=1.\linewidth]{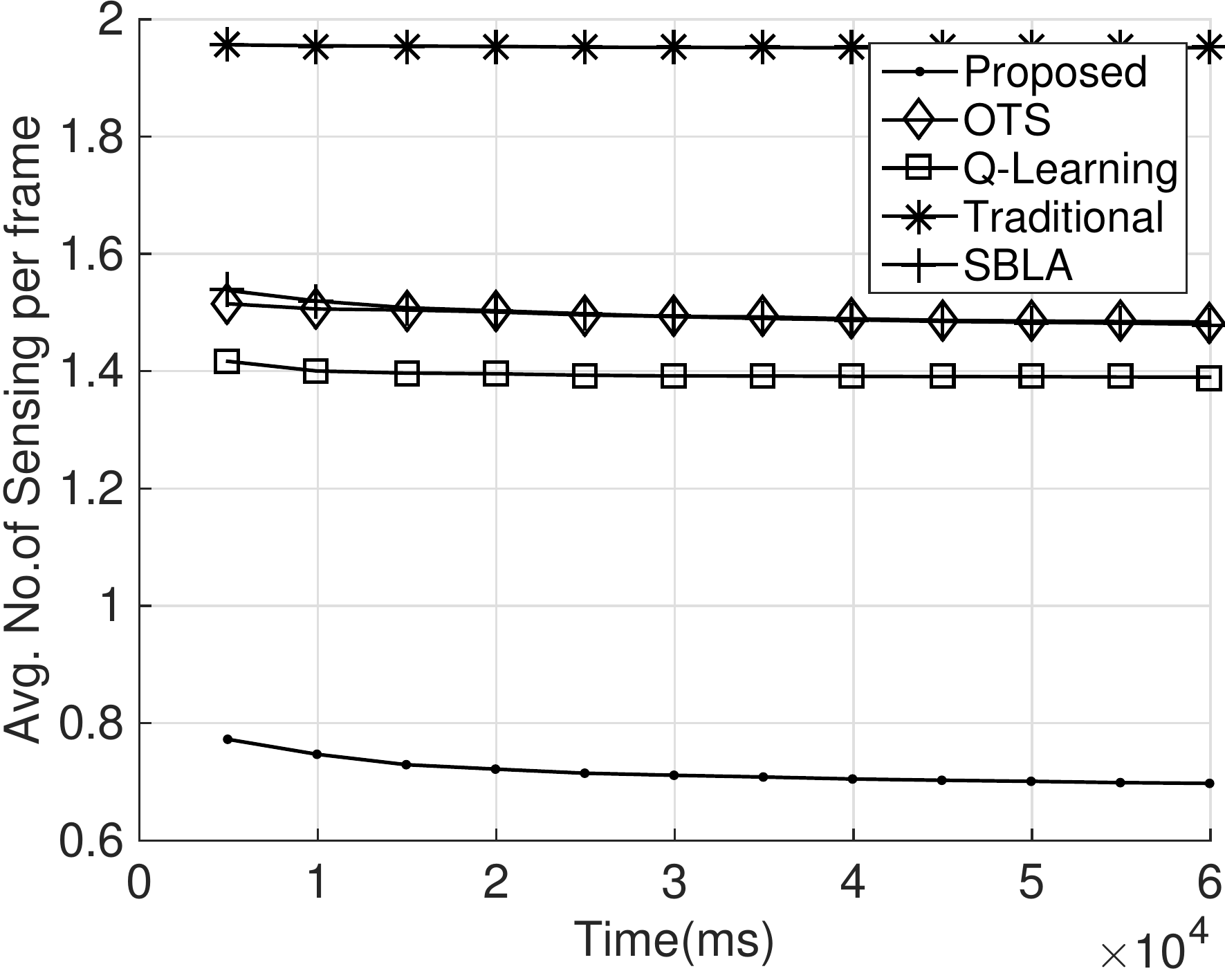}
                \caption{Avg. No.of Sensing per frame}
                \label{fig:Exp_C05_F050_S03_SE}
            \end{subfigure}%
            \begin{subfigure}{.33\textwidth}
                \includegraphics[width=1.\linewidth]{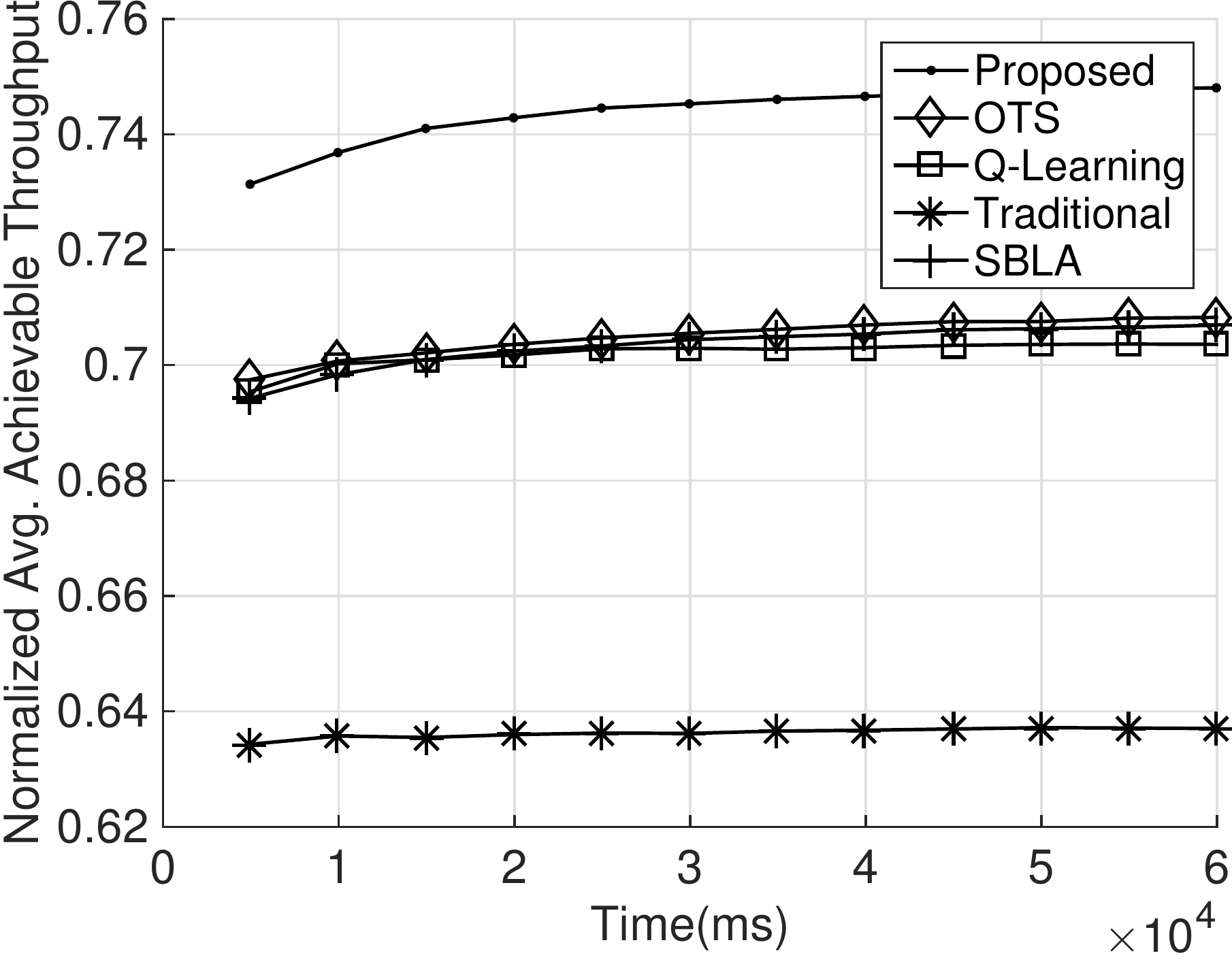}
                \caption{Avg. Achievable Throughput}
                \label{fig:Exp_C05_F050_S03_TP}
            \end{subfigure}%
            \begin{subfigure}{.33\textwidth}
                \includegraphics[width=1.\linewidth]{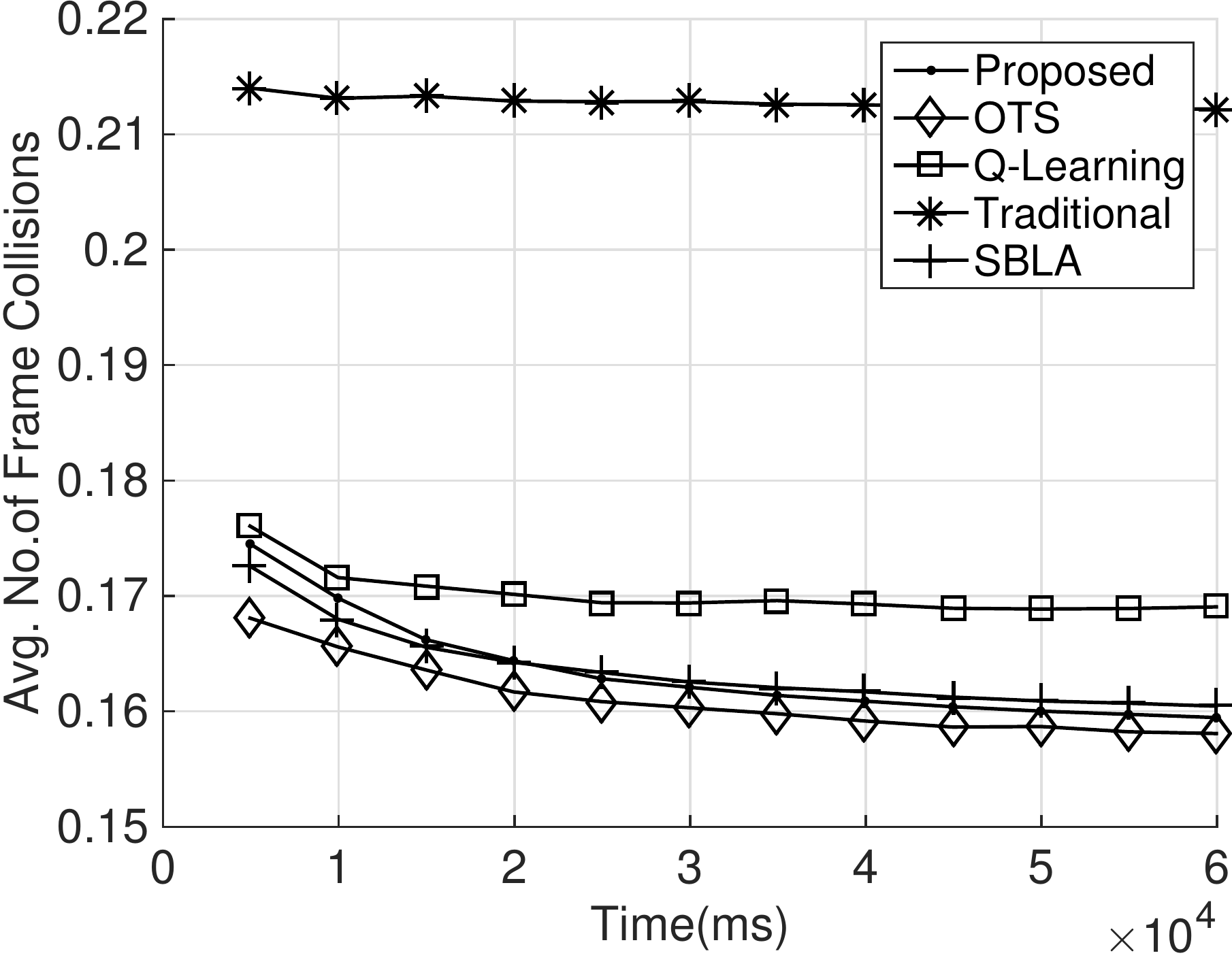}
                \caption{Avg. Frame Collision}
                \label{fig:Exp_C05_F050_S03_FC}
            \end{subfigure}%
            \caption{Results for Exponential Traffic Model for frame size $50$ms for $5$ channels.}
            \label{fig:Exp_C05_F050_S03}
        \end{figure*}
\else
\fi
        We also evaluate the proposed method in the Exponential traffic model which is capable of capturing bursts in traffic in a realistic fashion. Figure \ref{fig:Exp_C05_F050_S03} provides the results of the experiment. One immediate inference from the plots is that, due to the burstiness in traffic, the frame collisions is high when compared to the GPD and DTMC L-I and M-I models. This often presents a challenge to CR channel access algorithms as the channel can be occupied by a PU at the time of SU transmission even though it is recently sensed to be vacant. Due to our two-tier learning strategy we see that, the proposed algorithm is able to perform better in all the evaluation metrics. The proposed algorithm is able to cut-down the number of sensings required by more than half and is able to provide a $5\%$ improvement in throughput when compared to the best performing method in comparison. Interestingly, the frame collisions caused while following the proposed method is very close to the best algorithm in comparison. This improvement can be attributed to the design of the second layer of algorithm which assumed an exponential model of PU traffic.
        
\ifCLASSOPTIONonecolumn
        \begin{figure*}[h]
            \begin{subfigure}{.33\textwidth}
                \includegraphics[width=1.\linewidth]{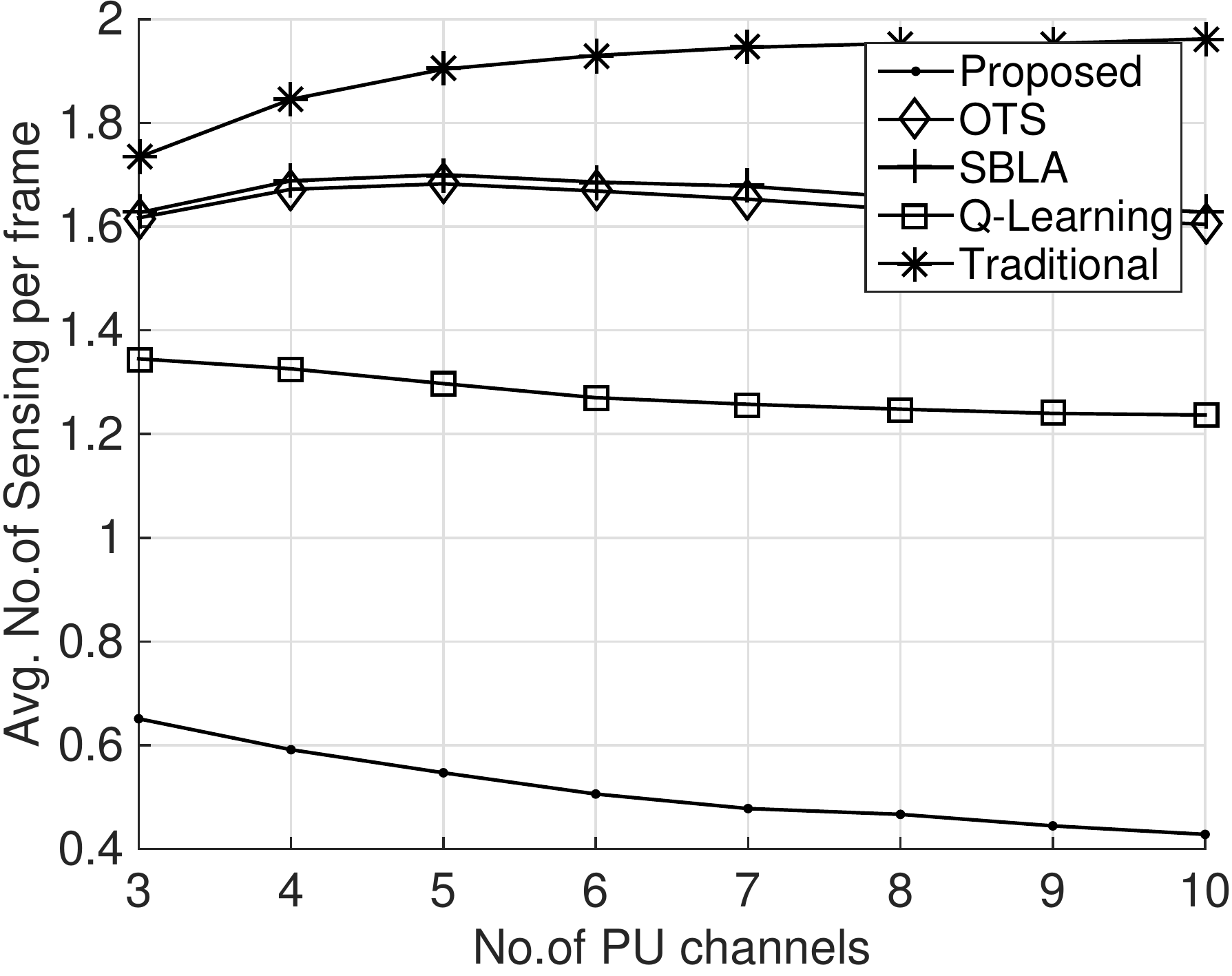}
                \caption{Avg. No.of Sensing per frame}
                \label{fig:CH_GPD_F050_S03_SE}
            \end{subfigure}%
            \begin{subfigure}{.33\textwidth}
                \includegraphics[width=1.\linewidth]{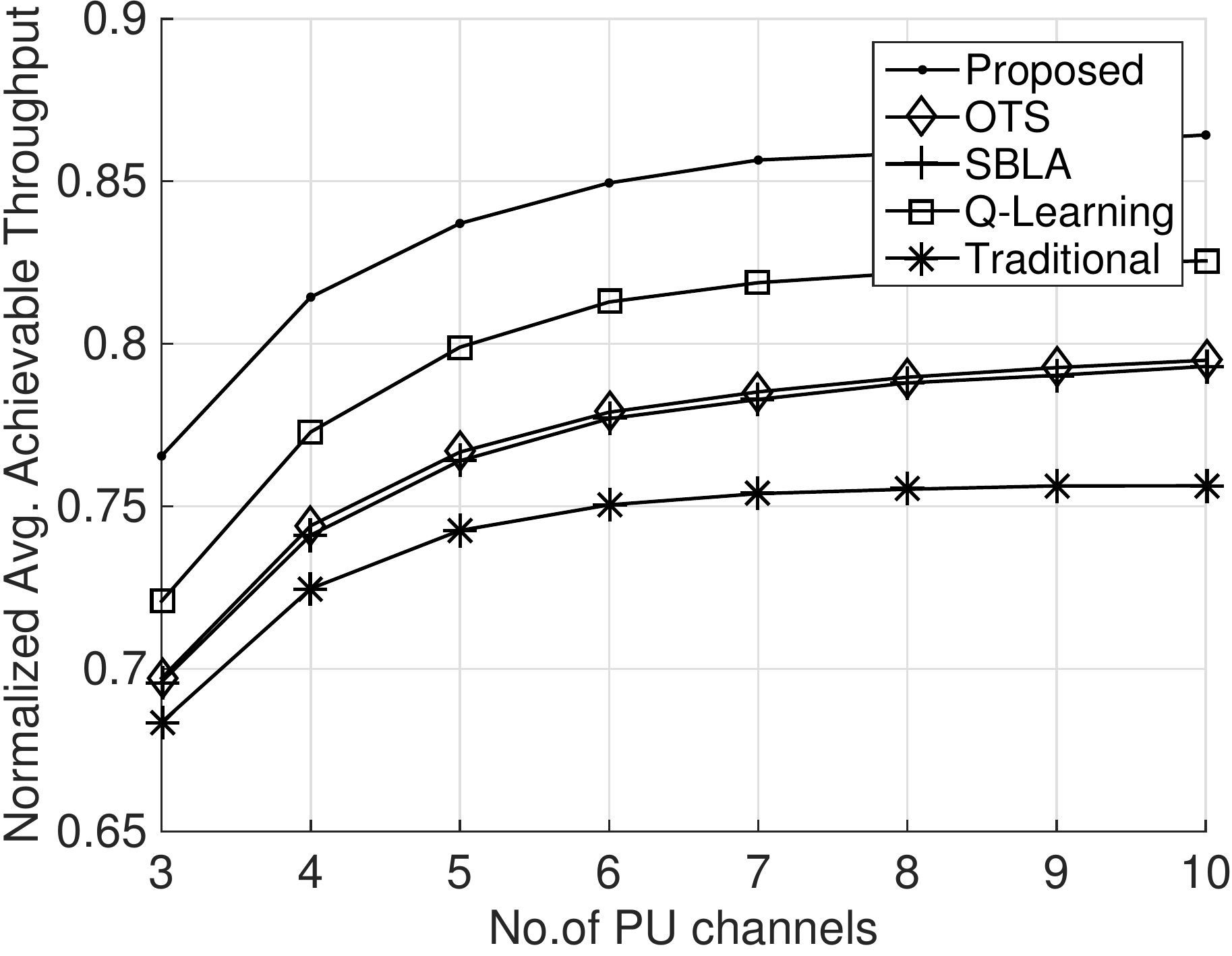}
                \caption{Avg. Achievable Throughput}
                \label{fig:CH_GPD_F050_S03_TP}
            \end{subfigure}%
            \begin{subfigure}{.33\textwidth}
                \includegraphics[width=1.\linewidth]{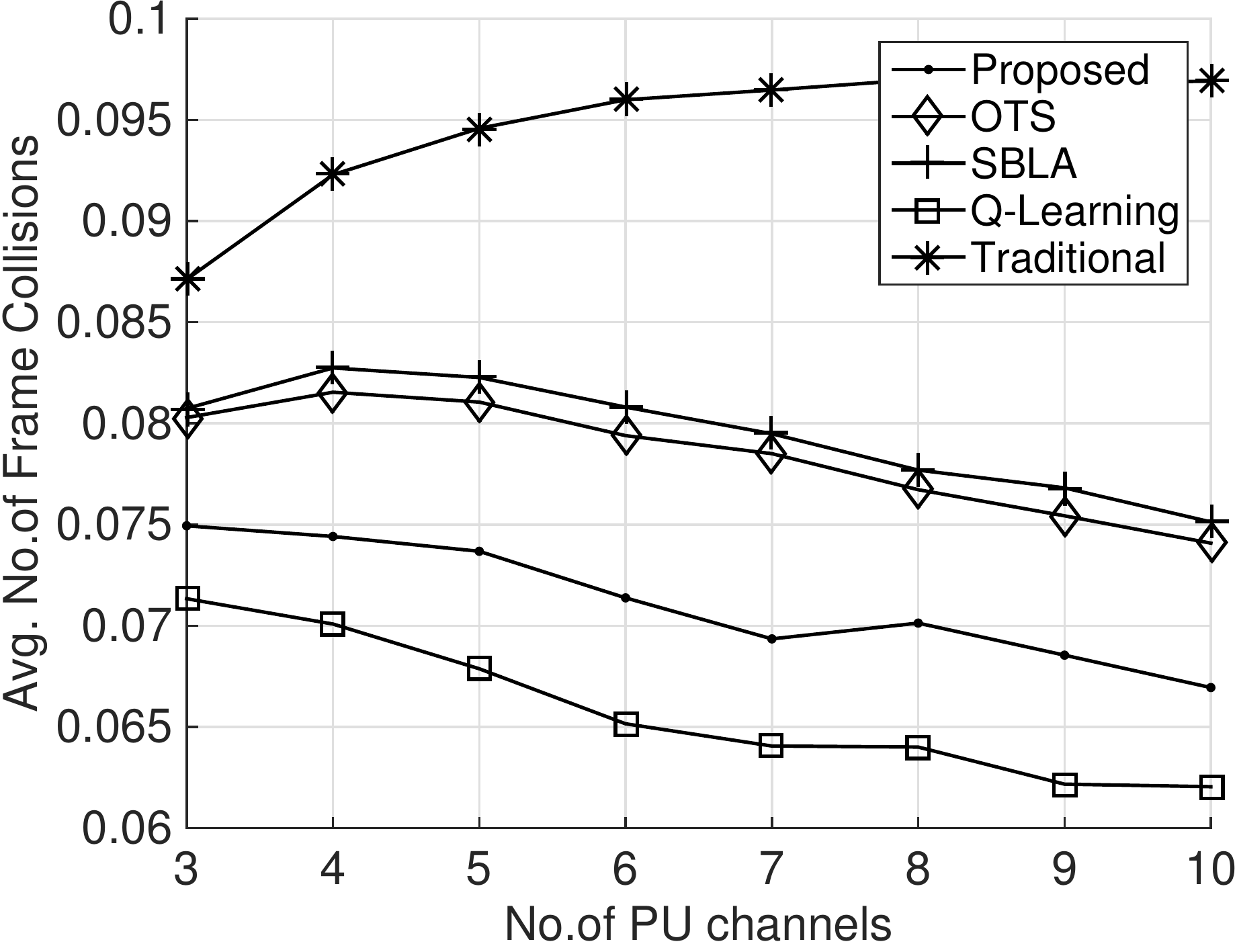}
                \caption{Avg. Frame Collision}
                \label{fig:CH_GPD_F050_S03_FC}
            \end{subfigure}%
            \caption{Results for GPD Traffic Model with varying number of channels for frame size $50$ms.}
            \label{fig:CH_GPD_F050_S03}
        \end{figure*}
\else
\fi
        \subsection{Effect of number of channels}
        In this section, we provide results for how each of the algorithms perform with varying number of available channels. In all the results provided below, we consider a system with a frame size of $50ms$ and a sensing period of $3ms$. Experiments were run for $60s$ and the reported values are the metrics measured at the end of the simulation time.
        
        Figure \ref{fig:CH_GPD_F050_S03} shows the effect of increasing the number of channels in GPD traffic model. We can observe that the proposed method is able to reduce the number of sensings required  atleast by $50\%$ and increase the throughput by atleast $5\%$ with only a negligible ($\sim0.05\%$) increase in PU traffic collision. Also we can observe that, as the number of available channel increases, all the methods are able to reduce the number of sensing required and increase the throughput with reduction in PU traffic collision. However, the proposed method still remains superior.

\ifCLASSOPTIONonecolumn
        \begin{figure*}[h]
            \begin{subfigure}{.33\textwidth}
                \includegraphics[width=1.\linewidth]{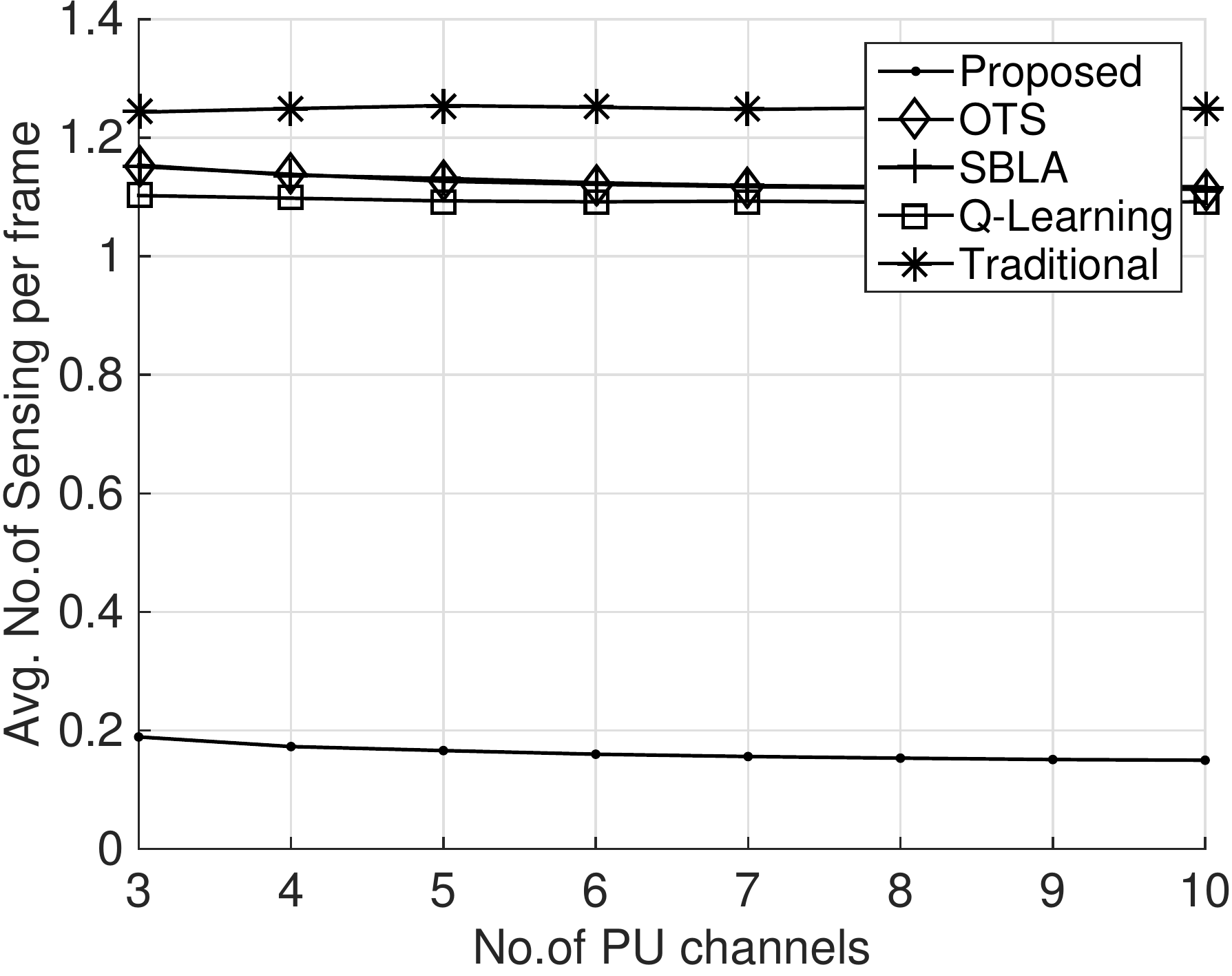}
                \caption{Avg. No.of Sensing per frame}
                \label{fig:CH_DTMC_Beta_L-I_F050_S03_SE}
            \end{subfigure}%
            \begin{subfigure}{.33\textwidth}
                \includegraphics[width=1.\linewidth]{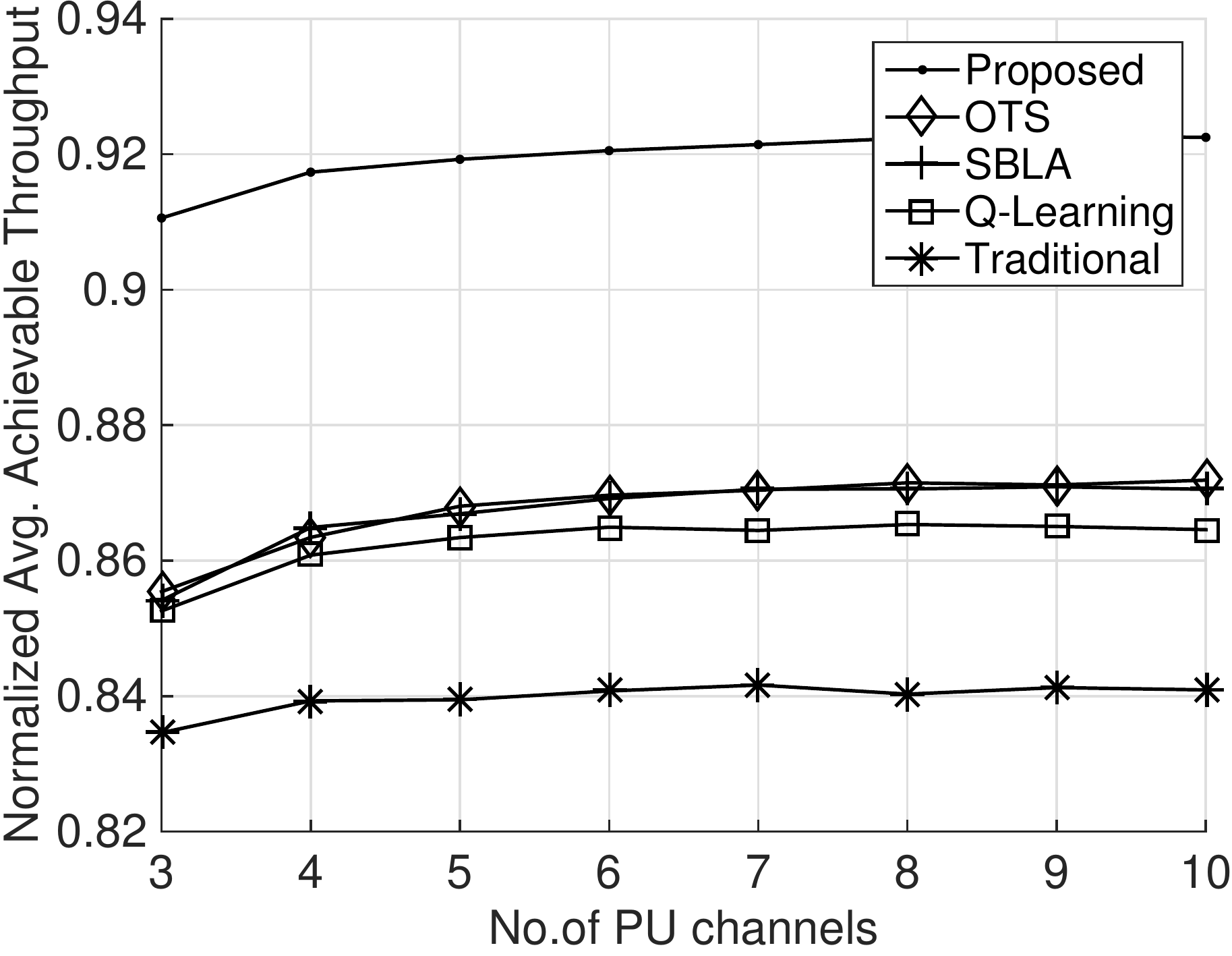}
                \caption{Avg. Achievable Throughput}
                \label{fig:CH_DTMC_Beta_L-I_F050_S03_TP}
            \end{subfigure}%
            \begin{subfigure}{.33\textwidth}
                \includegraphics[width=1.\linewidth]{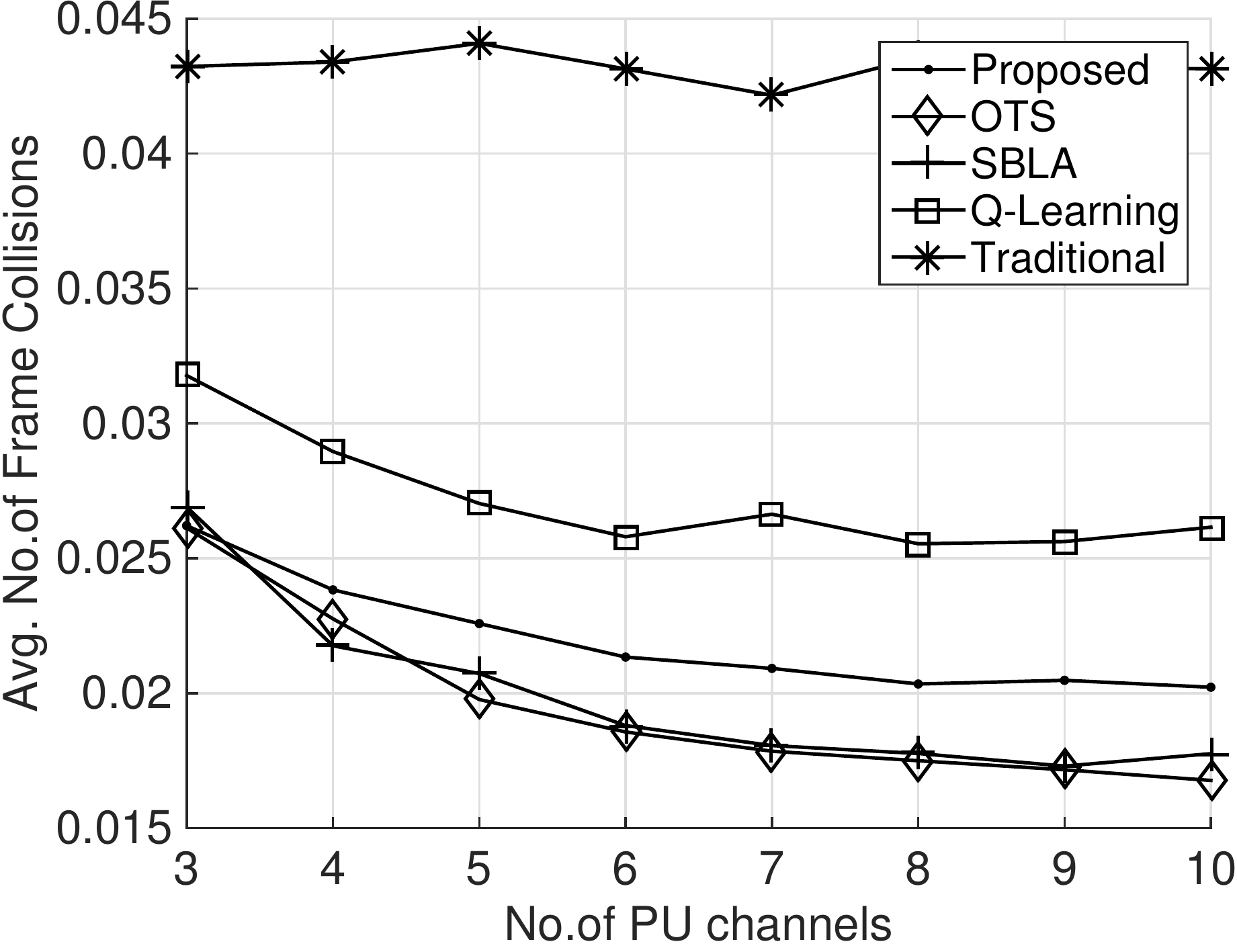}
                \caption{Avg. Frame Collision}
                \label{fig:CH_DTMC_Beta_L-I_F050_S03_FC}
            \end{subfigure}%
            \caption{Results for DTMC Low Traffic Model with varying number of channels for frame size $50$ms.}
            \label{fig:CH_DTMC_Beta_L-I_F050_S03}
        \end{figure*}
        
        \begin{figure*}[h]
            \begin{subfigure}{.33\textwidth}
                \includegraphics[width=1.\linewidth]{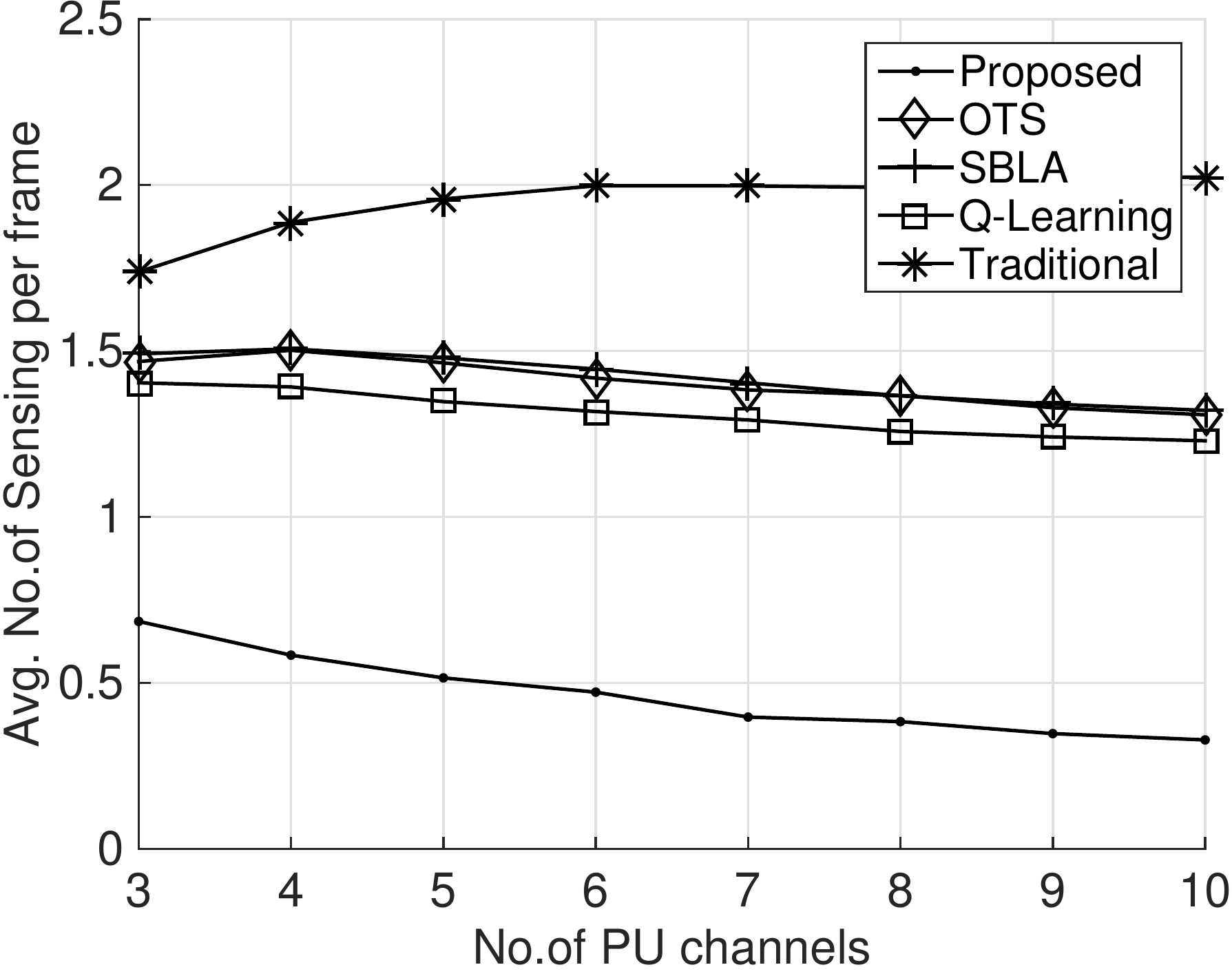}
                \caption{Avg. No.of Sensing per frame}
                \label{fig:CH_DTMC_Beta_M-I_F050_S03_SE}
            \end{subfigure}%
            \begin{subfigure}{.33\textwidth}
                \includegraphics[width=1.\linewidth]{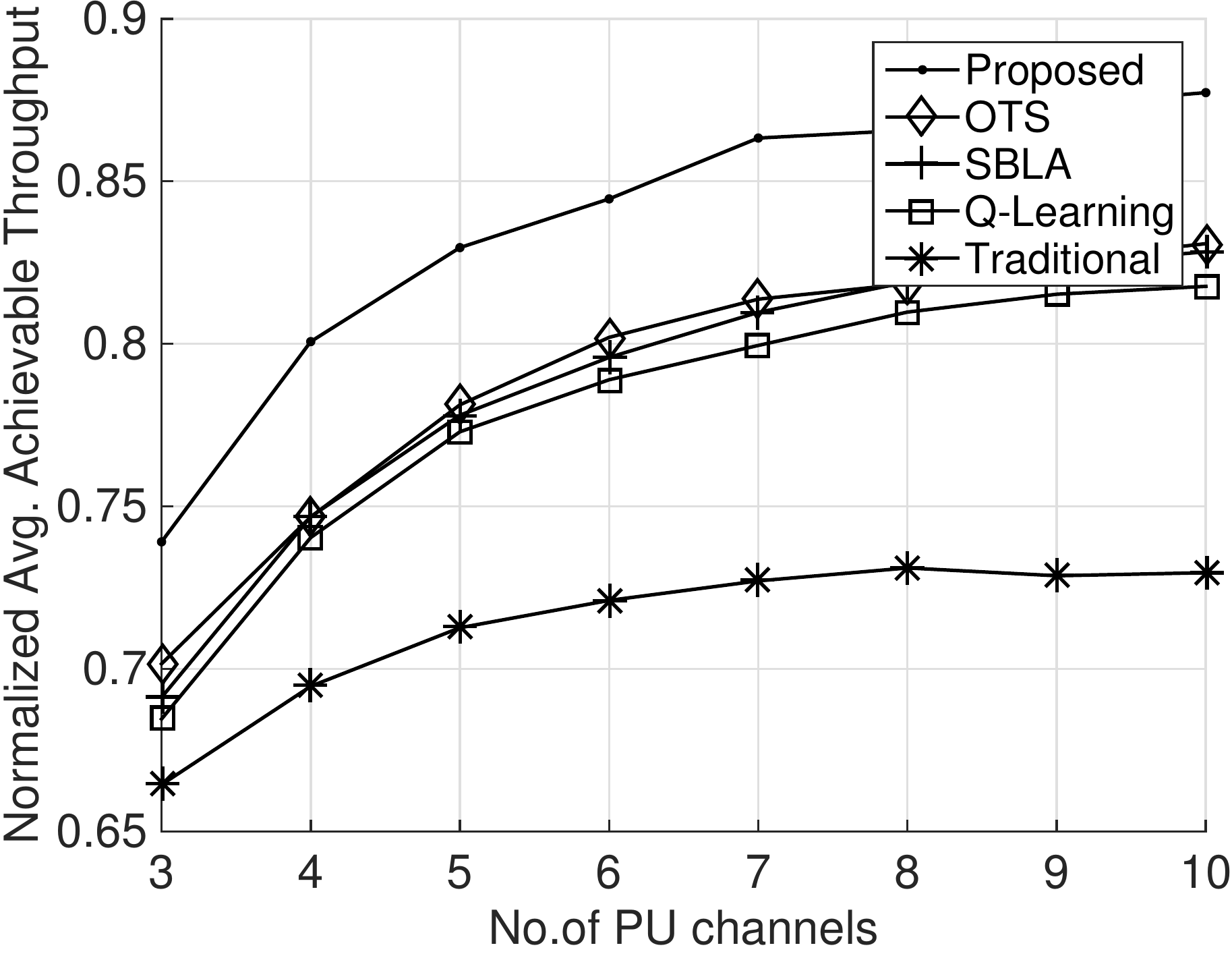}
                \caption{Avg. Achievable Throughput}
                \label{fig:CH_DTMC_Beta_M-I_F050_S03_TP}
            \end{subfigure}%
            \begin{subfigure}{.33\textwidth}
                \includegraphics[width=1.\linewidth]{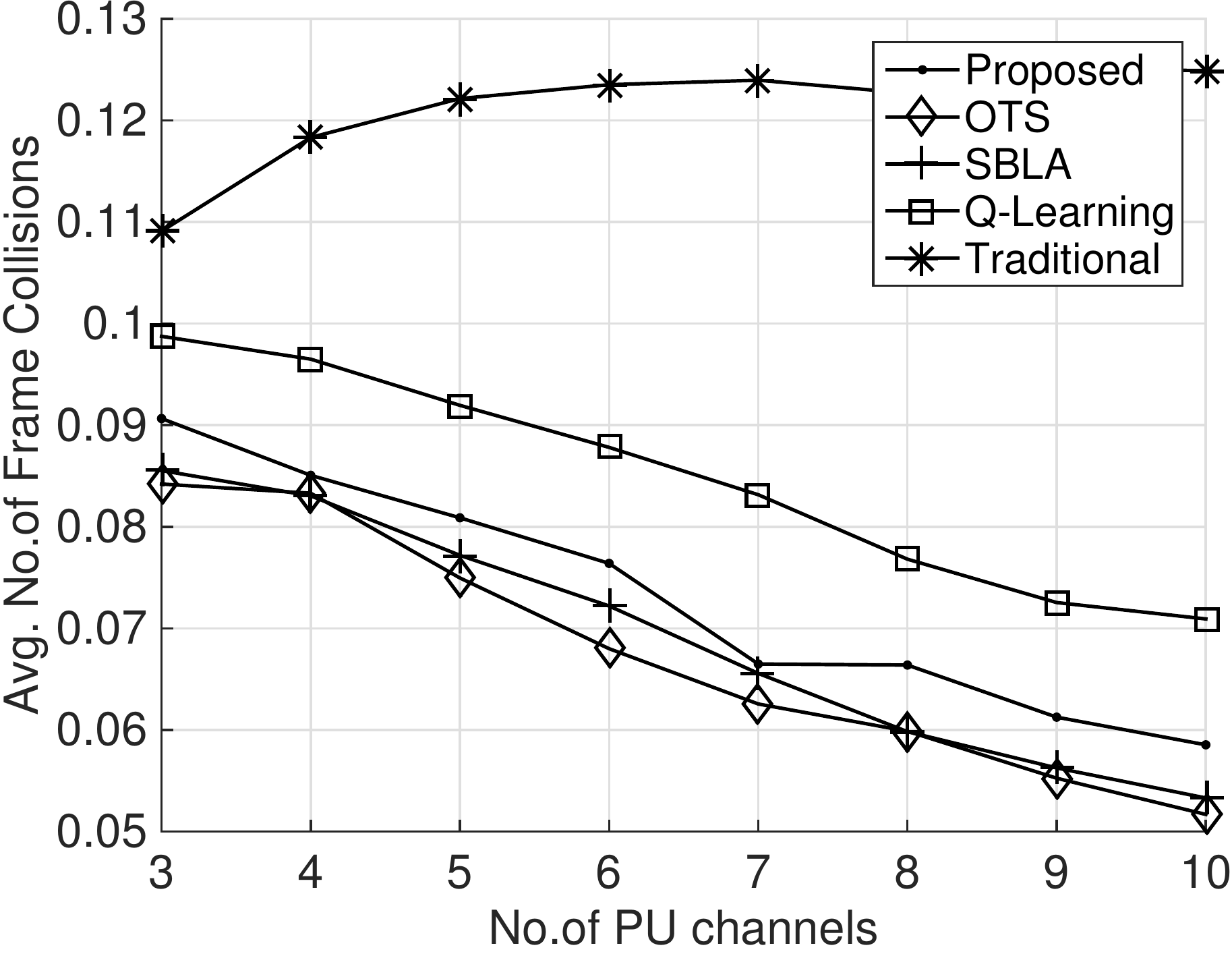}
                \caption{Avg. Frame Collision}
                \label{fig:CH_DTMC_Beta_M-I_F050_S03_FC}
            \end{subfigure}%
            \caption{Results for DTMC Medium Traffic Model with varying number of channels for frame size $50$ms.}
            \label{fig:CH_DTMC_Beta_M-I_F050_S03}
        \end{figure*}
        
\else
\fi

        Figures \ref{fig:CH_DTMC_Beta_L-I_F050_S03}-\ref{fig:CH_DTMC_Beta_M-I_F050_S03} show the results in a DTMC traffic model of two different traffic intensities with varying the number of channels. In both the cases, the proposed method is able to show a clear improvement in number of sensings required and throughput with negligible increase in PU collision compared to other RL techniques. 

        Figure \ref{fig:CH_Exp_F050_S03} shows the result of varying the number of channels in Exponential PU traffic model. In this model, again, the proposed method shows a clear improvement in number of sensings and throughput varying with number of channels. One of the important observations is that the frame collision with PU of the proposed method is almost same as the other best performing RL algorithms, even though the proposed method provides a clear improvement in other evaluation metrics.
        
        From all the simulation results provided, we can conclude that the proposed method is able to improve the throughput and number of sensings required by the SU with negligible increase in PU traffic collision. Although the second stage of our algorithm (estimating the number of frame periods to skip) assumes an exponential model for PU traffic, it performs well in other models also. Maybe one can further improve the performance with assumed model knowledge.
\ifCLASSOPTIONonecolumn
        \begin{figure*}[h]
            \begin{subfigure}{.33\textwidth}
                \includegraphics[width=1.\linewidth]{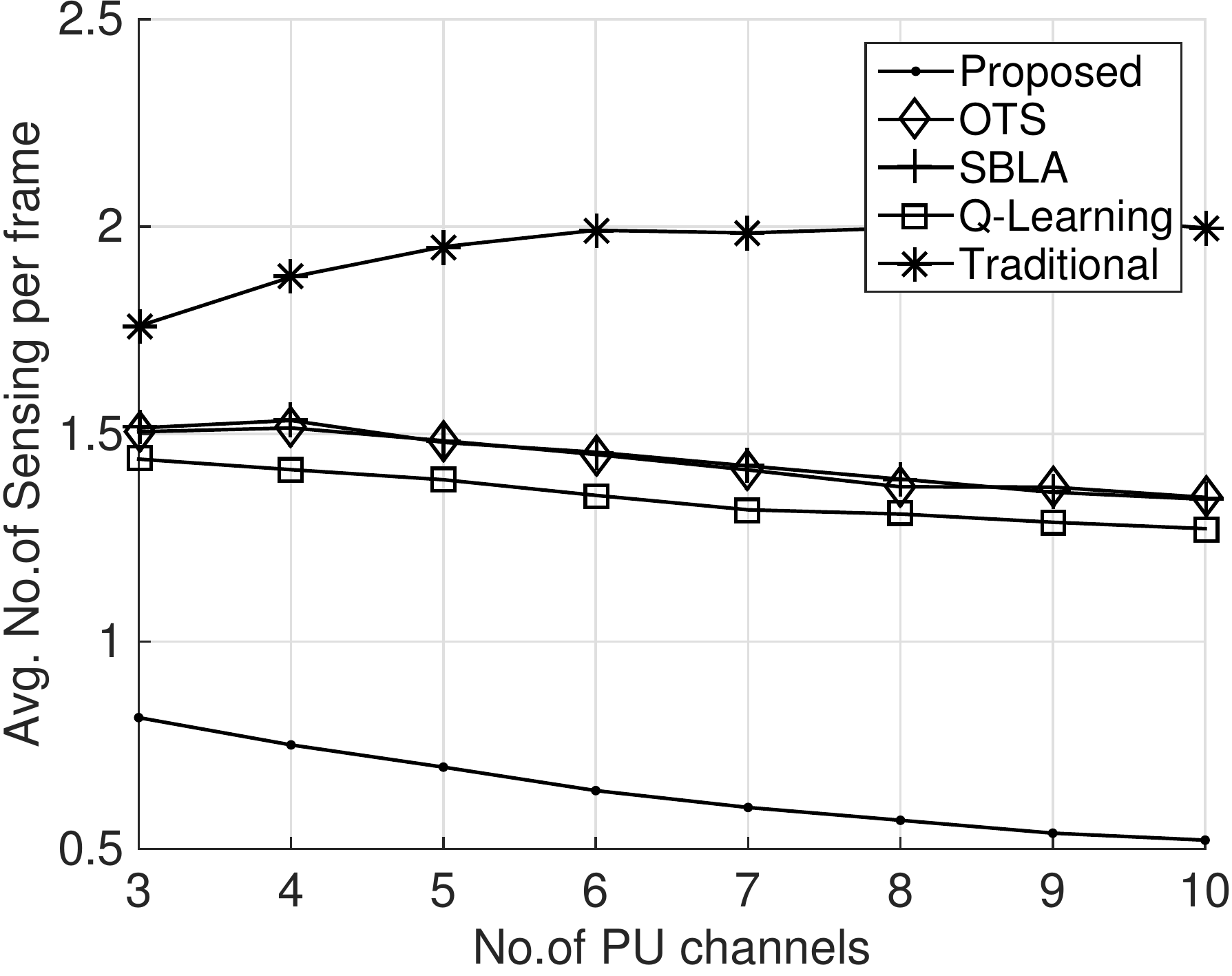}
                \caption{Avg. No.of Sensing per frame}
                \label{fig:CH_Exp_F050_S03_SE}
            \end{subfigure}%
            \begin{subfigure}{.33\textwidth}
                \includegraphics[width=1.\linewidth]{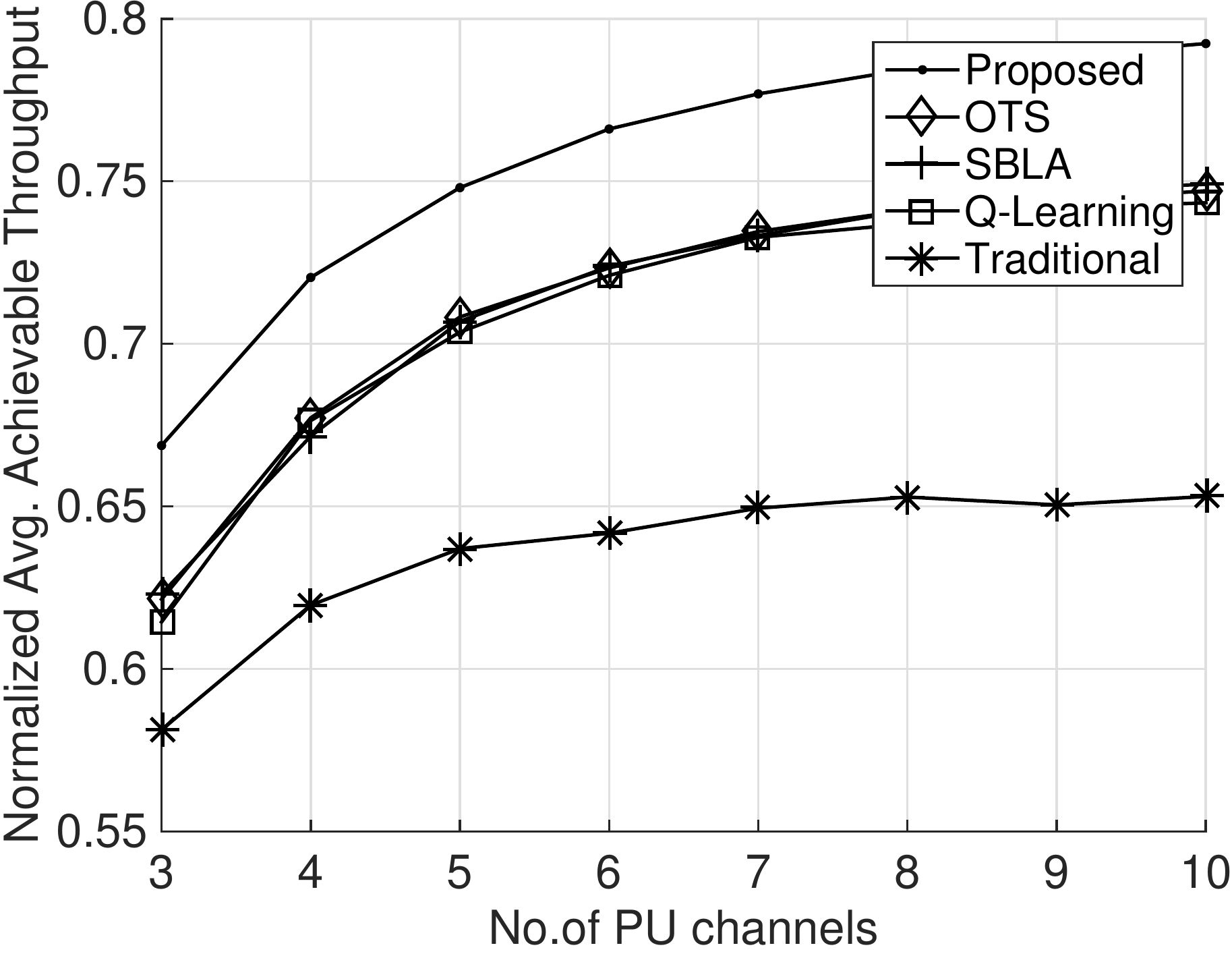}
                \caption{Avg. Achievable Throughput}
                \label{fig:CH_Exp_F050_S03_TP}
            \end{subfigure}%
            \begin{subfigure}{.33\textwidth}
                \includegraphics[width=1.\linewidth]{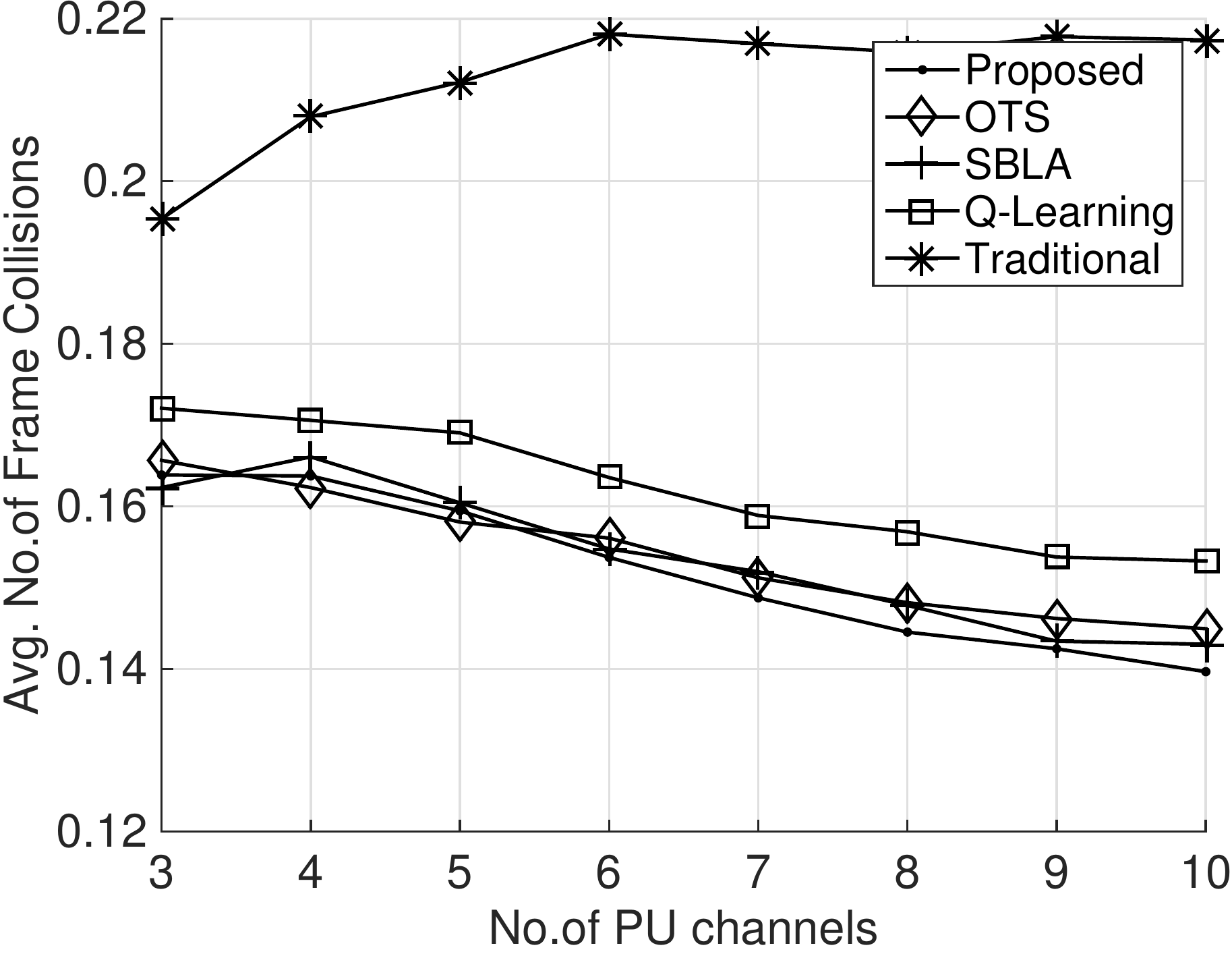}
                \caption{Avg. Frame Collision}
                \label{fig:CH_Exp_F050_S03_FC}
            \end{subfigure}%
            \caption{Results for Exponential Traffic Model with varying number of channels for frame size $50$ms.}
            \label{fig:CH_Exp_F050_S03}
        \end{figure*}
\else
\fi

 	\section{Conclusions} \label{sec:Conc}
 	    A two stage reinforcement learning approach is proposed for channel sensing and selection in cognitive radio networks. While the available RL literature in cognitive radio focuses on only deciding which channel to sense, we also try to learn how often to sense the selected channels. Judicious sensing as suggested by our approach leads to higher throughput, lower energy consumption (due to significant reduction in sensing) and better spectrum utilization with negligible increase in primary interference as compared to methods that employ sensing in every frame. Further, the focus of this work has been to reduce sensing, while keeping the interference to PU comparable to the existing algorithms. An interesting extension could be to incorporate a constraint on the amount of interference allowed to PU in our learning algorithm, so that SU can keep PU collisions at a fixed level. While our work considers a single user scenario, recently there has been significant interest in multi-user setting where a combinatorial bandit approach is employed. In this setting, the possible combination of channel allocations are considered as the arms of the bandit. However, to the best of our knowledge, the prospect of not sensing a frame altogether, when the PU statistics are known, is not considered. Another extension would be to propose a variation of our algorithm for the combinatorial bandit setting. We believe that even greater gains can be obtained by judicious sensing in the multi-user setting.
	
    \bibliographystyle{IEEEtran}
	\bibliography{IEEEabrv,library.bib}

\end{document}